\documentclass[12pt]{article}
\usepackage{epsfig}
\usepackage{amsmath}
\usepackage{amssymb}
\usepackage{amsthm}
\usepackage{epstopdf}
\usepackage{graphicx}
\usepackage{enumerate}
\usepackage{mathtools}
\usepackage{authblk}
\usepackage{array}
\usepackage{cite}
\usepackage{color}
\usepackage[margin=0.9in]{geometry}
\usepackage[caption=false]{subfig}
\usepackage{float}
\normalsize \baselineskip 16pt
\newcommand {\beq} {\begin{equation}}
\newcommand {\eeq} {\end{equation}}

\newcolumntype{P}[1]{>{\centering\arraybackslash}p{#1}}
\newcolumntype{M}[1]{>{\centering\arraybackslash}m{#1}}

\DeclareMathOperator{\sech}{sech}

\begin{document}
\title{Discrete breathers in a mechanical metamaterial }
\author[1]{Henry Duran}
\author[2,3]{Jes\'us Cuevas--Maraver}
\author[4]{Panayotis\ G.\ Kevrekidis}
\author[1]{Anna Vainchtein}
\affil[1]{\small Department of Mathematics, University of Pittsburgh, Pittsburgh, Pennsylvania 15260, USA}
\affil[2]{\small Grupo de F\'{\i}sica No Lineal. Departamento de F\'{\i}sica Aplicada I, Escuela Polit\'{e}cnica Superior, Universidad de Sevilla. C/Virgen de \'Africa, 7, Sevilla 41011, Spain}
\affil[3]{\small Instituto de Matem\'{a}ticas de la Universidad de Sevilla (IMUS). Edificio Celestino Mutis, Avda. Reina Mercedes s/n, 41012-Sevilla, Spain}
\affil[4]{\small Department of Mathematics and Statistics, University of
Massachusetts, Amherst, MA 01003-9305, USA}

\maketitle

\begin{abstract}

  We consider a previously experimentally realized discrete model that describes a mechanical metamaterial consisting of a chain of pairs of rigid units connected by flexible hinges. Upon analyzing the linear band structure of the model, we identify parameter regimes in which this system may possess discrete breather solutions with frequencies inside the gap between optical and acoustic dispersion bands. We compute numerically exact solutions of this type for several different parameter regimes and investigate their properties and stability. Our findings demonstrate that upon appropriate parameter tuning within experimentally tractable ranges, the system exhibits a plethora of discrete breathers, with multiple branches of solutions that feature period-doubling and symmetry-breaking bifurcations, in addition to other mechanisms of stability change such as saddle-center and
  Hamiltonian Hopf bifurcations. The relevant stability analysis is corroborated by direct numerical computations examining the dynamical
  properties of the system and paving the way for potential further experimental exploration of this rich nonlinear dynamical lattice setting.

\end{abstract}

\section{Introduction}
Mechanical metamaterials are engineered structures~\cite{Bertoldi17,Chen14,Christensen15,Wu21,Dalela21,Zadpoor16}  whose macroscopic properties are primarily controlled by their geometry and may differ considerably from those of their building blocks \cite{Clausen15,Hussein14,Kochmann17,Pishvar20,Xia19}. In recent years, there has been a lot of interest in nonlinear dynamic response of flexible mechanical metamaterials, a new class of engineered materials that exploit large deformation and mechanical instabilities of their components to yield a desired collective response \cite{Bertoldi17,Deng21}. Examples include metamaterials consisting of rotating rigid elements that are connected by flexible hinges \cite{Deng17,Deng2018}, multistable kirigami sheets \cite{Jin20}, chains of bistable shells \cite{Vasios21} and beams \cite{Raney16}, as well as origami-inspired \cite{Yasuda16,Yasuda19} and linkage-based \cite{Zareei20} deployable structures. These metamaterials can be designed to enable potential applications that include morphing surfaces, soft robotics, reconfigurable devices, mechanical logic and controlled energy absorption \cite{Chen18,Deng20,Fang17,Novelino20,Preston19,Rafsanjani18,Anzel14}. Recent studies have demonstrated that metamaterials of this type can be designed to control propagation of a variety of nonlinear waves \cite{Deng2018,Deng21,Raney16,Rafsanjani19,Yasuda16,Yasuda20,Zareei20}.

In this work we consider the flexible mechanical metamaterial that was recently studied experimentally and theoretically in \cite{Deng2018,Deng19,Deng21}. The experimentally realized system, schematically shown in Fig.~\ref{fig:chain}, consists of a chain of pairs of cross-type rigid units made of LEGO bricks and connected by thin flexible polyester or plastic hinges \cite{Deng2018,Deng19}. Under certain assumptions, the system can be described by a discrete model that assigns two degrees of freedom to each pair of rigid units: horizontal displacement and rotation. This system, in turn, can be approximated at the continuum level by a Klein-Gordon equation with cubic nonlinearity, a nonlinear wave-bearing system that possesses both soliton and cnoidal wave solutions \cite{Deng21}. In \cite{Deng2018}, the authors use a combination of experiments, direct numerical simulations of the discrete system and analysis of the continuum model to investigate traveling waves in this system that correspond to elastic vector solitons on the continuum level. They demonstrate that the metamaterial lattice may be designed to exhibit amplitude gaps where soliton propagation is forbidden, which, in turn, enables the design of soliton splitters and diodes. In \cite{Deng19} the anomalous nature of the soliton collisions in this system is explored. These developments clearly illustrate
the promise of this type of nonlinear lattice in regards to the wave dynamics and interactions.

In this work we demonstrate that in certain parameter regimes the discrete system derived in \cite{Deng2018} also exhibits a plethora of spatially localized,
time-periodic patterns in the form of discrete breathers. These structures
arise in terms of the angle and strain (relative displacement) variables. Similar to discrete breathers observed in other settings, including Josephson junction arrays \cite{Binder00,Trias00}, forced-damped arrays of coupled pendula \cite{Cuevas09}, electrical lattices \cite{Palmero11,English13,remoissenet}, micromechanical systems \cite{Sato03,Sato04,Sato06} and granular chains \cite{Boechler10,Chong14,Zhang17,granularBook}, they emerge as a result of the interplay of (discrete) dispersion and nonlinearity \cite{Aubry97,Aubry06,Flach08}
and appear to be generic in the gaps of the linear excitation spectrum, as we will
show below.

To construct such solutions for the metamaterial system, we start by analyzing the dispersion relation, which features optical and acoustic branches. We show that when the angle $\phi_0$ measuring the vertical offset between the neighboring horizontal hinges, takes values in certain parameter-dependent intervals, there is a frequency gap between the optical and acoustic branches that enables existence of discrete breathers. We then use the iterations of Newton's  method with a suitable initial guess and
(once converged to a member of a solution family)
parameter continuation to compute branches of discrete breather solutions that have frequency inside the gap and either bifurcate from or exist near the edges of the optical and acoustic bands. Stability of the obtained solutions is investigated using Floquet analysis.

As our first example, we consider the system parameters from \cite{Deng2018} and show that in this case a branch of discrete breather solutions bifurcates from the edge of the optical band provided that the offset angle $\phi_0$ is above a certain threshold. Floquet analysis reveals that this branch eventually undergoes period-doubling bifurcations, and we compute the corresponding double-period solutions and investigate their stability.

As a second example, we consider a different set of parameters that enables existence of breathers for small offset angles $\phi_0$ in a certain interval. Choosing two different values in this interval, we compute branches of solutions that exist in the gap between the optical and acoustic bands. Here, our computation reveals a complex bifurcation diagram in the energy-frequency plane involving branches of symmetric and asymmetric discrete breather solutions and emergence of instability modes associated with real and complex Floquet multipliers. In particular, we find that the onset of real instability can take place via collisions of complex multipliers, as well as symmetry-breaking and period-doubling bifurcations. Another mechanism involves critical points of the breather's energy as a function of its frequency (effectively, a saddle-center
bifurcation), in line with the stability criterion established in \cite{Kevrekidis2016} for discrete breathers in Fermi-Pasta-Ulam and Klein-Gordon lattices. We investigate the fate of some of the unstable solutions by perturbing them along the corresponding eigenmodes and show that in each case the ensuing dynamic evolution leads to a discrete breather that is effectively stable if one neglects the presence of small-magnitude complex eigenvalues. The computed primary branches have a snake-like form with multiple turning points, and the solution profiles often evolve in a nontrivial way along a branch, e.g., via the emergence of additional peaks or troughs in the strain and angle variables describing a discrete breather with even symmetry. Some features of the obtained bifurcation diagrams are reminiscent of the ``snake-and-ladder" patterns observed in other nonlinear systems \cite{Beck2009,Chong2009,TAYLOR201014}, although a detailed exploration of such
a phenomenology is outside the scope of the present work.

The rest of the paper is organized as follows. In Sec.~\ref{sec:formulation} we introduce the discrete model and formulate the problem. Analysis of the dispersion relation for the linearized system is presented in Sec.~\ref{sec:dispersion}. In Sec.~\ref{sec:period_doubling} we discuss a solution branch bifurcating from the edge of the optical mode for the parameter values in \cite{Deng2018} and exhibiting period-doubling bifurcations. In Sec.~\ref{sec:snaking} we consider another set of parameters and describe the complex bifurcation diagrams involving branches that exist in the gap between the optical and acoustic bands. Concluding remarks can be found in Sec.~\ref{sec:conclusions}.

\section{Problem formulation}
\label{sec:formulation}
\begin{figure}[htb!]
\centering
\includegraphics[width=\textwidth]{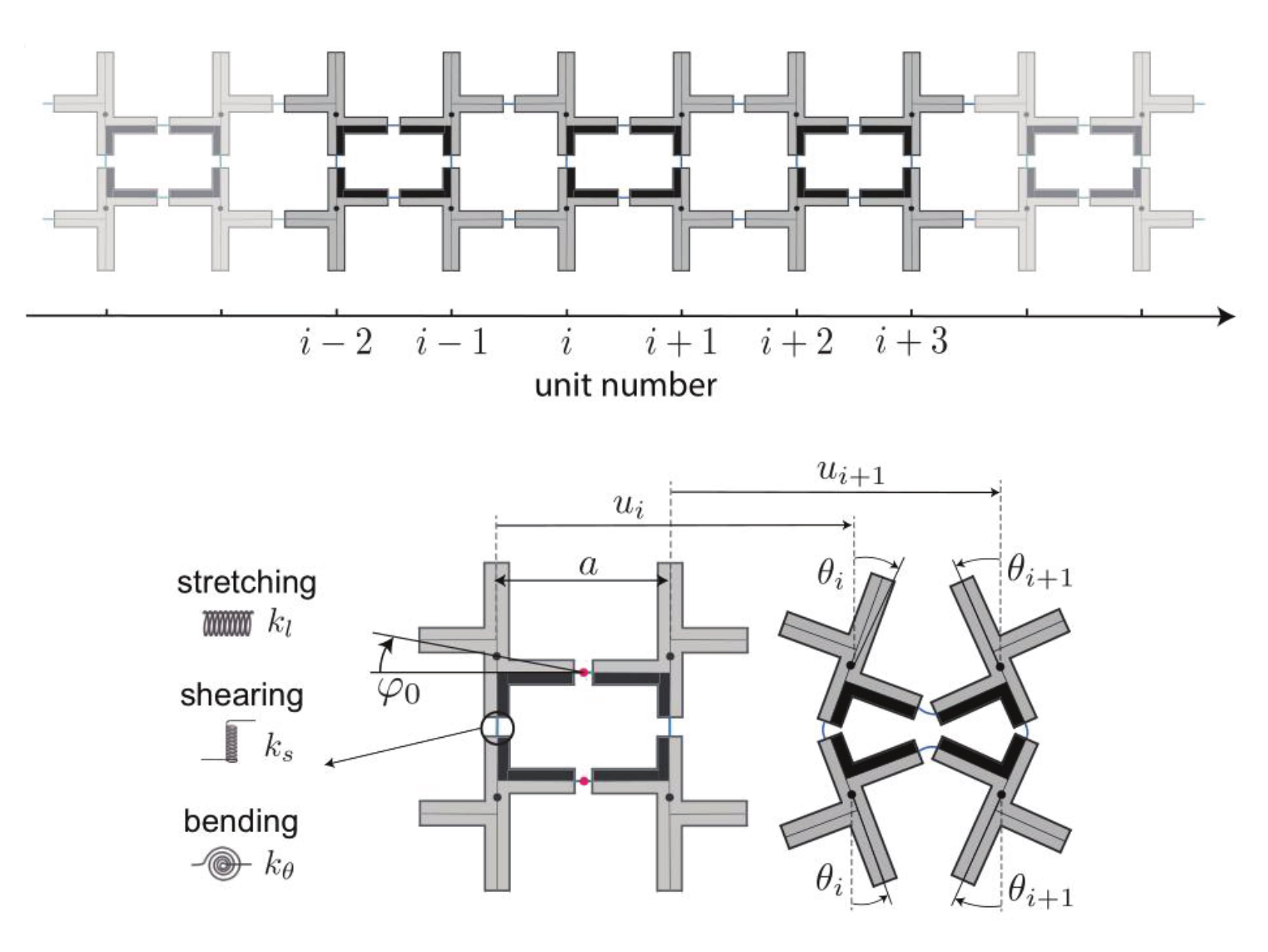}
\caption{\footnotesize Top panel: discrete chain of cross-shaped rigid units. Bottom panel: kinematic variables and parameters. Adapted from Supplementary Figure 6 in \cite{Deng2018}.}
\label{fig:chain}
\end{figure}
Motivated by experimental and theoretical investigations in \cite{Deng2018}, we consider a chain that consists of $2 \times N$ cross-type rigid units of mass $m$ and moment of inertia $J$ connected by thin flexible hinges, as shown in Fig.~\ref{fig:chain}. The neighboring horizontal hinges are shifted in the vertical direction by $a\tan\phi_0$, where $a$ is the center-to-center horizontal distance between the neighboring units (see the bottom panel of Fig.~\ref{fig:chain}). The hinges are modeled as a combination of three linear springs, with stiffness parameters $k_l$, $k_s$ and $k_\theta$ corresponding to longitudinal stretching, shearing and bending, respectively. Following \cite{Deng2018}, we describe the dynamics of the system by two degrees of freedom for $n$-th vertical pair of rigid units: the longitudinal displacement $u_n(t)$ and the rotation angle $\theta_n(t)$ at time $t$. Here it is assumed \cite{Deng2018} that the two rigid units in each vertical pair have the same displacement and rotate by the same amount but in the opposite directions, with positive direction of rotation defined as shown in the bottom panel of Fig.~\ref{fig:chain}. Introducing dimensionless variables
\[
\tilde{u}_n=\dfrac{u_n}{a}, \qquad \tilde{t}=t\sqrt{\dfrac{k_l}{m}}
\]
and parameters
\[
\alpha=\dfrac{a}{2\cos\phi_0}\sqrt{\dfrac{m}{J}}, \qquad K_s=\dfrac{k_s}{k_l}, \qquad K_\theta=\dfrac{4k_\theta\cos^2\phi_0}{k_l a^2},
\]
one obtains \cite{Deng2018}
\beq
\begin{split}
&\ddot{u}_n =
u_{n+1} - 2u_n + u_{n-1} - \frac{\cos(\theta_{n+1}+\phi_0) - \cos(\theta_{n-1}+\phi_0)}{2\cos(\phi_0)} \\
&\dfrac{1}{\alpha^2}\ddot{\theta}_n=
-K_{\theta}(\theta_{n+1}+4\theta_n+\theta_{n-1}) + K_s\cos(\theta_n+\phi_0)\bigg(\sin(\theta_{n+1}+\phi_0) + \sin(\theta_{n-1}+\phi_0)\\
&-2\sin(\theta_n+\phi_0)\bigg)-\sin(\theta_n+\phi_0) \bigg(2\cos(\phi_0)(u_{n+1}-u_{n-1}) + 4\cos(\phi_0)-\cos(\theta_{n+1}+\phi_0)\\
&-2\cos(\theta_n+\phi_0)-\cos(\theta_{n-1}+\phi_0)\bigg),
\end{split}
\label{eq:EoM}
\eeq
where we dropped the tildes in the rescaled displacement and time variables, and double dot denotes second time derivative. The total energy of the system is \cite{Deng2018}
\beq
H = \sum_{n=1}^N \left[(\Delta_n^l)^2 + K_s(\Delta_n^s)^2 + \frac{K_{\theta}}{8\cos^2(\phi_0)}(2(\delta_n^h)^2+(\delta_n^v)^2)+\dot{u}_n^2 +\frac{1}{4\alpha^2\cos^2(\phi_0)}\dot{\theta}_n^2\right],
\label{eq:TotalEnergy}
\eeq
where
\[
\begin{split}
\delta_n^h &= \theta_{n+1} + \theta_n, \qquad \delta_n^v = 2\theta_n,\\
\Delta_n^l &= u_{n+1}-u_n + \frac{1}{2\cos(\phi_0)}\left[2\cos(\phi_0)-\cos(\phi_0 + \theta_n) - \cos(\phi_0 + \theta_{n+1})\right],\\
\Delta_n^s &= \frac{1}{2\cos(\phi_0)}\left[\sin(\phi_0+\theta_{n+1}) - \sin(\phi_0 + \theta_n) \right]
\end{split}
\]
characterize the deformation associated with horizontal ($\delta_n^h$, $\Delta_n^l$, $\Delta_n^s$) and vertical ($\delta_n^v$) hinges.

We consider discrete breather (DB) solutions of \eqref{eq:EoM}. These are time-periodic nonlinear waves with frequency $\omega$ and the corresponding period $T=2\pi/\omega$,
\beq
u_n(t+T)=u_n(t), \qquad \theta_n(t+T)=\theta_n(t)
\label{eq:breather}
\eeq
that are spatially localized in terms of strain
\beq
w_n(t)=u_{n+1}(t)-u_n(t)
\label{eq:strain}
\eeq
and angle $\theta_n(t)$ variables.

\section{Dispersion Relation}
\label{sec:dispersion}
To obtain conditions for existence of DB solutions bifurcating from the linear modes,
we need to study the linear spectrum of the problem first. To that effect,
we linearize \eqref{eq:EoM} around the undeformed configuration. This yields
\beq
\begin{split}
&\ddot{u}_n = u_{n+1} - 2u_n + u_{n-1} +\dfrac{1}{2}\tan\phi_0(\theta_{n+1}-\theta_{n-1}) \\
&\dfrac{1}{\alpha^2}\ddot{\theta}_n=(K_s\cos^2\phi_0-\sin^2\phi_0-K_\theta)(\theta_{n+1}+\theta_{n-1})\\
&-2(K_s\cos^2\phi_0+\sin^2\phi_0+2K_\theta)\theta_n-\sin(2\phi_0)(u_{n+1}-u_{n-1}).
\end{split}
\label{eq:linearized}
\eeq
Considering plane-wave solutions $u_n(t) = U e^{i(kn-\omega t)}$, $\theta_j(t) = \Theta e^{i(kn-\omega t)}$ of \eqref{eq:linearized} in the limit of an infinite chain ($N \to \infty$), we obtain the following solvability condition:
\[
\begin{split}
&\left(\omega^2 - 4\sin^2\frac{k}{2}\right)\left[\frac{\omega^2}{\alpha^2} - 2(K_{\theta}-K_s\cos^2\phi_0+\sin^2\phi_0)\cos(k) - 2(2K_{\theta}+K_s\cos^2\phi_0+\sin^2\phi_0)\right]\\
&-2\tan(\phi_0)\sin(2\phi_0)\sin^2(k)=0,
\end{split}
\]
which yields explicit (but cumbersome) expressions for the acoustic, $\omega_-(k)$, and optical, $\omega_+(k)$, branches of the dispersion relation between the wave number $k$ and the frequency $\omega$. The two branches satisfy
\beq
\omega_{-}(0)=0, \qquad \omega_+(0)=\alpha\sqrt{2(3 K_\theta +2 \sin^2\phi_0)}>0.
\label{eq:k_zero}
\eeq

\begin{figure}[!htb]
\centering
\subfloat[$\phi_0 = 0.55<\phi_0'$]
{{\includegraphics[width=0.33\textwidth]{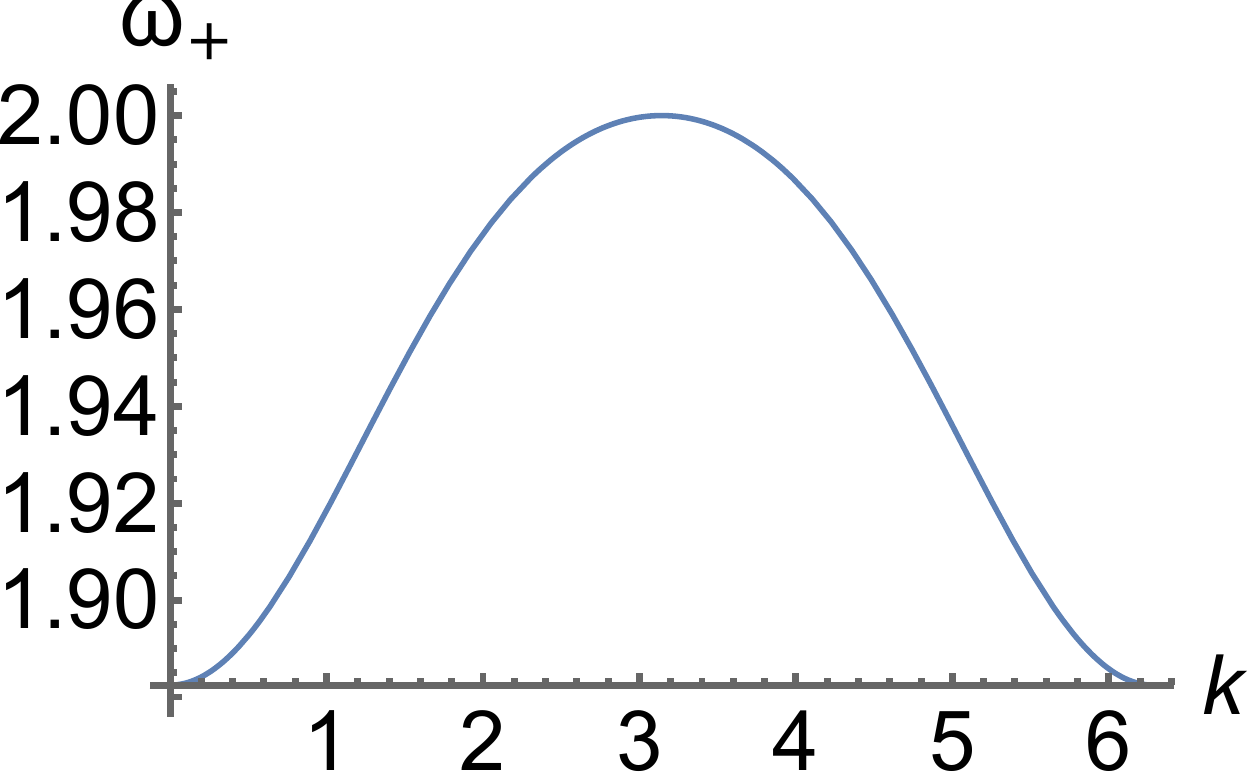}}}
\subfloat[$\phi_0=\phi_0'=0.5736$]
{{\includegraphics[width=0.33\textwidth]{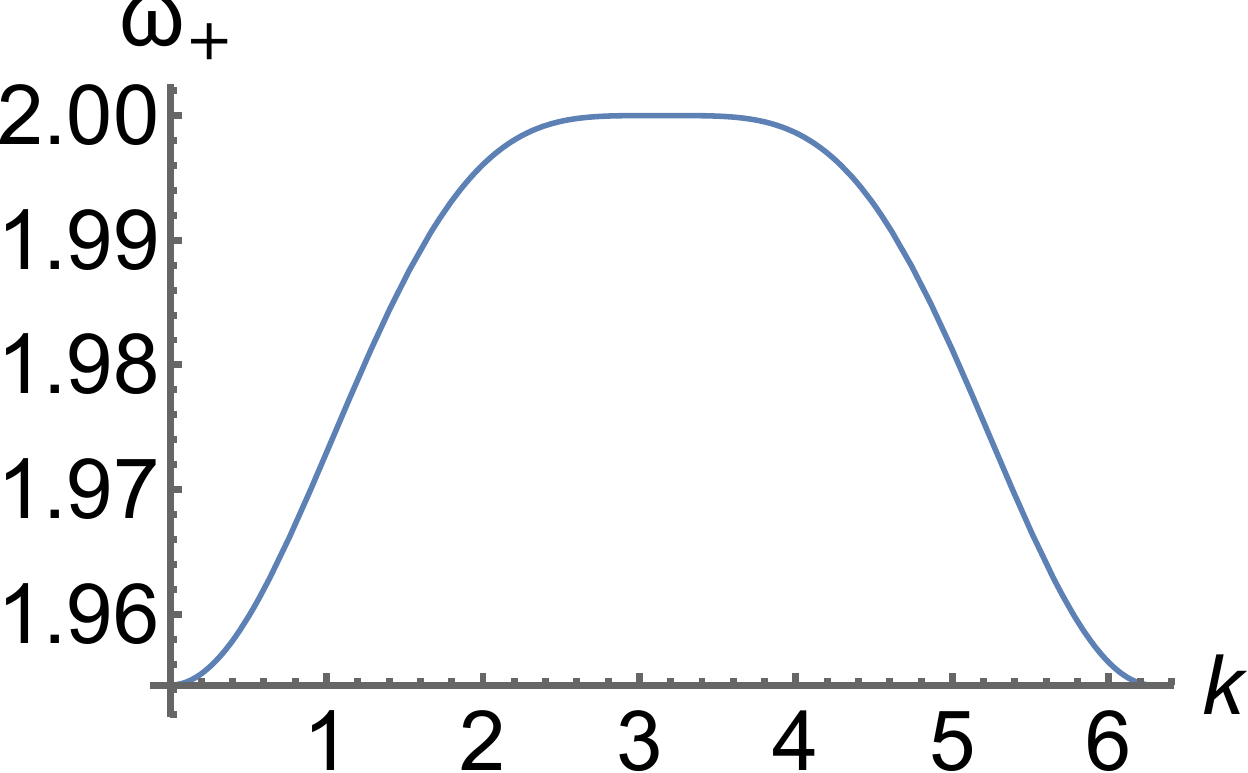}}}
\subfloat[$\phi_0'<\phi_0=0.58<\phi_0''$]
{{\includegraphics[width=0.33\textwidth]{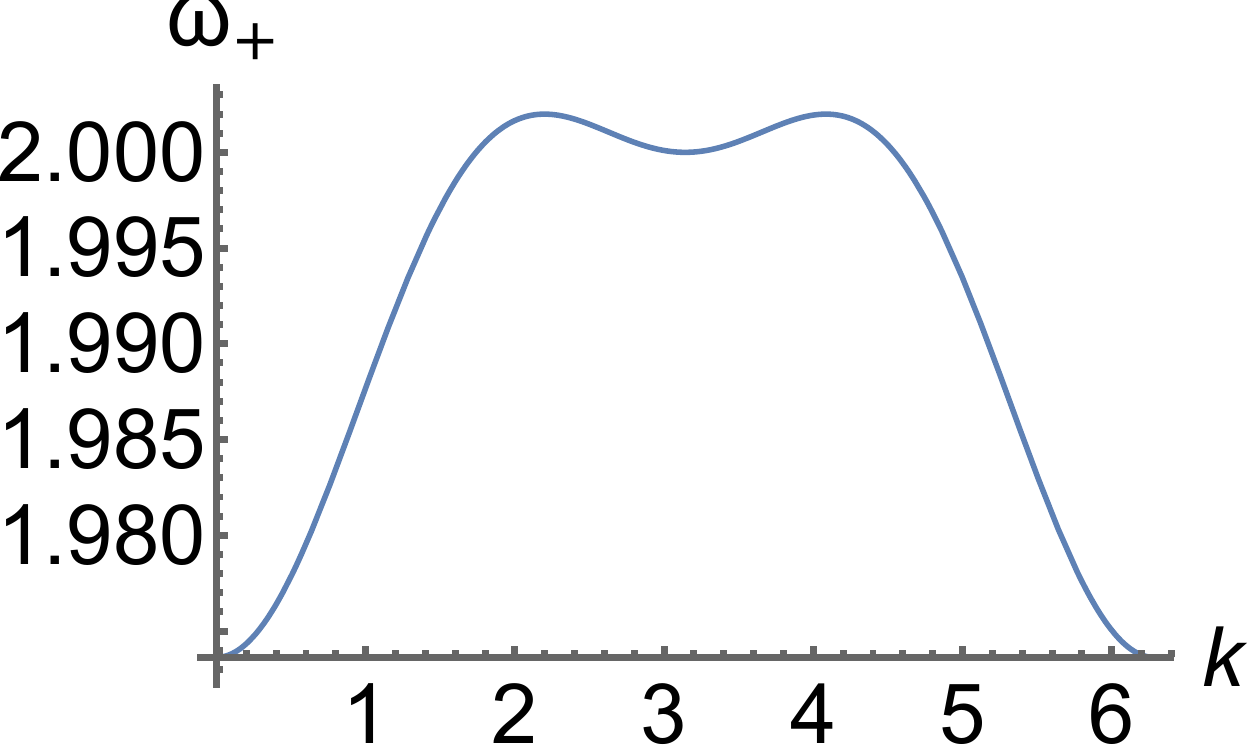}}}\\
\subfloat[$\phi_0=\phi_0''=0.5888$]
{{\includegraphics[width=0.33\textwidth]{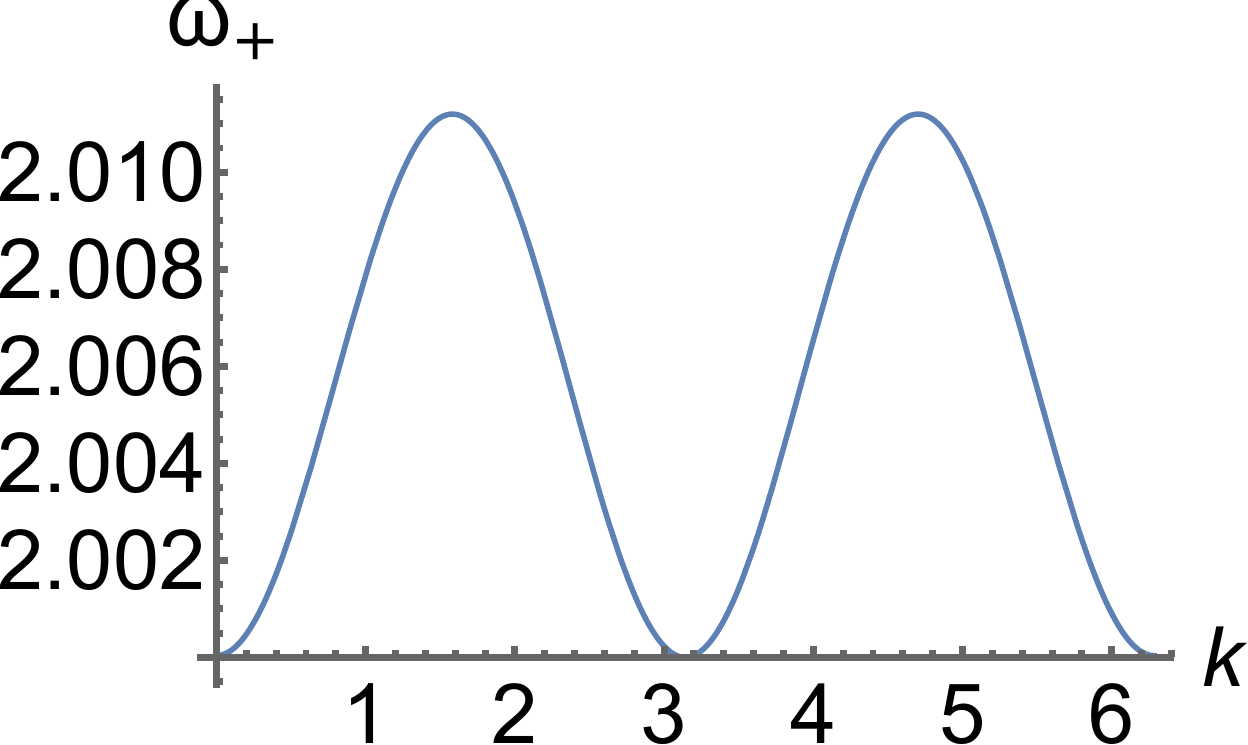}}}
\subfloat[$\phi_0''<\phi_0=0.595<\phi_0'''$]
{{\includegraphics[width=0.33\textwidth]{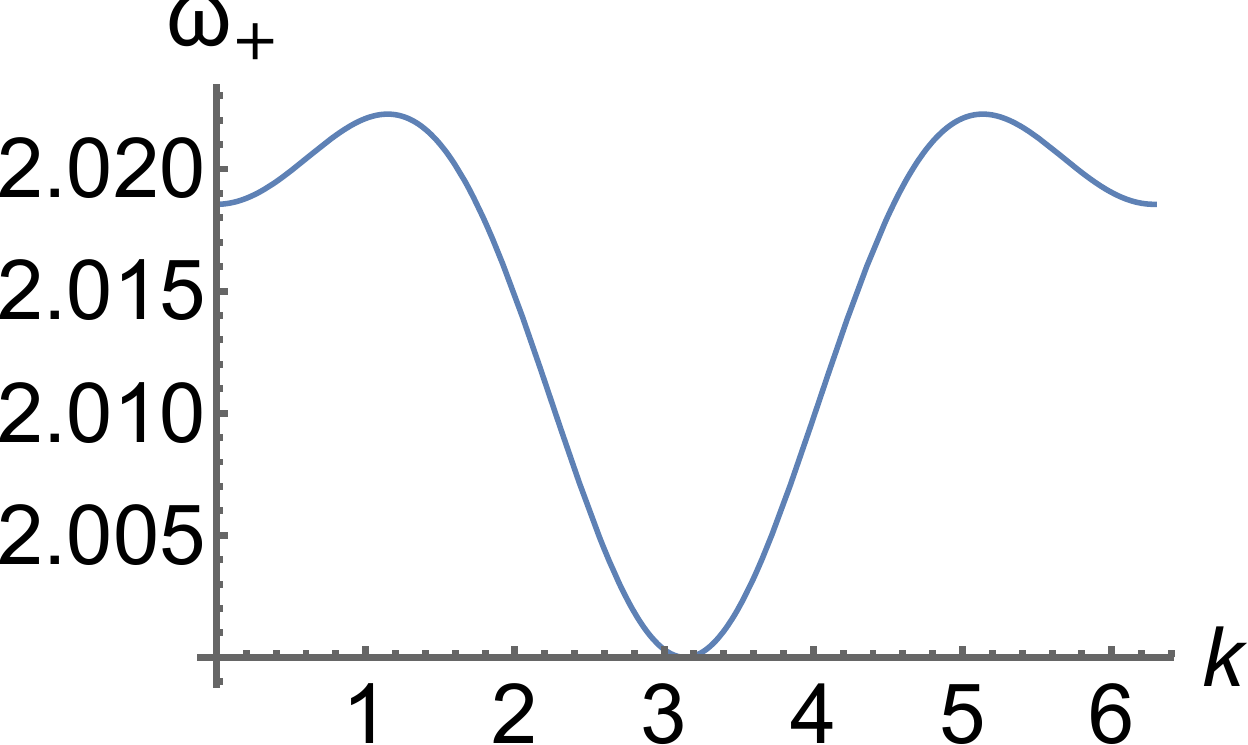}}}
\subfloat[$\phi_0=\phi_0'''=0.6032$]
{{\includegraphics[width=0.33\textwidth]{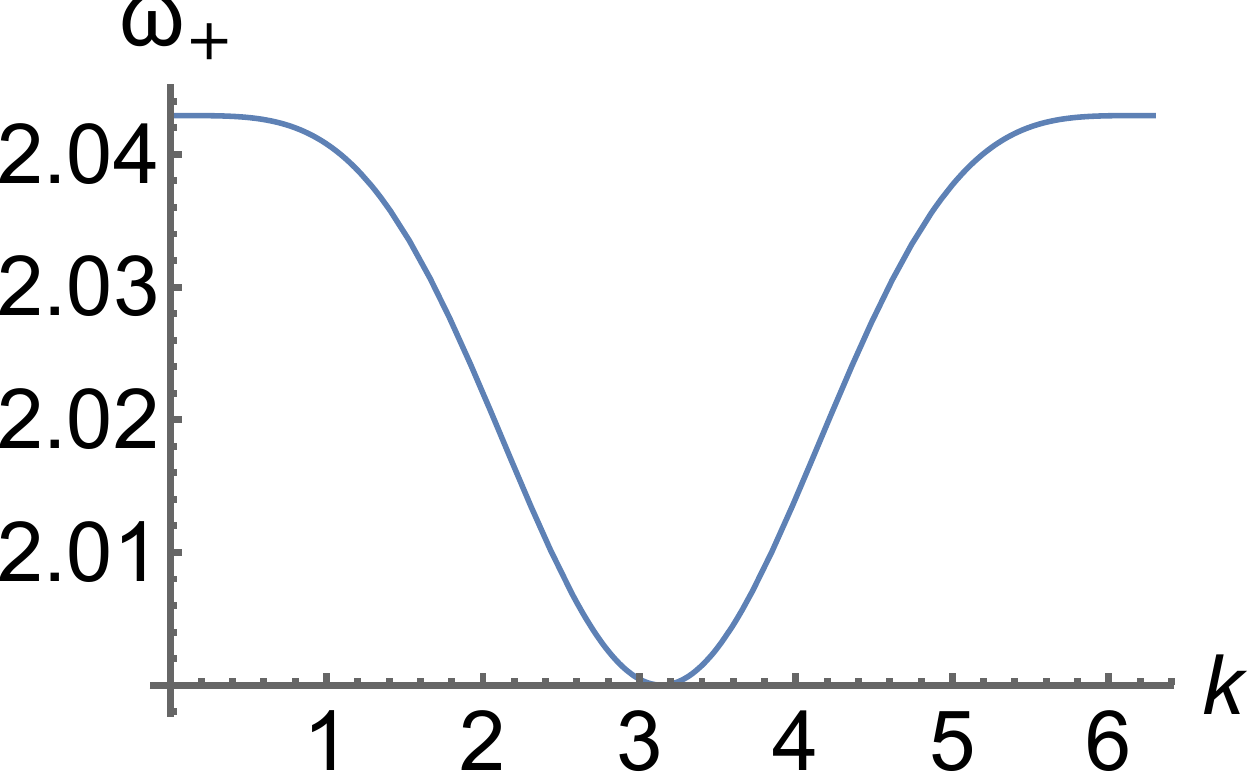}}}\\
\subfloat[$\phi_0=0.61>\phi_0'''$]
{{\includegraphics[width=0.33\textwidth]{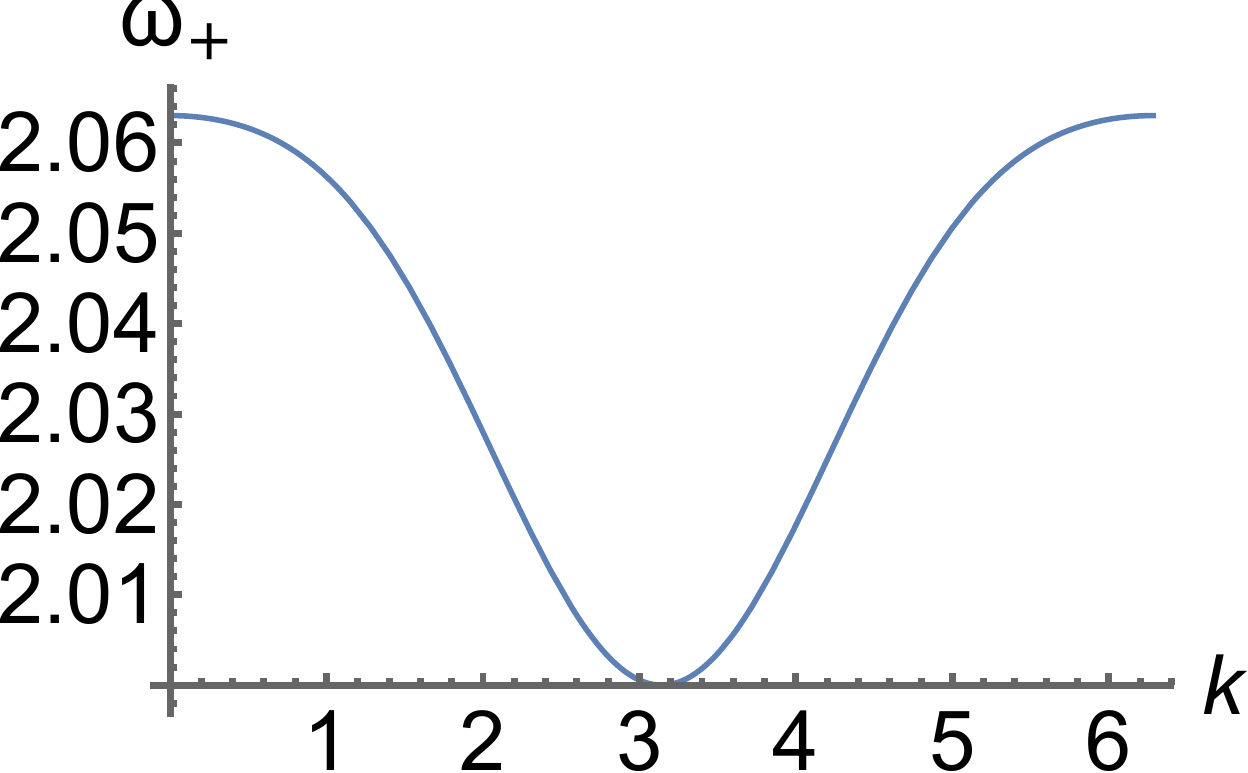}}}
\caption{\footnotesize Optical branch of the dispersion relation at different values of $\phi_0$. For each panel the corresponding value of $\phi_0$ is
  given; see also the discussion in the text. Here $\alpha = 1.8$, $K_s = 0.02$, $K_{\theta} = 1.5 \times 10^{-4}$.}
\label{fig:optical}
\end{figure}

We now examine the evolution of the dispersion relation when the parameters $\alpha$, $K_s$ and $K_\theta$ are fixed, while $\phi_0$ is varied. Due to $2\pi$-periodicity and even symmetry about $k=\pi$, it suffices to consider wave numbers $k$ in $[0,\pi]$.  In what follows, we consider two sets of parameters $\alpha$, $K_s$ and $K_\theta$. In the first representative example, we set $\alpha = 1.8$, $K_s = 0.02$, and $K_{\theta} = 1.5 \times 10^{-4}$ from \cite{Deng2018}. In the second case we keep the same value of $K_s$ and set $\alpha=5$ and $K_\theta=0.01$. In both cases
\beq
\alpha^2(K_{\theta}+2K_s\cos^2\phi_0) < 2
\label{eq:assume}
\eeq
is satisfied for all $\phi_0$, and we thus have
\beq
\omega_{-}(\pi)=\alpha\sqrt{2(K_\theta + 2K_s\cos^2\phi_0)}<\omega_{+}(\pi)=2.
\label{eq:k_pi}
\eeq
Furthermore, one can show that for these parameter values the acoustic branch $\omega_{-}(k)$ has the maximum value at $k=\pi$ given in \eqref{eq:k_pi} for all $\phi_0$. Meanwhile, as shown in Fig.~\ref{fig:optical}(a), the optical branch $\omega_{+}(k)$ has a maximum at $k=\pi$ and a minimum at $k=0$ for $0 \leq \phi_0<\phi'_0$, where
\beq
\phi'_0 = \arccos\left(\sqrt{\frac{1+\frac{K_{\theta}}{2}-\frac{1}{\alpha^2}}{1-K_s}} \right)
\label{eq:phi1}
\eeq
is obtained by setting $\omega''_+(\pi)=0$ at $\phi_0=\phi'_0$ and using \eqref{eq:assume}. The corresponding inflection point at $k=\pi$ is shown in Fig.~\ref{fig:optical}(b). For $\phi_0'<\phi_0<\phi_0''$, where
\beq
\phi''_0 = \arcsin\left(\sqrt{\frac{1}{\alpha^2} - \frac{3}{2}K_{\theta}}\right),
\label{eq:phi2}
\eeq
$k=\pi$ becomes a local minimum, and $\omega_+(k)$ reaches its maximum at $k=k_{max}$ in $(0,\pi)$ and a global minimum at $k=0$ (see Fig.~\ref{fig:optical}(c)). At $\phi_0=\phi_0''$, the case shown in Fig.~\ref{fig:optical}(d), we have $\omega_+(0)=\omega_+(\pi)=2$, which together with the second expression in \eqref{eq:k_zero} yields \eqref{eq:phi2}. For $\phi_0>\phi''_0$, the optical branch has a global minimum $\omega_+(\pi)=2$ at $k=\pi$. As shown in Fig.~\ref{fig:optical}(e), it has a local minimum at $k=0$ and the maximum at $k=k_{max}$ in $(0,\pi)$ until $\phi_0$ reaches the value
\beq
\begin{split}
&\phi_0'''=
\arccos
\bigg(
\frac{1}{2\alpha}
\bigg[
\Big(-2 + \alpha^2 ( K_s (3 K_{\theta}+2)+5 K_{\theta}+4)\\
&-\sqrt{\alpha^4 (K_s (3 K_{\theta}+2)+K_{\theta})^2
-4 \alpha^2 (K_s (3 K_{\theta}-2)+5 K_{\theta})+4}\Big)/(K_s+1)\bigg]^{1/2}
\bigg),
\end{split}
\label{eq:phi3}
\eeq
where $k=0$ becomes an inflection point ($\omega_{+}''(0)=0$); see Fig.~\ref{fig:optical}(f). For $\phi_0>\phi_0'''$ the optical branch is inverted and has the maximum value at $k=0$ and the minimum value at $k=\pi$, as shown in Fig.~\ref{fig:optical}(g).
We obtain $\phi_0^{'} = 0.5736$, $\phi_0^{''} = 0.5888$ and $\phi_0^{'''} = 0.6032$ for the parameters $\alpha = 1.8$, $K_s = 0.02$, $K_{\theta} = 1.5 \times 10^{-4}$. In the case $\alpha = 5$, $K_s = 0.02$, $K_{\theta} = 0.01$, the evolution of the optical branch is similar to Fig.~\ref{fig:optical} but the critical values are $\phi_0^{'} = 0.1240$, $\phi_0^{''} = 0.1588$ and $\phi_0^{'''} = 0.1959$.

\begin{figure}[!htb]
\centering
\subfloat[]
{{\includegraphics[width=0.33\textwidth]{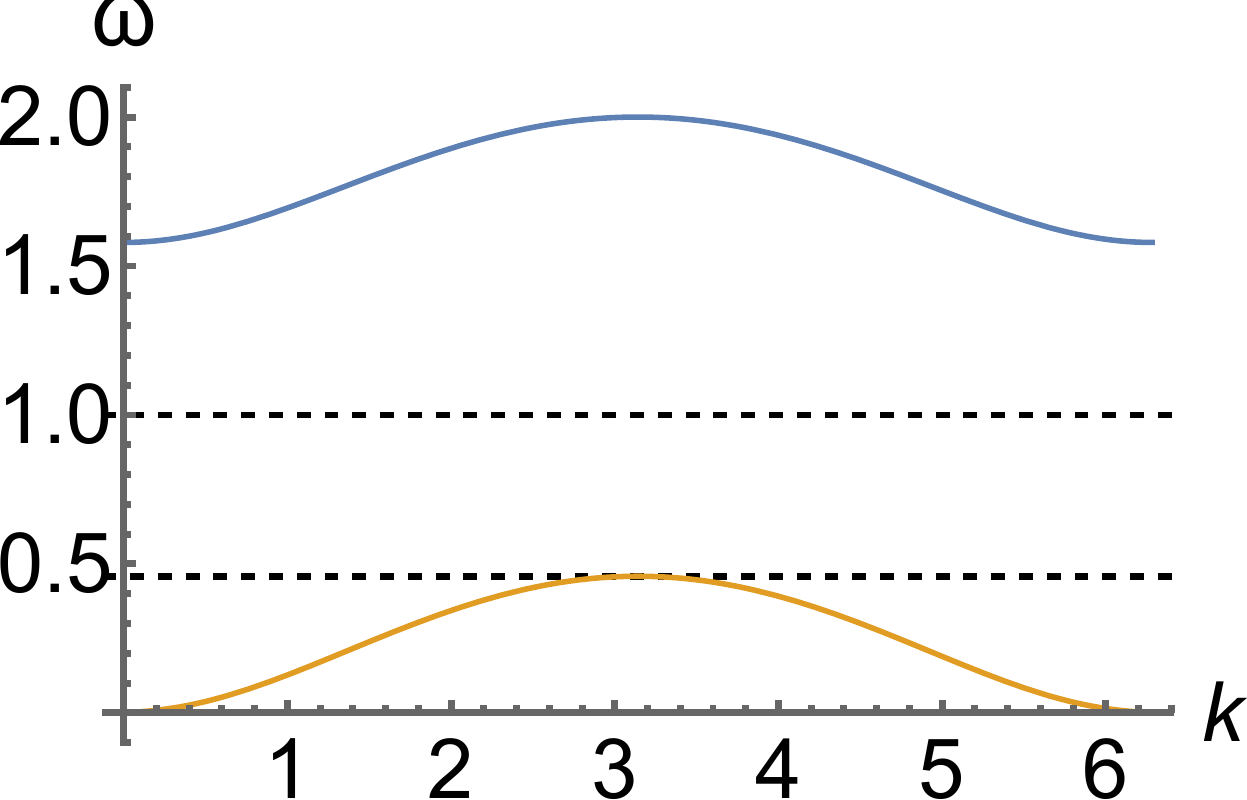}}}
\subfloat[]
{{\includegraphics[width=0.33\textwidth]{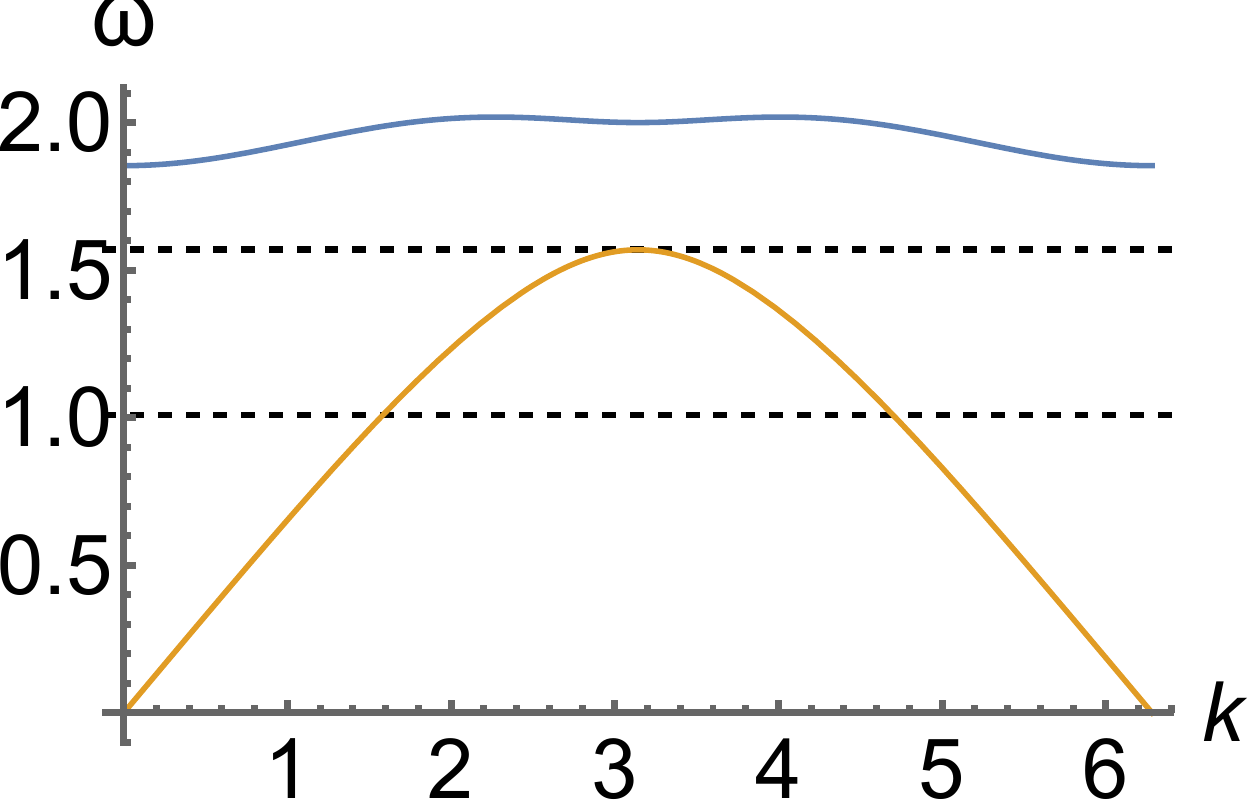}}}
\subfloat[]
{{\includegraphics[width=0.33\textwidth]{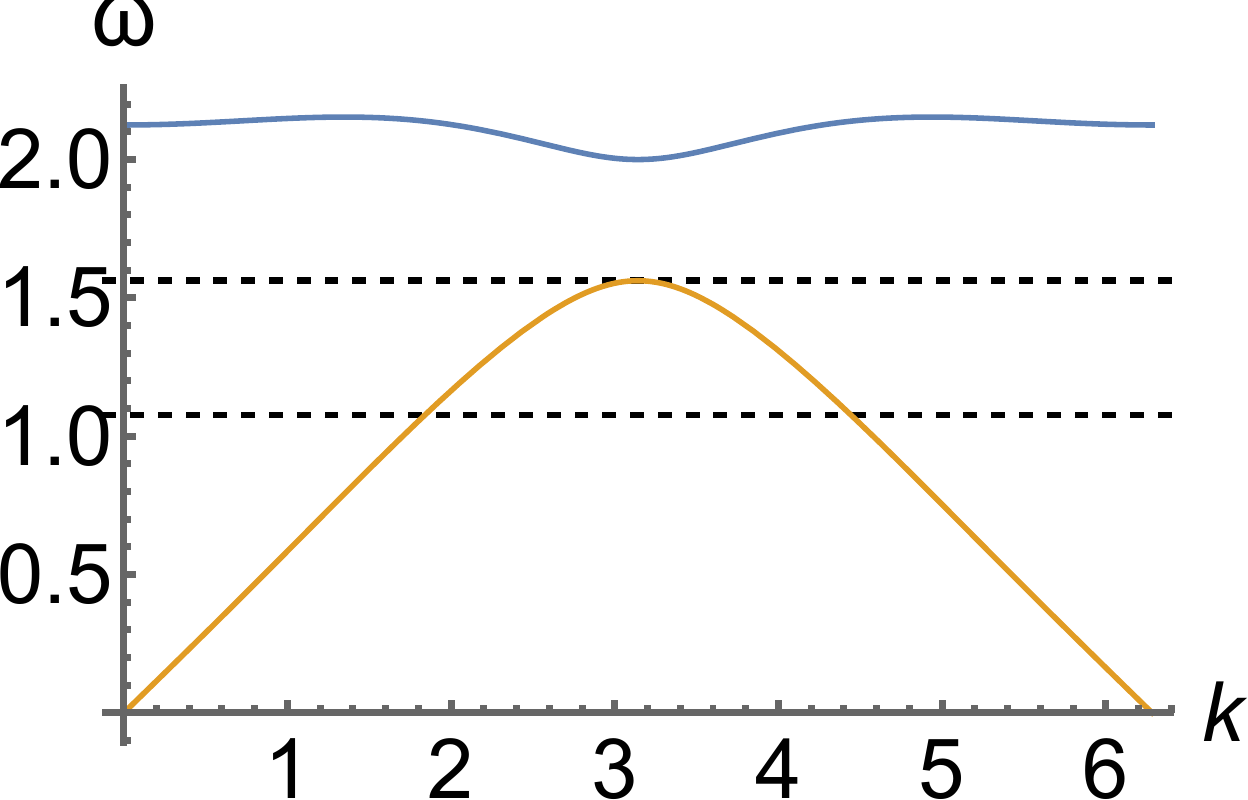}}}
\caption{\footnotesize Optical (blue) and acoustic (orange) branches for (a) $\phi_0=26\pi/180 \approx 0.4538$, $\alpha = 1.8$, $K_s = 0.02$, $K_{\theta} = 1.5 \times 10^{-4}$; (b) $\phi_0 = 8\pi/180\approx 0.1396$, $\alpha=5$, $K_s=0.02$, $K_\theta=0.01$; (c) $\phi_0 = 10\pi/180 \approx 0.1745 $, $\alpha=5$, $K_s=0.02$, $K_\theta=0.01$. The horizontal lines indicate the maximum $\omega_{-}(\pi)$ of the acoustic branch and $\omega_+(k_{max})/2$, half of the maximum of the optical branch. When the optical branch is above $\omega_{-}(\pi)$, \eqref{eq:Gap} holds, and when it is above $\omega_+(k_{max})/2$, \eqref{eq:OpticalGap} holds.}
\label{fig:OpticalAcousticGap}
\end{figure}

Let $k_{min}$ denote the wave number where the optical branch $\omega_+(k)$ reaches its minimum value. From the above discussion it follows that $k_{min}=0$ for $0 \leq \phi_0 \leq \phi_0''$, with the minimum value $\omega_+(0)=\alpha(6 K_\theta +4 \sin^2\phi_0)^{1/2}$, and $k_{min}=\pi$ for $\phi_0>\phi_0''$, with the minimum value $\omega_+(\pi)=2$. Recalling that the acoustic branch has a maximum at $k=\pi$, we find that when
\beq
G=\omega_+(k_{min}) - \omega_-(\pi)>0,
\label{eq:Gap}
\eeq
there is a \emph{band gap} between the two branches. See Fig.~\ref{fig:OpticalAcousticGap} for examples of such a gap.
A DB solution with frequency $\omega$ inside the gap, i.e., $\omega_{-}(\pi)<\omega<\omega_+(k_{\min})$, may exist provided that
\beq
S=\omega_+(k_{min}) - \frac{1}{2}\omega_+(k_{max})>0
\label{eq:OpticalGap}
\eeq
holds in addition to \eqref{eq:Gap} and $\omega>\omega_+(k_{max})/2$. Here $k_{max}$ is the wavenumber where the optical branch $\omega_+(k)$ reaches its maximum value. The fact that $\omega$ does not coincide with either optical or acoustic values for any wave number means that the breather is not in resonance with any linear modes, while the condition \eqref{eq:OpticalGap} eliminates the second harmonic resonances by ensuring that $2\omega>\omega_{+}(k)$ for all wave numbers. This enables both the spatial localization (due to its presence in the bandgap)
and the non-resonance of the breather, as discussed, e.g., in~\cite{macaub}.

\begin{figure}[!htb]
\centering
\subfloat[]
{{\includegraphics[width=0.5\textwidth]{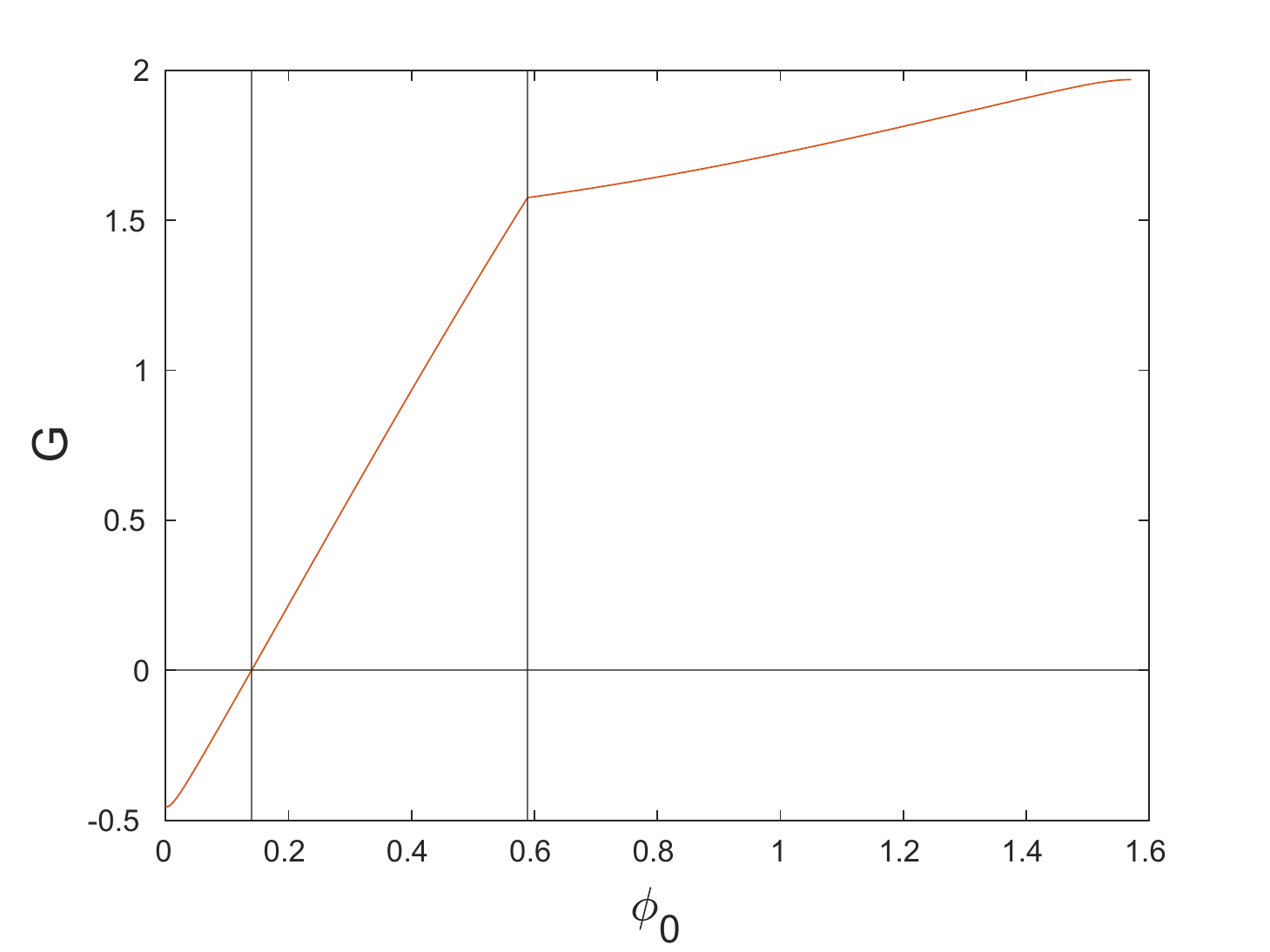}}}
\subfloat[]
{{\includegraphics[width=0.5\textwidth]{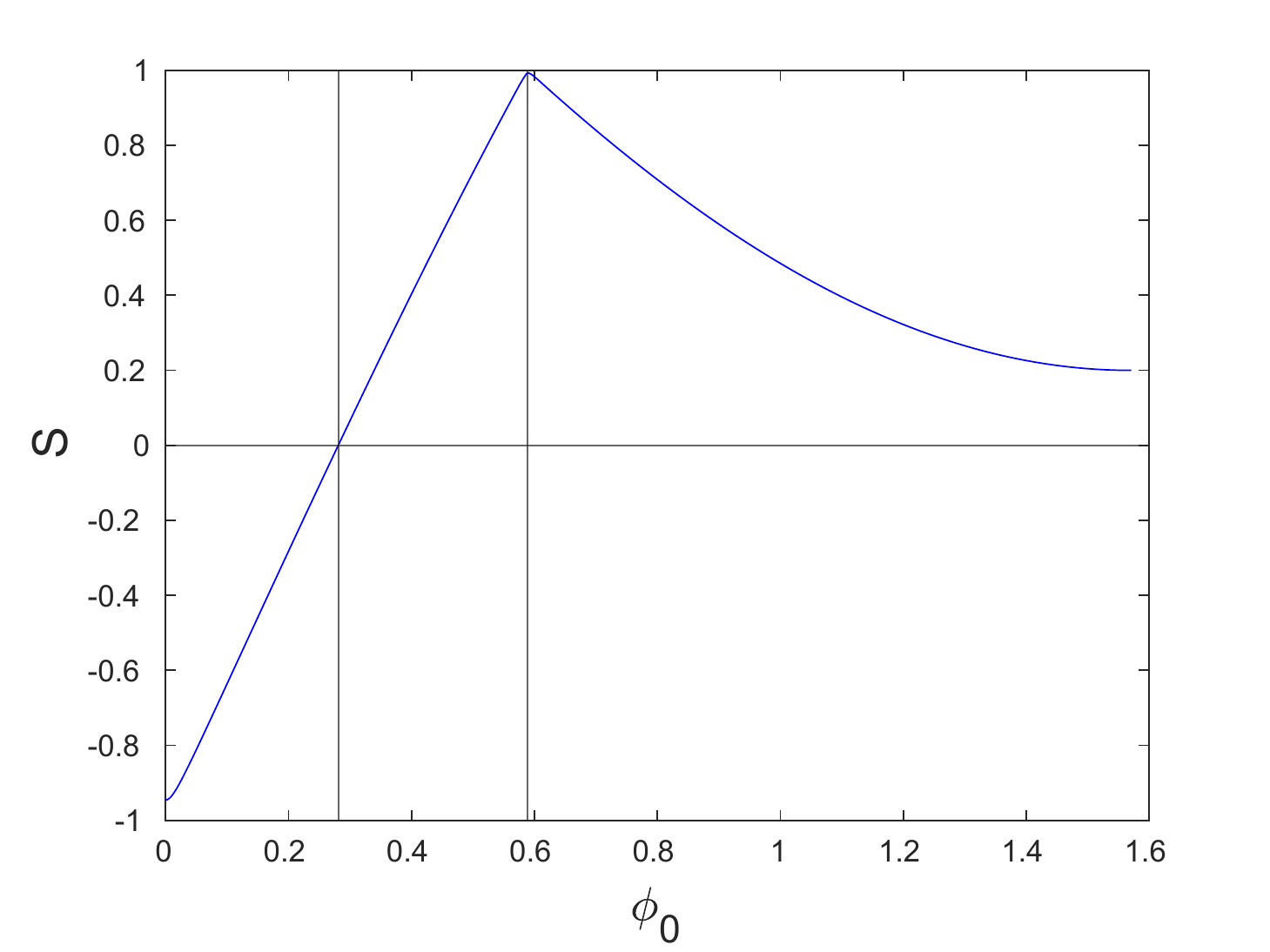}}}
\caption{\footnotesize (a) $G$ defined in \eqref{eq:Gap} as a function of $\phi_0$. The horizontal line is $G=0$, and the two vertical lines indicate $\phi_0 = \phi_0^{*} = 0.1400$ and $\phi_0 = \phi_0^{''} = 0.5888$. (b) $S$ defined in \eqref{eq:OpticalGap} as a function of $\phi_0$. The horizontal line is $S=0$ and the two vertical lines indicate $\phi_0 = \phi_0^{**} = 0.2811$ and $\phi_0 = \phi_0^{''} = 0.5888$. Here $\alpha = 1.8$, $K_s = 0.02$, $K_{\theta} = 1.5 \times 10^{-4}$.}
\label{fig:Gapvphi0_paperpara}
\end{figure}

Fig.~\ref{fig:Gapvphi0_paperpara} shows $G$ and $S$ defined in \eqref{eq:Gap} and \eqref{eq:OpticalGap}, respectively, as functions of $\phi_0$ for the first parameter set.
Both functions have a corner at $\phi_0=\phi_0''$ where $k_{min}$ changes from $0$ to $\pi$.  Noting that $G$ changes sign from negative to positive for $\phi_0<\phi_0''$, when $k_{min}=0$, we set
\[
G=\omega_+(0) - \omega_-(\pi) = \alpha\left(\sqrt{6K_{\theta}+4\sin^2(\phi_0)} - \sqrt{2K_{\theta}+4K_s\cos^2(\phi_0)}\right)=0
\]
to find the critical angle
\beq
\phi_0^{*} = \arccos\sqrt{\frac{1+K_{\theta}}{1+K_s}}
\label{eq:Phi0Star}
\eeq
above which \eqref{eq:Gap} holds. The function $S$ in Fig.~\ref{fig:Gapvphi0_paperpara}(b) also changes sign for $\phi_0<\phi_0'$, where $k_{min}=0$ and $k_{max}=\pi$, so that
\[
S=\omega_+(0) - \frac{1}{2}\omega_+(\pi) = \alpha\sqrt{6K_{\theta}+4\sin^2(\phi_0)} - 1=0
\]
at
\beq
\phi_0^{**} = \arcsin\bigg(\sqrt{\frac{1}{4\alpha^2} - \frac{3}{2}K_{\theta} }\bigg),
\label{eq:Phi0DoubleStar}
\eeq
and hence \eqref{eq:OpticalGap} holds for $\phi_0>\phi_0^{**}$. We find that $\phi_0^{*} = 0.1400 $ and $\phi_0^{**} = 0.2811$ in this case.
Thus for $\phi_0 > 0.2811$, both \eqref{eq:Gap} and \eqref{eq:OpticalGap} hold, and DB solutions may exist with frequencies $\omega$ in the interval $(\omega_{+}(k_{\max})/2,\omega_+(k_{\min}))$; otherwise, first or second resonances set in. The example at $\phi_0=26\pi/180 \approx 0.4538$, where \eqref{eq:Gap} and \eqref{eq:OpticalGap} hold for $1<\omega<1.57906$, is shown in Fig.~\ref{fig:OpticalAcousticGap}(a). As shown in Fig.~\ref{fig:Gapvphi0_paperpara}(b), the frequency gap increases until $\phi_0''=0.5888$ and then starts decreasing. Note that for $\phi_0<\phi_0''$, DB solutions bifurcating from the optical band emerge from $k=0$ mode, while for $\phi_0$ above this threshold the breathers bifurcate from the $k=\pi$ mode.

\begin{figure}[!htb]
\centering
\subfloat[]
{{\includegraphics[width=0.5\textwidth]{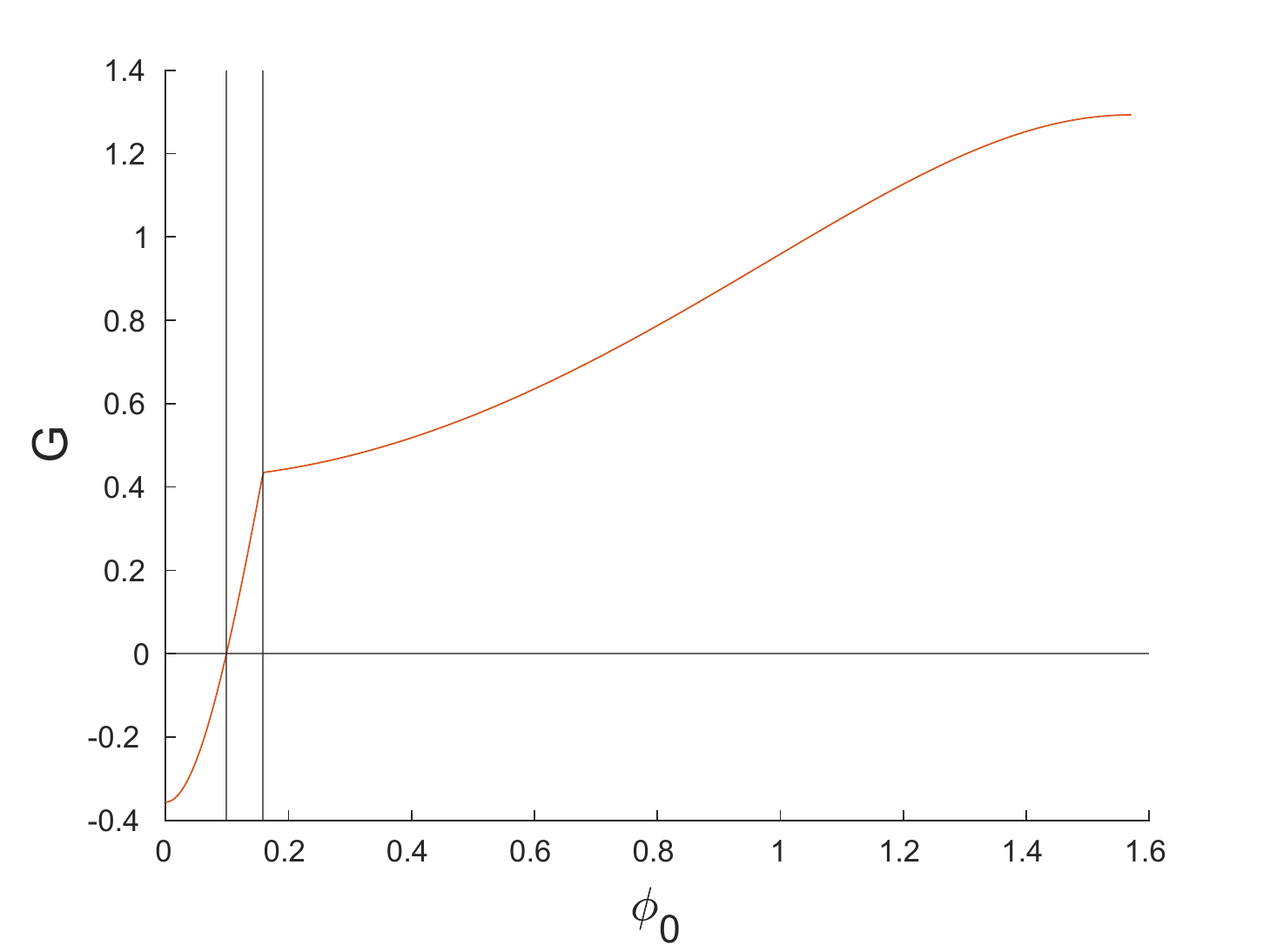}}}
\subfloat[]
{{\includegraphics[width=0.5\textwidth]{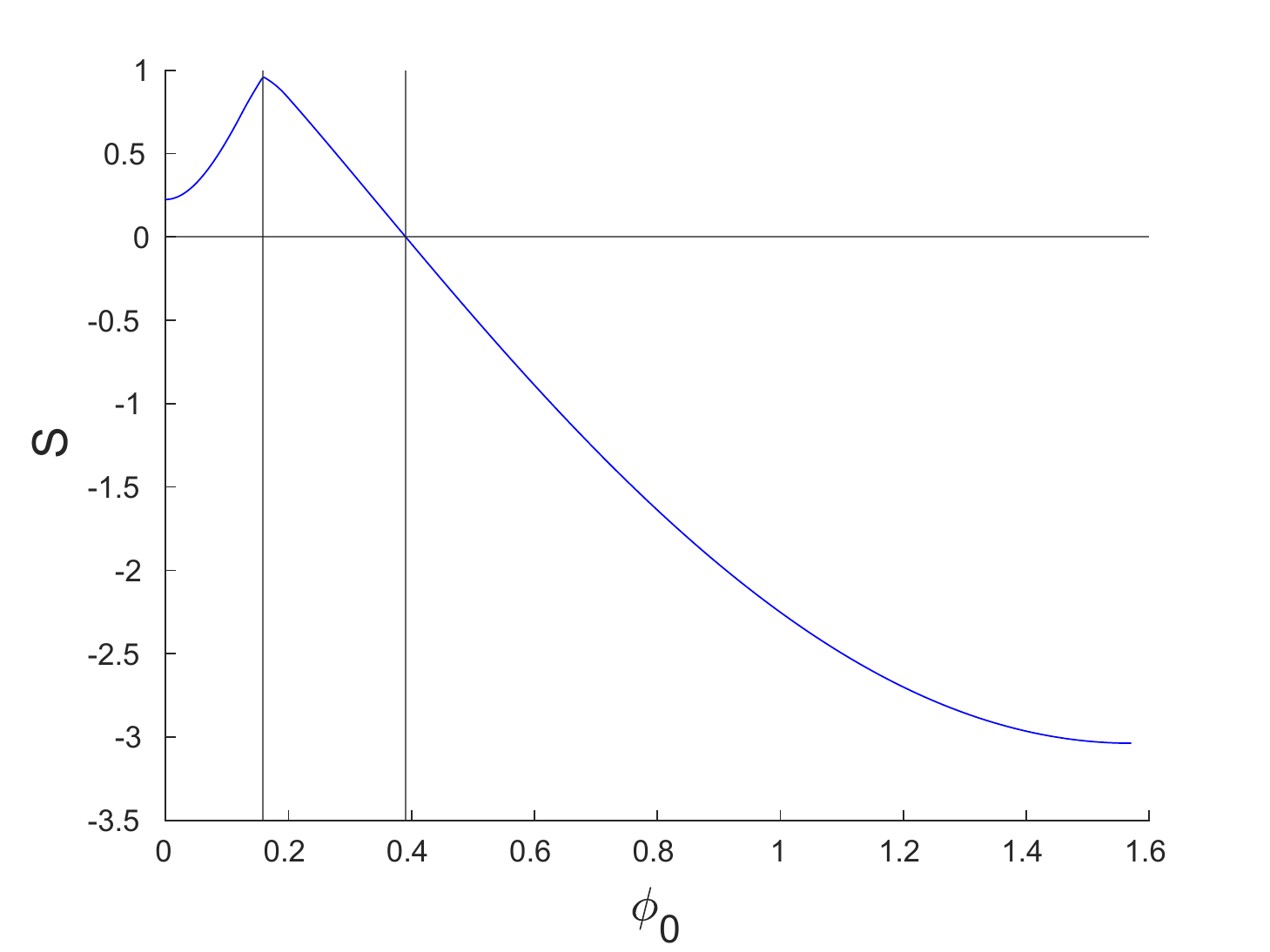}}}
\caption{\footnotesize (a) $G$ defined in \eqref{eq:Gap} as a function of $\phi_0$. The horizontal line is $G=0$ and the two vertical lines indicate $\phi_0 = \phi_0^{*} = 0.0992$ and $\phi_0 = \phi_0^{''} = 0.1588$. (b) $S$ defined in \eqref{eq:OpticalGap} as a function of $\phi_0$. The horizontal line is $S=0$ and the two vertical lines indicate $\phi_0 = \phi_0^{''} = 0.1588$ and $\phi_0 = \phi_0^{***} = 0.3906$. Here $\alpha = 5$, $K_s = 0.02$, $K_{\theta} = 0.01$. }
\label{fig:Gapvphi0_newpara}
\end{figure}

The functions $G(\phi_0)$ and $S(\phi_0)$ for the second parameter set are shown in Fig.~\ref{fig:Gapvphi0_newpara}. Recall that in this case \eqref{eq:phi1}, \eqref{eq:phi2} and \eqref{eq:phi3} yield $\phi_0'=0.1240$, $\phi_0''=0.1588$ and $\phi_0'''=0.1959$. One can see that \eqref{eq:Gap} holds ($F(\phi_0)>0$) for $\phi_0>\phi_0^*$, where $\phi_0^*=0.0992$ is found from \eqref{eq:Phi0Star}. Meanwhile, $S(\phi_0)$ is positive for $0 \leq \phi_0<\phi_0^{***}$. To find this value, we observe that it is above $\phi_0'''$, which means that $k_{\min}=\pi$ and $k_{\max}=0$ in \eqref{eq:OpticalGap}. Thus,
\[
S=2-\dfrac{1}{2}\alpha\sqrt{6K_{\theta}+4\sin^2(\phi_0)}=0
\]
must hold at $\phi_0=\phi_0^{***}$, which yields
\beq
\phi_0^{***} = \arcsin\bigg(\sqrt{\frac{4}{\alpha^2} - \frac{3}{2}K_{\theta} }\bigg).
\label{eq:Phi0TripleStar}
\eeq
We obtain $\phi_0^{***}=0.3906$ for the second parameter set. Thus, in this case \eqref{eq:Gap} and \eqref{eq:OpticalGap} both hold when
$0.0992<\phi_0<0.3906$. Examples of dispersion relations with band gaps for this parameter regime are shown in panels (b) and (c) of Fig.~\ref{fig:OpticalAcousticGap}. Note that in both cases the maximum of the acoustic branch lies above the half of the maximum of the optical one, and hence the frequency range where DB solutions may exist includes the entire gap between the two bands. This is in contrast to the example shown in Fig.~\ref{fig:OpticalAcousticGap}(a) for the first parameter set, where the breather frequency must exceed $\omega_{+}(\pi)/2=1$.

\section{Period-doubling bifurcation}
\label{sec:period_doubling}
We first discuss DB solutions bifurcating from the optical $k=0$ mode for $\phi_0<\phi_0''$ for the parameters considered in \cite{Deng2018} and associated with the experimental implementation of the metamaterial in that work: $\alpha = 1.8$, $K_s = 0.02$, $K_{\theta} = 1.5 \times 10^{-4}$. We set $\phi_0=26\pi/180 \approx 0.4538$, which enables existence of DB solutions with frequency $\omega$ in $(1,1.57906)$. The corresponding dispersion relation is shown in Fig.~\ref{fig:OpticalAcousticGap}(a).

To obtain the breathers with frequency $\omega$ and corresponding period $T = 2\pi/\omega$, we consider a chain of $N=200$ elements and solve
iteratively using Newton's method the following equations:
\[
\begin{pmatrix}
\mathbf{u}(T)-\mathbf{u}(0) \\
\dot{\mathbf{u}}(T)-\dot{\mathbf{u}}(0) \\
\boldsymbol{\theta}(T)- \boldsymbol{\theta}(0)\\
\dot{\boldsymbol{\theta}}(T)-\dot{\boldsymbol{\theta}}(0)
\end{pmatrix}
 =
\mathbf{0},
\]
where the vector functions $\mathbf{u}(t)$, $\dot{\mathbf{u}}(t)$, $\boldsymbol{\theta}(t)$, and $\dot{\boldsymbol{\theta}}(t)$ have the components $u_n(t)$, $\dot{u}_n(t)$, $\theta_n(t)$, and $\dot{\theta}_n(t)$, $n=1,\dots,N$, respectively. We perform numerical continuation in the frequency $\omega$, starting with $\omega = 1.57$, just below the edge of the optical band at $k=0$.
The initial guess is of the form
\beq
u_n = \varepsilon_u\tanh[\delta(n - N/2)], \quad \theta_n = \varepsilon_\theta\sech[\delta(n - N/2)],
\label{eq:initial_guess_k0}
\eeq
where $\varepsilon_u$, $\varepsilon_\theta$ and $\delta$ are small. The dynamical
evolution of Eq.~\eqref{eq:EoM} (over the prescribed period $T$)
is performed using the symplectic fourth-order Runge-Kutta-Nystr\"{o}m algorithm \cite{calvo93} with free-end boundary conditions.

To study the linear stability of the obtained solutions, we use Floquet analysis. Setting $u_n(t) = \hat{u}_n(t) +\epsilon v_n(t)$ and $\theta_n(t) = \hat{\theta}_n(t) + \epsilon \gamma_n(t)$ in \eqref{eq:EoM}, where $\hat{u}_n(t)$ and $\hat{\theta}_n(t)$ comprise the DB solutions, and considering $O(\epsilon)$ terms, we obtain the linearized system
\[
\begin{split}
&\ddot{v}_n = v_{n+1} + v_{n-1} - 2v_n - \left[\frac{-\sin(\hat{\theta}_{n+1}+\phi_0)\gamma_{n+1} + \sin(\hat{\theta}_{n-1}+\phi_0)\gamma_{n-1}}{2\cos(\phi_0)}\right] \\
&\frac{1}{\alpha^2}\ddot{\gamma}_n = -K_{\theta}(\gamma_{n+1} + 4\gamma_n + \gamma_{n-1}) + K_s[\cos(\hat{\theta}_n+\phi_0)\cos(\hat{\theta}_{n+1}+\phi_0)\gamma_{n+1} \\
&- \sin(\hat{\theta}_n + \phi_0)\sin(\hat{\theta}_{n+1}+\phi_0)\gamma_n +
     \cos(\hat{\theta}_n + \phi_0)\cos(\hat{\theta}_{n-1}+\phi_0)\gamma_{n-1} \\
& - \sin(\hat{\theta}_n + \phi_0)\sin(\hat{\theta}_{n-1} + \phi_0)\gamma_n
     -2(\cos^2(\hat{\theta}_n + \phi_0) - \sin^2(\hat{\theta}_n + \phi_0))\gamma_n] \\
& -[2\sin(\hat{\theta}_n + \phi_0)\cos(\phi_0)(v_{n+1} - v_{n-1}) + 2\cos(\hat{\theta}_n + \phi_0)\cos(\phi_0)(\hat{u}_{n+1} - \hat{u}_{n-1})\gamma_n \\
& + 4\cos(\hat{\theta}_n + \phi_0)\cos(\phi_0)\gamma_n
     -(\cos(\hat{\theta}_n + \phi_0)\cos(\hat{\theta}_{n+1}+ \phi_0)\gamma_n\\
& - \sin(\hat{\theta}_n + \phi_0)\sin(\hat{\theta}_{n+1} + \phi_0)\gamma_{n+1})-2(\cos^2(\hat{\theta}_n + \phi_0) - \sin^2(\hat{\theta}_n + \phi_0))\gamma_n \\
& -(\cos(\hat{\theta}_n + \phi_0)\cos(\hat{\theta}_{n-1} + \phi_0)\gamma_n - \sin(\hat{\theta}_n + \phi_0)\sin(\hat{\theta}_{n-1}+\phi_0)\gamma_{n-1})],
\end{split}
\]
which is used to compute the monodromy matrix $\mathcal{F}$ defined by
\[
\begin{pmatrix}
\mathbf{v}(T) \\
\dot{\mathbf{v}}(T) \\
\boldsymbol{\gamma} (T) \\
\dot{\boldsymbol{\gamma}}(T)
\end{pmatrix}
 =
\mathcal{F}
\begin{pmatrix}
\mathbf{v}(0) \\
\dot{\mathbf{v}}(0) \\
\boldsymbol{\gamma}(0) \\
\dot{\boldsymbol{\gamma}}(0)
\end{pmatrix},
\]
where the vector functions $\mathbf{v}(t)$, $\dot{\mathbf{v}}(t)$, $\boldsymbol{\gamma}(t)$, and $\dot{\boldsymbol{\gamma}}(t)$ have the components $v_n(t)$, $\dot{v}_n(t)$, $\gamma_n(t)$, and $\dot{\gamma}_n(t)$, $n=1,\dots,N$, respectively. The Floquet multipliers $\mu$ are the eigenvalues of the matrix $\mathcal{F}$. The existence of a Floquet multiplier $\mu$ satisfying $|\mu| > 1$ indicates the presence of instability. When the relevant instability-inducing
multiplier is real, we refer to the instability as \emph{exponential}, given the exponential
nature of the associated growth. When such real multipliers arise, they come
in pairs $(\mu,1/\mu)$ (one of which is outside, while the other is inside the unit circle).
In the case of a complex multiplier quartet $(\mu,1/\mu,\bar{\mu},1/\bar{\mu})$ with $|\mu|>1$, the instability is referred to as
\emph{oscillatory}, given that oscillations accompany the exponential growth due
to the imaginary part of the associated multipliers. The fact that the
multipliers come in real pairs or complex quartets is a generic by-product
of the Hamiltonian nature of the underlying lattice dynamical system.

\begin{figure}[!htb]
\centering
\subfloat[]
{\includegraphics[width=0.33\textwidth]{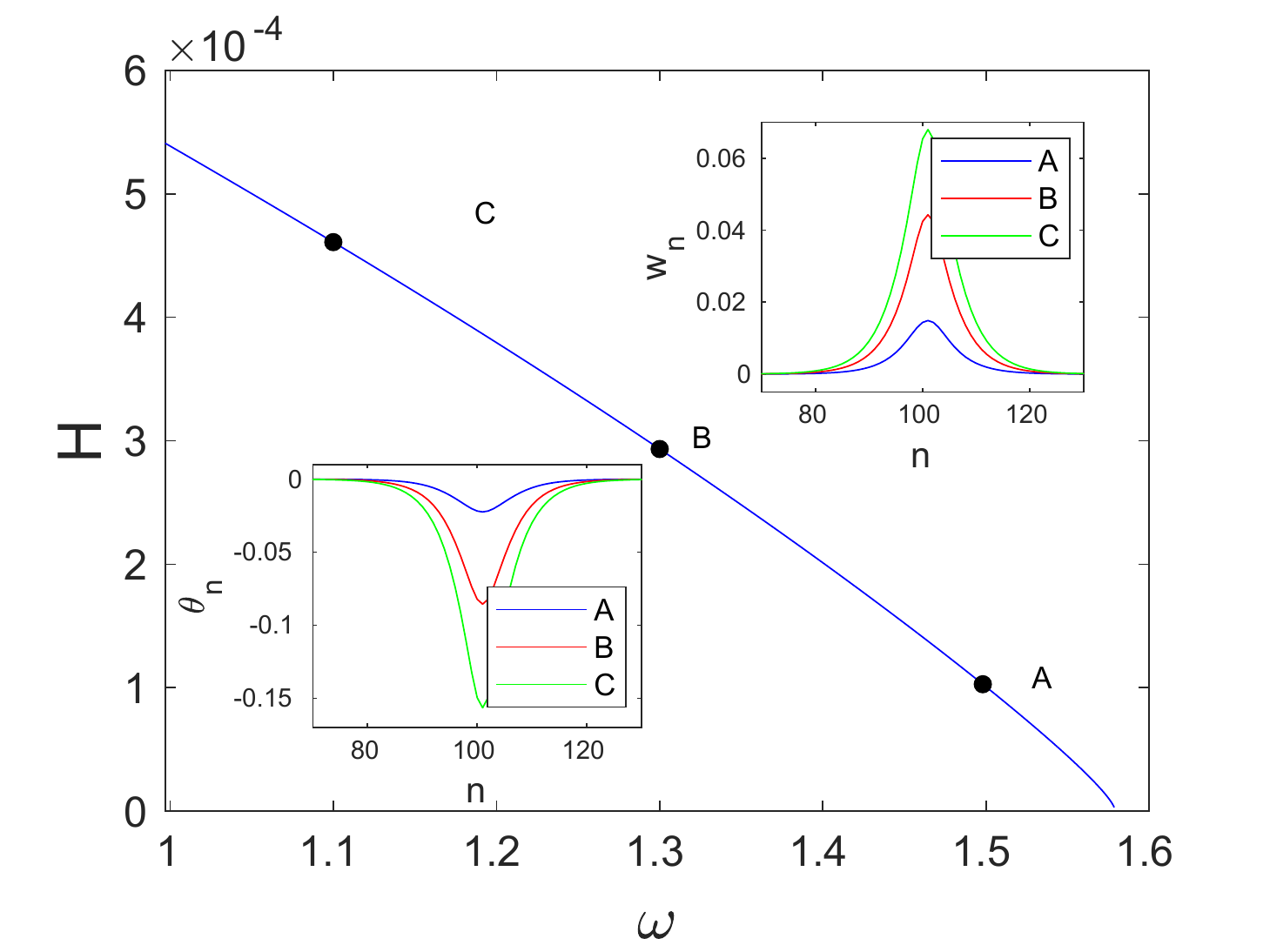}}
\subfloat[]
{\includegraphics[width=0.33\textwidth]{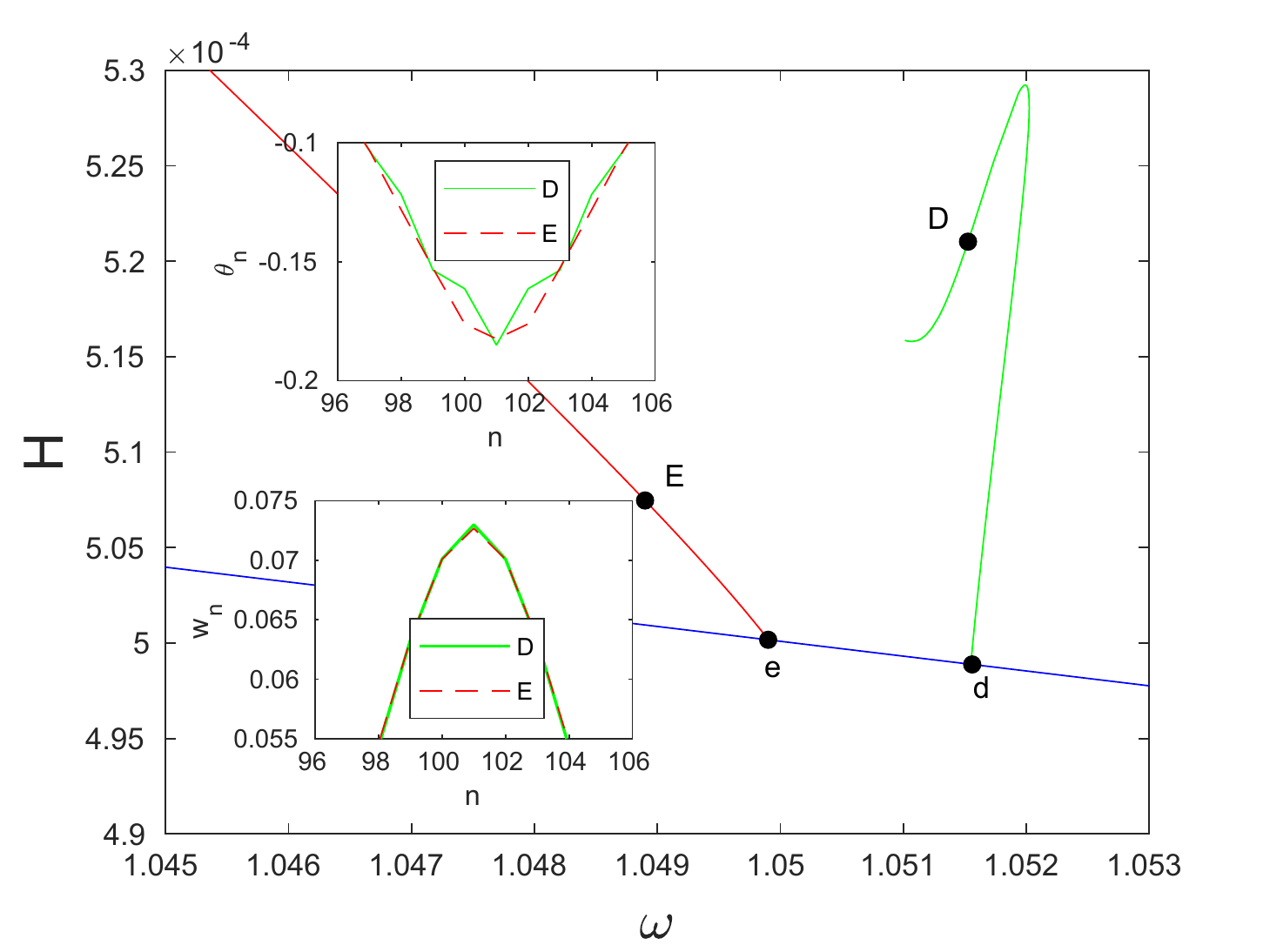}}
\subfloat[]
{\includegraphics[width=0.33\textwidth]{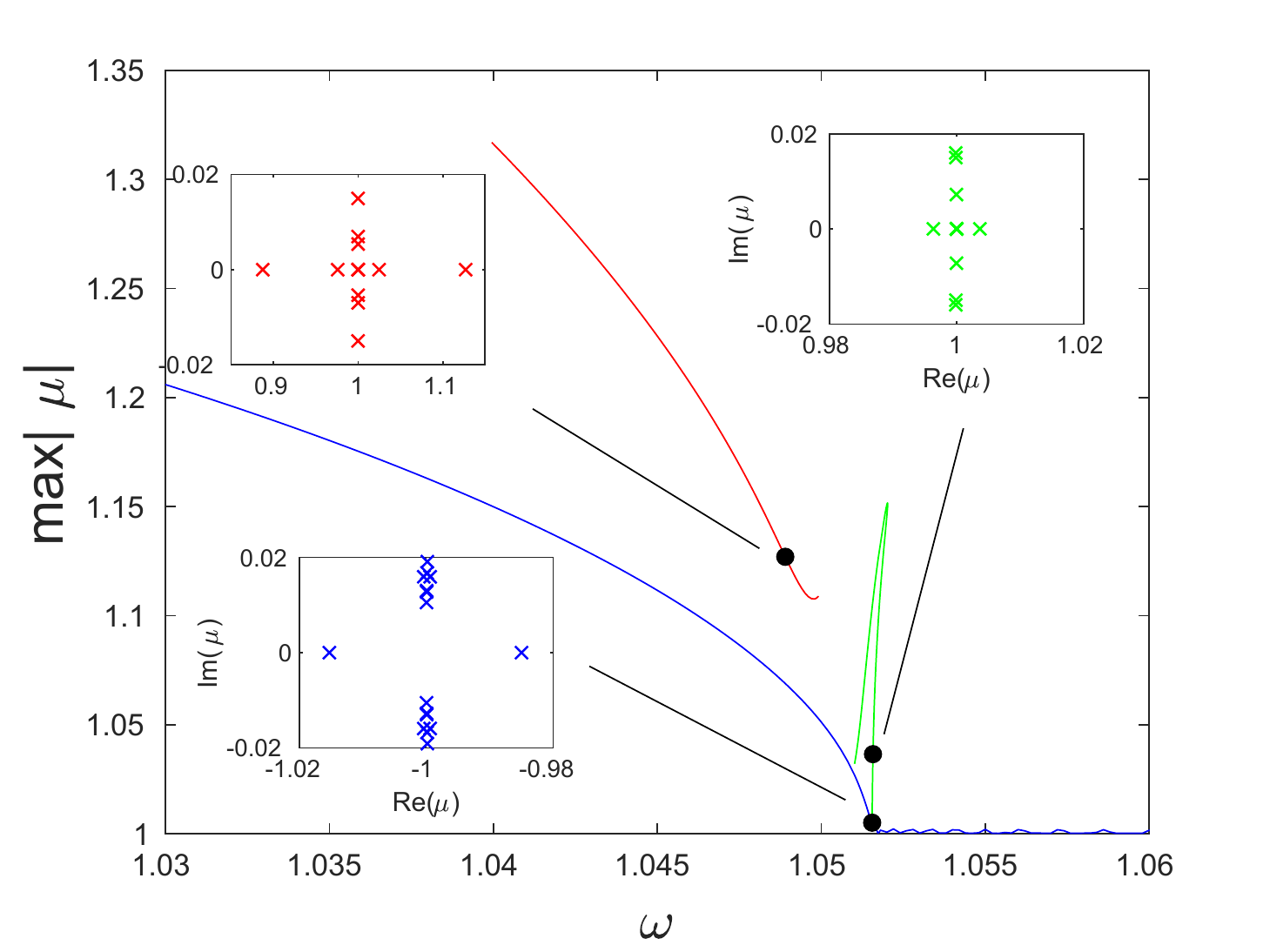}}\\
\subfloat[]
{\includegraphics[width=0.33\textwidth]{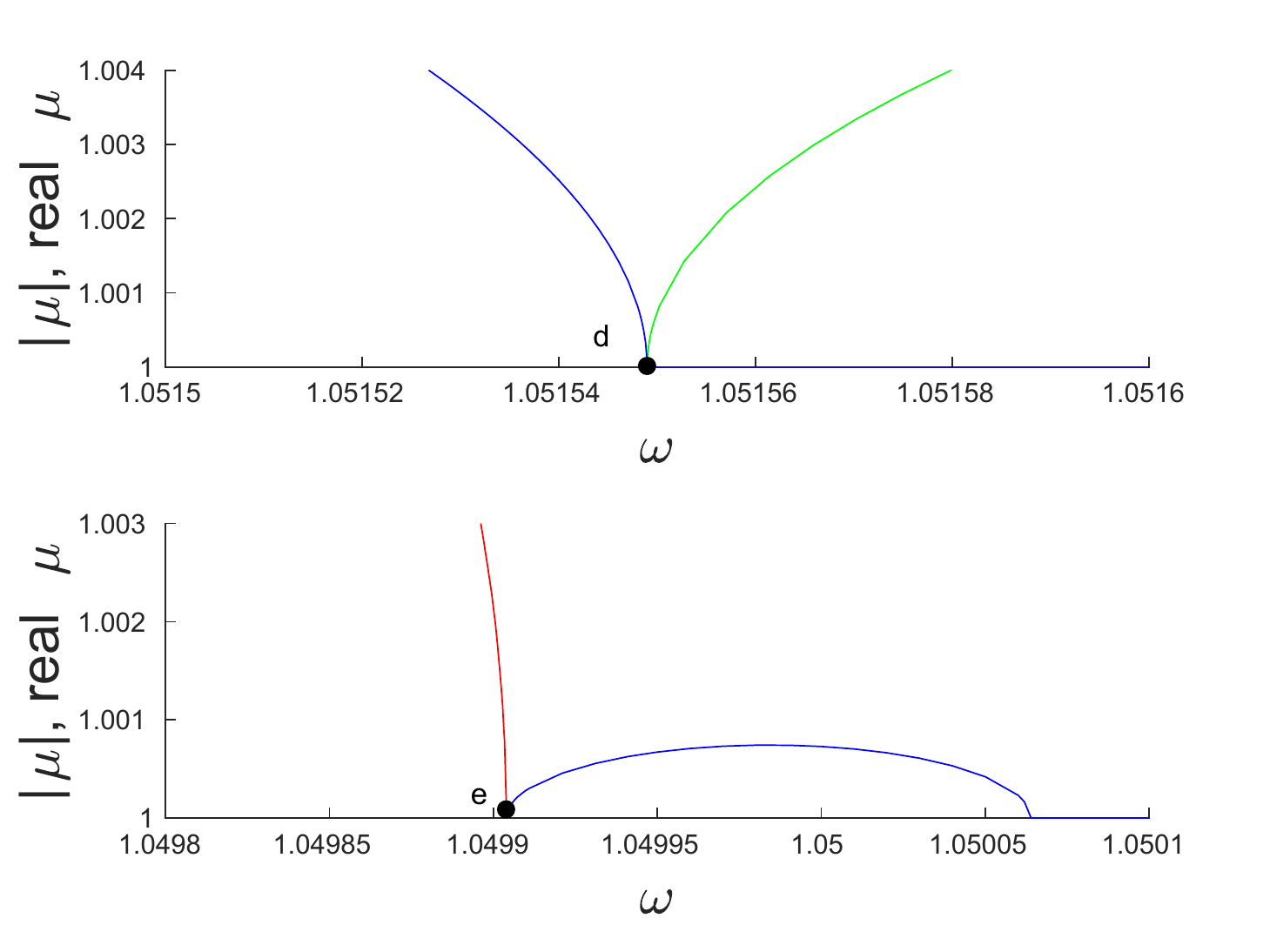}}
\subfloat[]
{\includegraphics[width=0.33\textwidth]{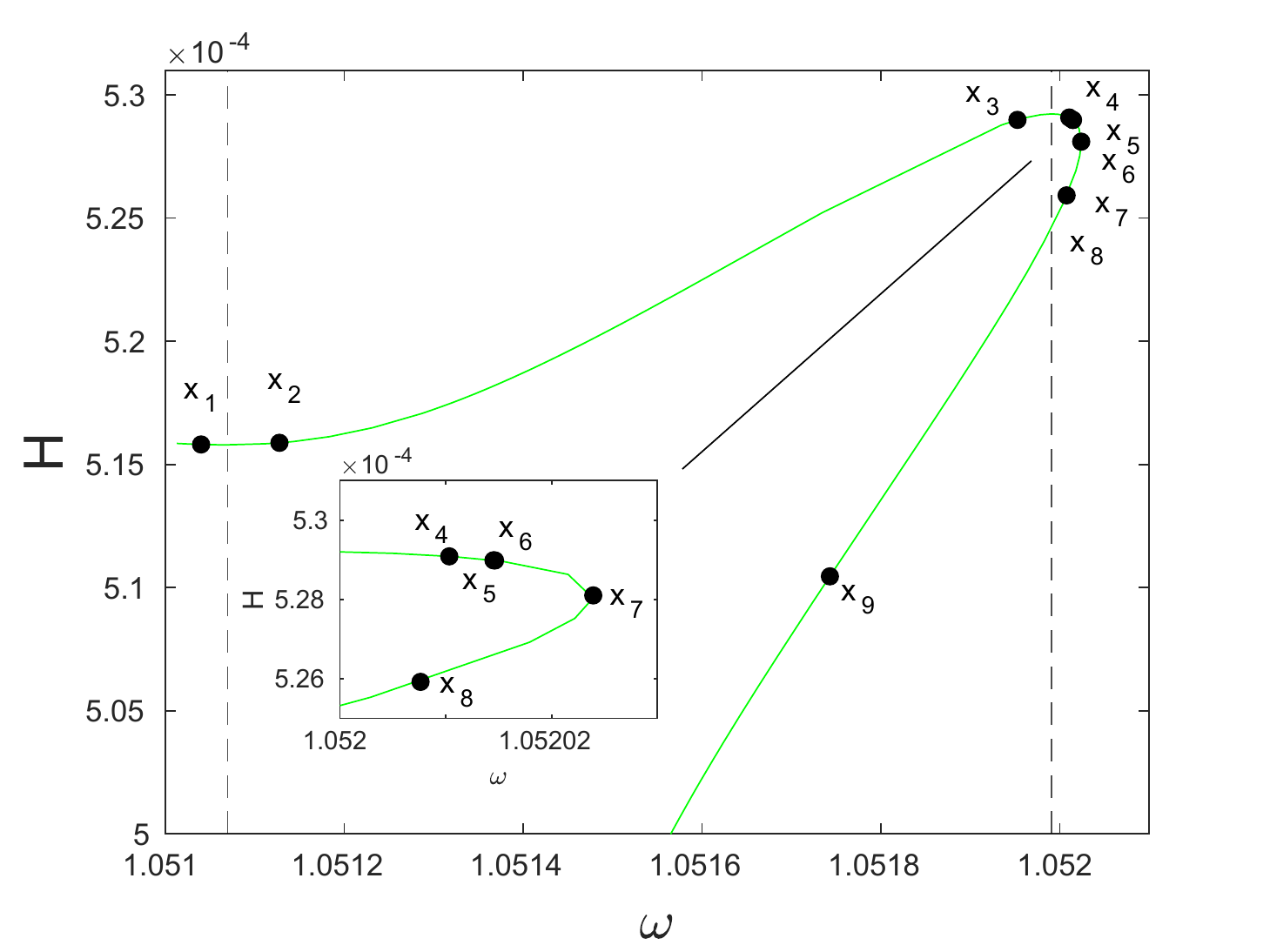}}
\subfloat[]
{\includegraphics[width=0.33\textwidth]{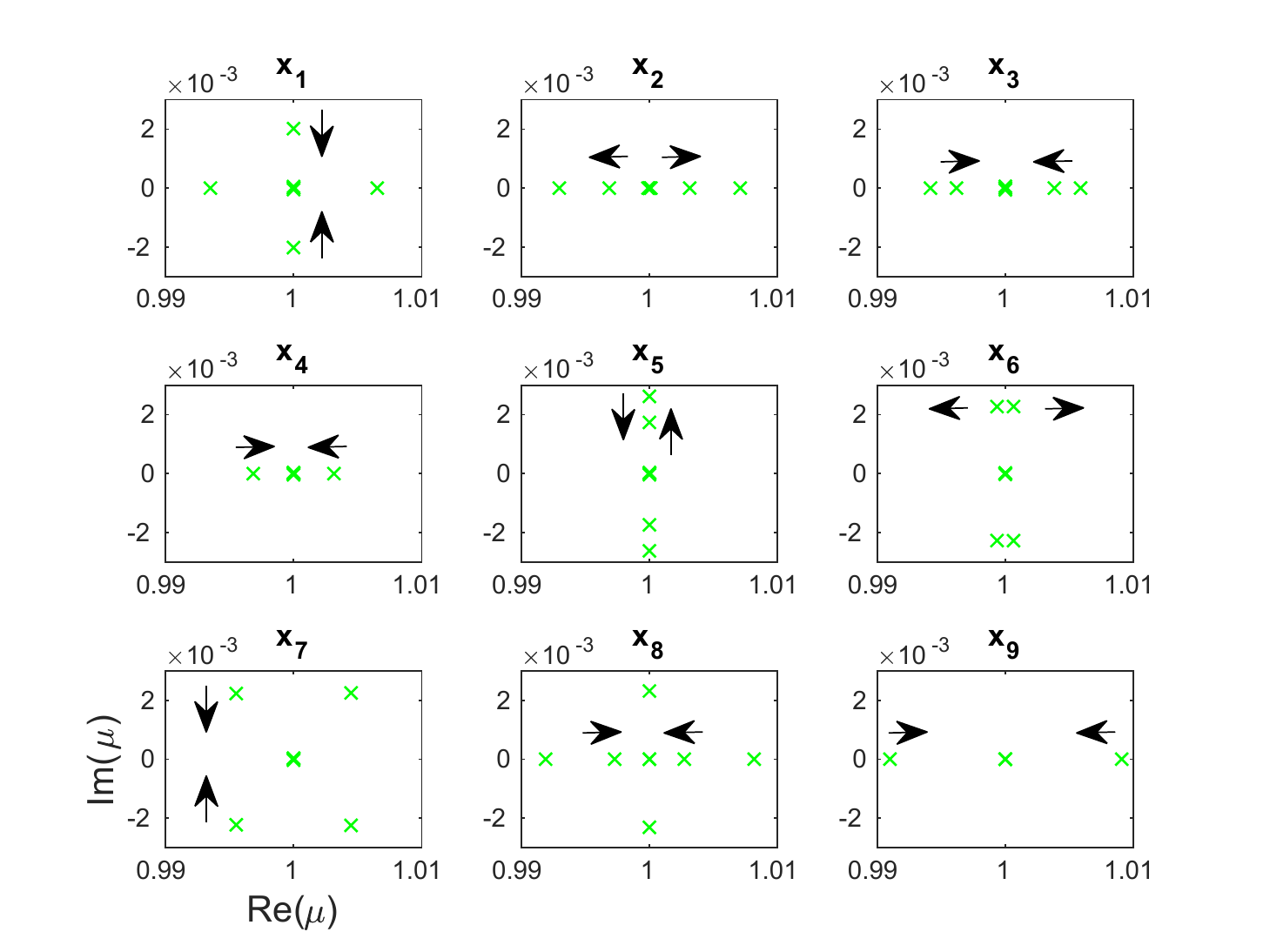}}
\caption{\footnotesize (a) Energy $H$ as a function of frequency $\omega$. The two insets show the strain \eqref{eq:strain} and angle variables at the points $A$, $B$, and $C$ along the solution curve. (b) $H(\omega)$ for the single-period (blue curve) and double-period (red and green curves) solution branches at twice their frequency. The bifurcation points are marked by $d$ and $e$. The two insets show the strain and angle variables at the points $D$ and $E$ along the double-period solution curves. (c) Maximum modulus $|\mu|$ of Floquet multipliers versus frequency $\omega$ along the single-period (blue) and double-period (red and green) solution branches. The insets show the corresponding Floquet multipliers near the unit circle. While the double-period solution along the red curve coincides with the single period solution (blue curve) at the bifurcation point $e$, the Floquet multipliers for the double-period solution are squares of those for the single-period one, resulting in the gap between the blue and red curves. (d) Upper panel: largest modulus $|\mu|$ of the real Floquet multipliers as a function of frequency $\omega$ along the blue single-period and green double-period solution curves near the bifurcation point $d$. Lower panel: second largest modulus $|\mu|$ of the real Floquet multipliers as a function of $\omega$ along the blue single-period and red double-period solution curves near the bifurcation point $e$. Note that these real Floquet multipliers are negative for the blue curve and positive for the red and green curves. (e) Enlarged view of $H(\omega)$ along the green double-period solution curve. The dashed vertical lines indicate the local minimum (left) and maximum (right). The points $x_1,\dots,x_9$ correspond to the Floquet multiplier panels shown in (f). The inset shows an enlarged view around the cluster of points. (f) The Floquet multipliers near $\mu=1$ for the points marked in the panel (e). The arrows indicate the motion of the Floquet multipliers. Here and in the remainder of this section we have $\alpha = 1.8$, $K_s = 0.02$, $K_{\theta} = 1.5 \times 10^{-4}$, $N = 200$, and $\phi_0 = 26\pi/180$.
}
\label{fig:EnergyFloquet_phi026pi}
\end{figure}

\begin{figure}[!htb]
\centering
\subfloat[]
{\includegraphics[width=0.33\textwidth]{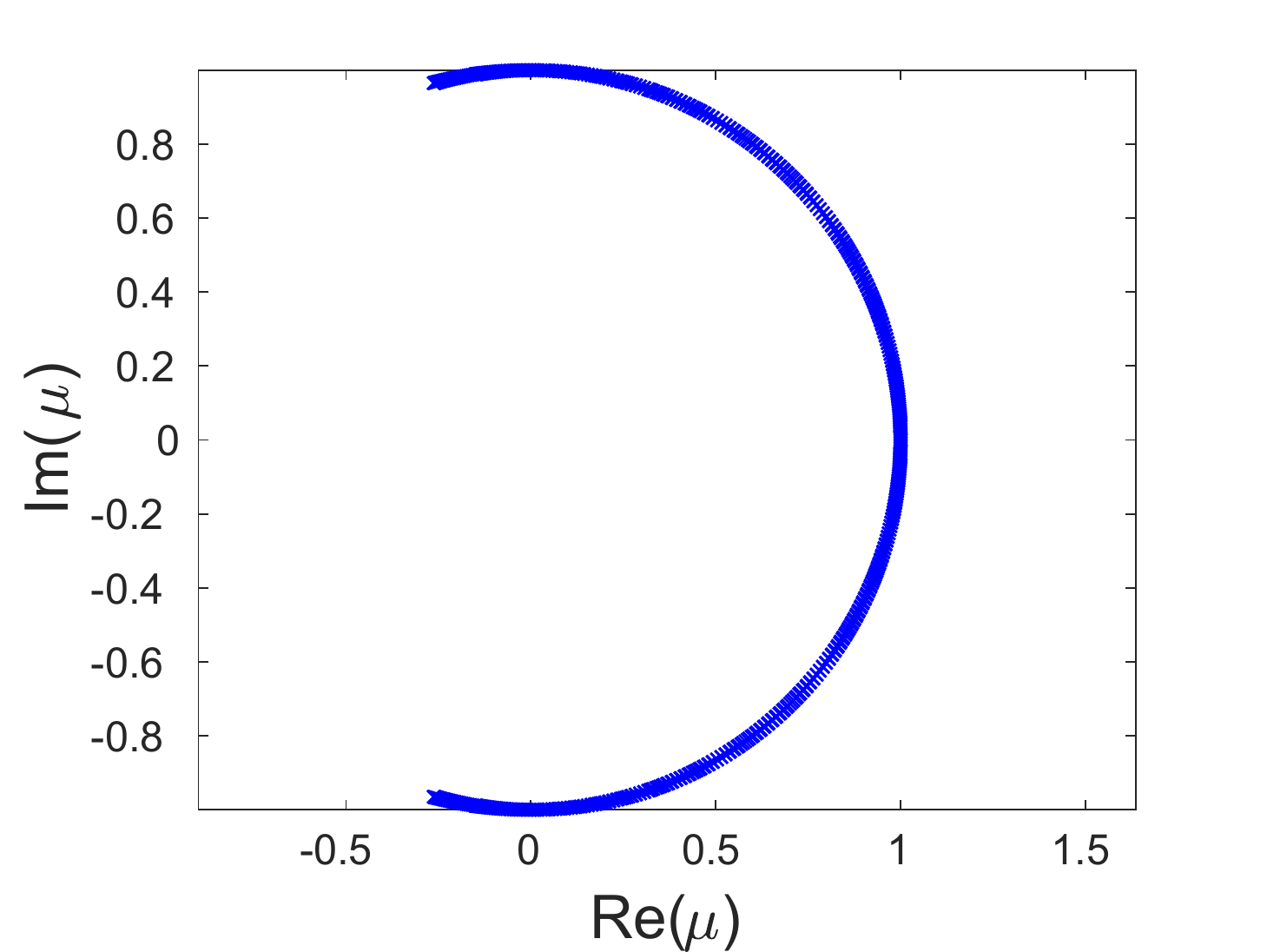}}
\subfloat[]
{\includegraphics[width=0.33\textwidth]{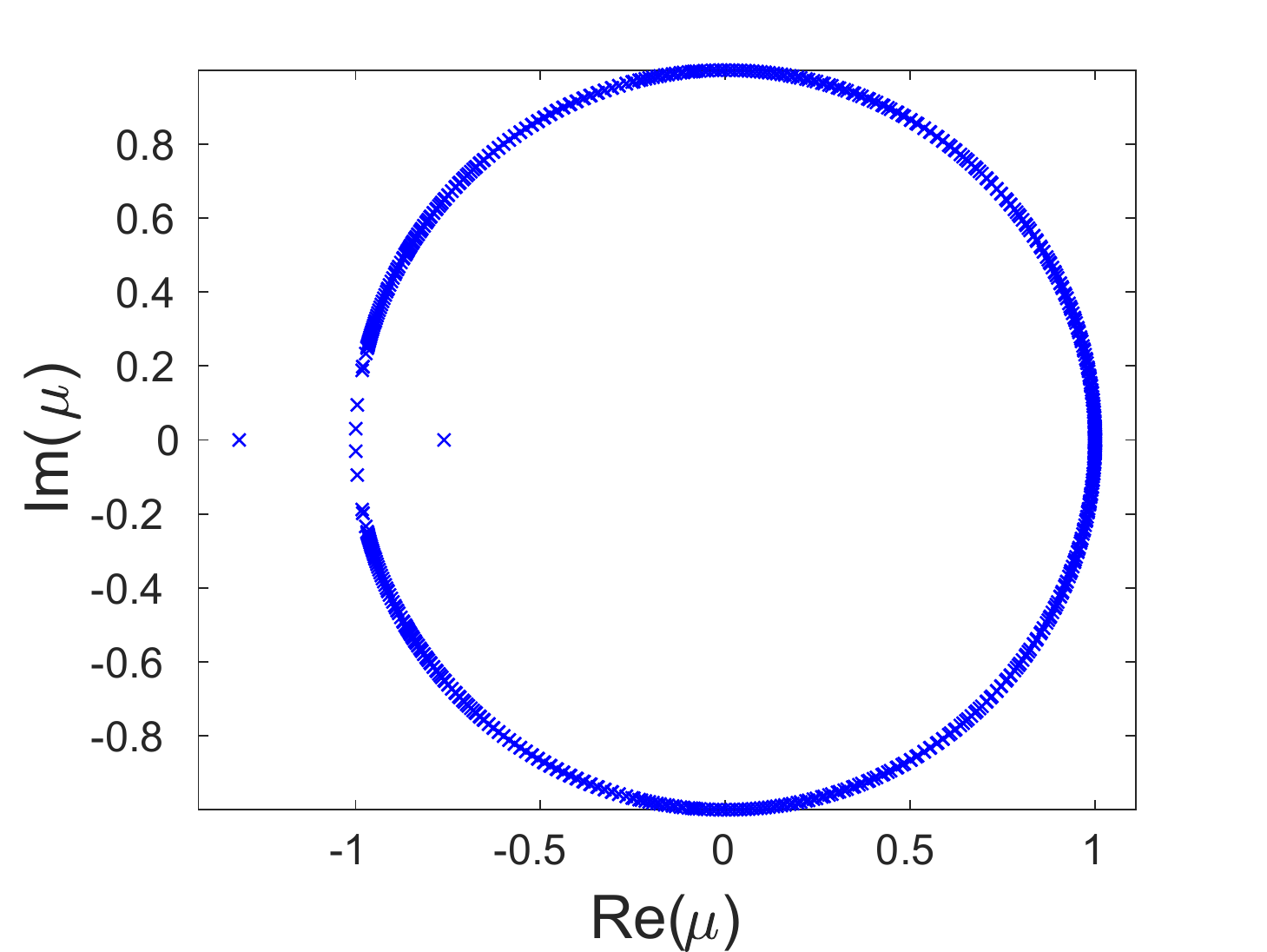}}
\subfloat[]
{\includegraphics[width=0.33\textwidth]{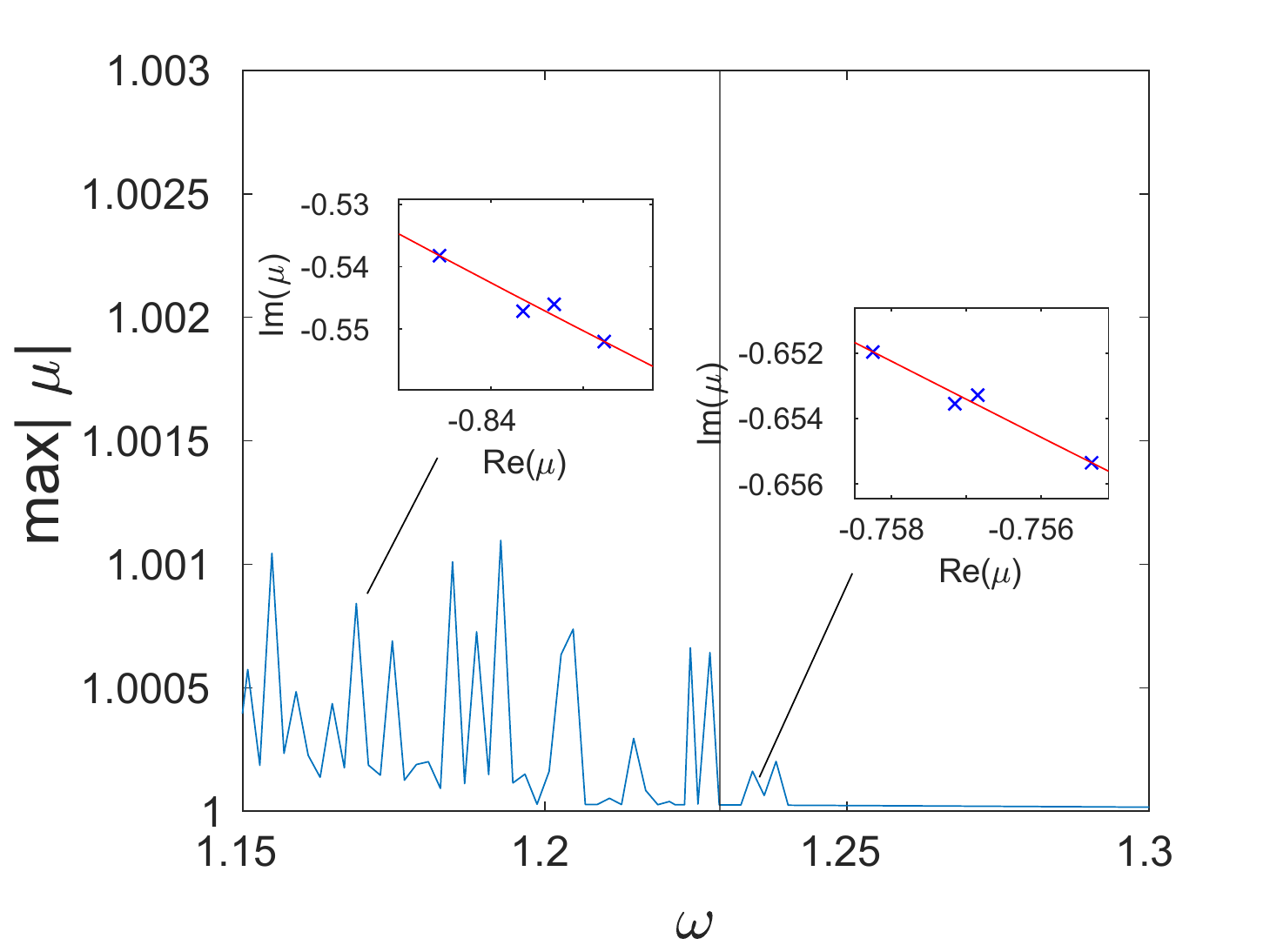}}
\caption{\footnotesize  Panels (a) and (b) show Floquet multipliers $\mu$ at the start ($\omega = 1.57$, panel (a)) and the end ($\omega = 0.9972$, panel (b)) of the continuation. Panel (c) shows an enlarged view of Fig.~\ref{fig:EnergyFloquet_phi026pi}(b). The vertical line indicates the frequency $\omega = 1.229$, at which the top optical and bottom acoustic arcs shown in Fig.~\ref{fig:AcousticOpticalContribution} below first intersect (see the text for details). Here $|\mu|>1$ corresponds to oscillatory instabilities, as shown in the insets, where the red curve is part of the unit circle.}
\label{fig:Multipliers_phi026pi}
\end{figure}

Fig.~\ref{fig:EnergyFloquet_phi026pi}(a) shows the energy $H$ of the breathers bifurcating from the $k=0$ mode as a function of the frequency $\omega$. As we will see below, using this pair of independent and dependent variables to illustrate our bifurcation diagrams allows us to connect the change in monotonicity of the energy-frequency curve with a potential stability change \cite{Kevrekidis2016}.
As illustrated in the insets, the amplitude of both the strain \eqref{eq:strain} and angle variables increases as the frequency is decreased away from the edge of the optical band, i.e., as the strength of the nonlinear contribution increases. The maximum modulus of the Floquet multipliers computed for this solution branch is shown by the blue curve in Fig.~\ref{fig:EnergyFloquet_phi026pi}(c). One can see that it exceeds unity and rapidly increases near the end of the continuation. As illustrated in the inset (see also panels (a) and (b) of Fig.~\ref{fig:Multipliers_phi026pi}, which show the Floquet multipliers at the beginning and the end of the continuation, respectively), this is due to the emergence of a pair $(\mu,1/\mu)$ of real Floquet multipliers from $\mu = -1$ at $\omega=1.05155$. One of these
has modulus greater than one and hence leads to an {exponential} instability, at the point $d$ in Fig.~\ref{fig:EnergyFloquet_phi026pi}(b), which corresponds to a \emph{period-doubling bifurcation}~\cite{Flach08}. A second pair of real Floquet multipliers emerges from $\mu =-1$ at $\omega = 1.05006$, leading to another exponential instability mode (not further explored). As the frequency is decreased, these multipliers first move away from the unit circle along the real line and then start moving back toward it, eventually colliding at $\mu=-1$ at $\omega = 1.0499$, which corresponds to the point $e$ in Fig.~\ref{fig:EnergyFloquet_phi026pi}(b) and is associated with another period-doubling bifurcation.

To compute the double-period solutions that arise as a result of the bifurcations at the points $d$ and $e$ along the single-period solution branch, we used the same iterative procedure as discussed above with the initial guess consisting of a single-period solution with twice the frequency perturbed along the corresponding unstable mode. Solutions along the bifurcating branches were then obtained using parameter continuation in frequency or energy. The resulting energy as a function of frequency for the double-period solutions (red and green curves) is shown in Fig.~\ref{fig:EnergyFloquet_phi026pi}(b) for each case together with the single-period solution branch (blue curve) discussed above. The double-period solution curves are plotted at twice their actual frequency in order to facilitate the comparison with the single-period solution curve. Insets in Fig.~\ref{fig:EnergyFloquet_phi026pi}(b) show examples of the symmetric breather solutions along the different double-period solution curves. As the insets of Fig.~\ref{fig:EnergyFloquet_phi026pi}(c) reveal, the Floquet spectra of the double-period and single-period solution branches are markedly different. While the single-period solutions, as noted above, are characterized by an exponential period-doubling instability associated with a Floquet multiplier $\mu<-1$ for frequencies below the value at the bifurcation point $d$, the double-period branches exhibit an exponential instability associated with a Floquet multiplier satisfying $\mu>1$. As the bifurcation points are approached, the corresponding pairs of real multipliers collide at $\mu=-1$ for the parent single-period branch and at $\mu=1$ for the bifurcating branches.

To examine the nature of these bifurcations further, we plot in the top panel of Fig.~\ref{fig:EnergyFloquet_phi026pi}(d) the largest modulus of real Floquet multipliers $\mu$ as a function of $\omega$ along the green and blue curves near the bifurcation point $d$. One can see that at the period-doubling bifurcation point $d$ the single-period branch develops an exponential instability associated with a Floquet multiplier $\mu<-1$ via a subcritical pitchfork bifurcation of the double-period branch, which has a pair of real multipliers $(\mu,1/\mu)$ with $\mu>1$. In the bottom panel of Fig.~\ref{fig:EnergyFloquet_phi026pi}(d), we show the second largest modulus of the real Floquet multipliers near the bifurcation point $e$, where the second pair of real multipliers emerges near $\mu=-1$ for the single-period branch and near $\mu=1$ for the bifurcating red branch. Due to the presence of the first pair of real multipliers, all solutions are unstable near the bifurcation point $e$, as indicated in Fig.~\ref{fig:EnergyFloquet_phi026pi}(c).

Note that the upper branch of the multivalued energy-frequency function corresponding to the unstable green double-period solution curve bifurcating from the point $d$ has a local minimum and a local maximum, marked by the dashed vertical lines in Fig.~\ref{fig:EnergyFloquet_phi026pi}(e). As illustrated in the first four panels in Fig.~\ref{fig:EnergyFloquet_phi026pi}(f), these extrema are associated with a change of multiplicity of the Floquet multiplier at $\mu=1$ along this branch and subsequent emergence or collision of a second pair of real Floquet multipliers. The change in multiplicity of the unit Floquet multiplier when $H'(\omega)$ changes sign is consistent with the energy-based stability criterion proved in \cite{Kevrekidis2016} for discrete breathers in Fermi-Pasta-Ulam and Klein-Gordon lattices. Note, however, that in this case the change in multiplicity does not lead to a stability change due to the presence of an additional pair of non-unit real multipliers at these frequency values. As we trace the solution curve toward the point $d$, this pair collides at $\mu=1$ on the unit circle at a bifurcation point and subsequently briefly remains on it (see panels 4 and 5 in Fig.~\ref{fig:EnergyFloquet_phi026pi}(f)), while the solutions are still unstable due to the presence of complex multipliers $\mu$ satisfying $|\mu|>1$ (not shown in panel 5). However, as illustrated in panels 7 and 8 in Fig.~\ref{fig:EnergyFloquet_phi026pi}(f), two pairs of real multipliers subsequently emerge on the real axis via collisions of complex conjugate pairs of multipliers. One of the pairs eventually collides on the unit circle at another bifurcation point, leaving a single pair (panel 9), which in turn collides at $\mu=1$ at the point $d$.

\begin{figure}[!htb]
\centering
\subfloat[]
{\includegraphics[width=0.33\textwidth]{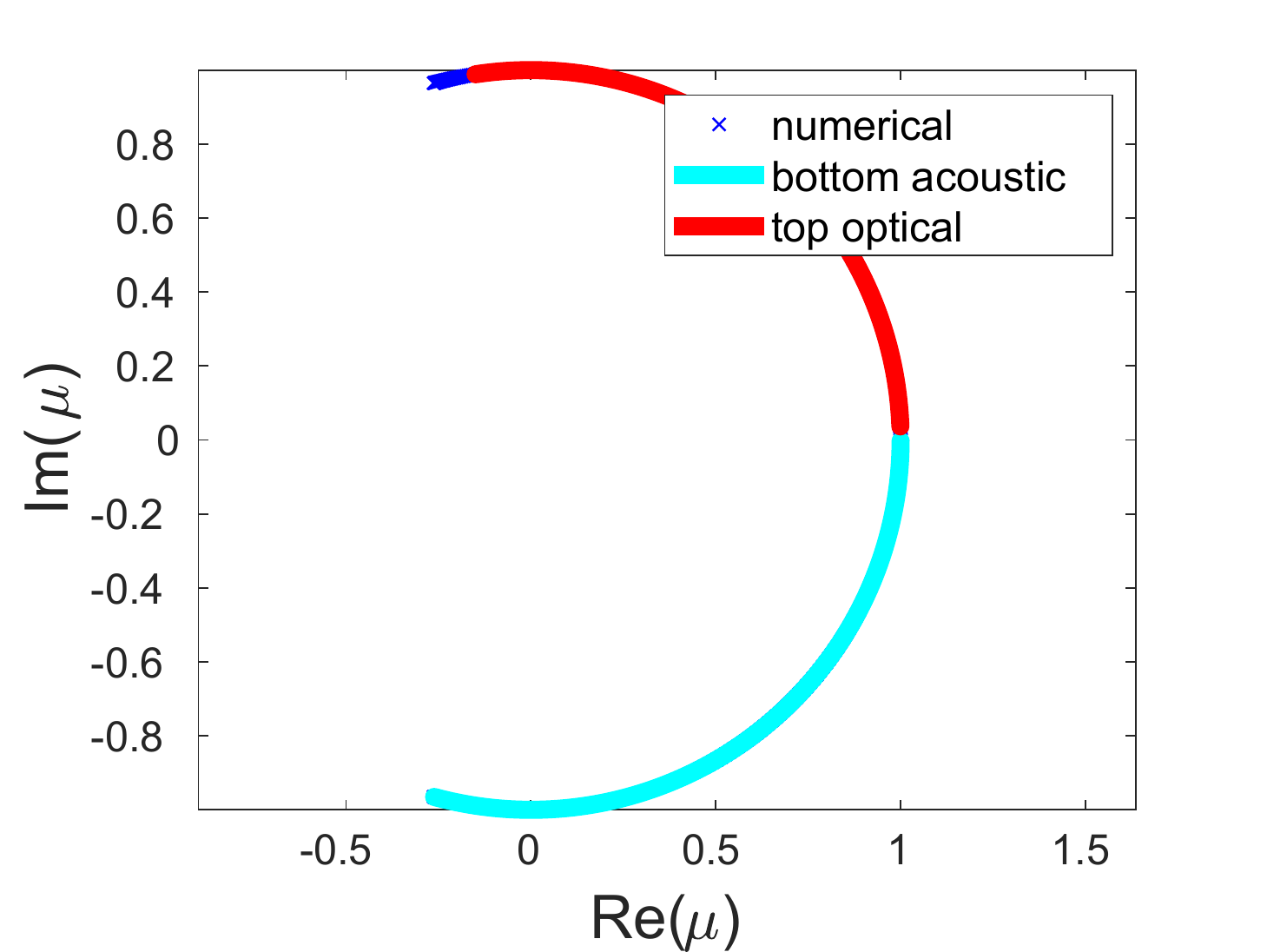}}
\subfloat[]
{\includegraphics[width=0.33\textwidth]{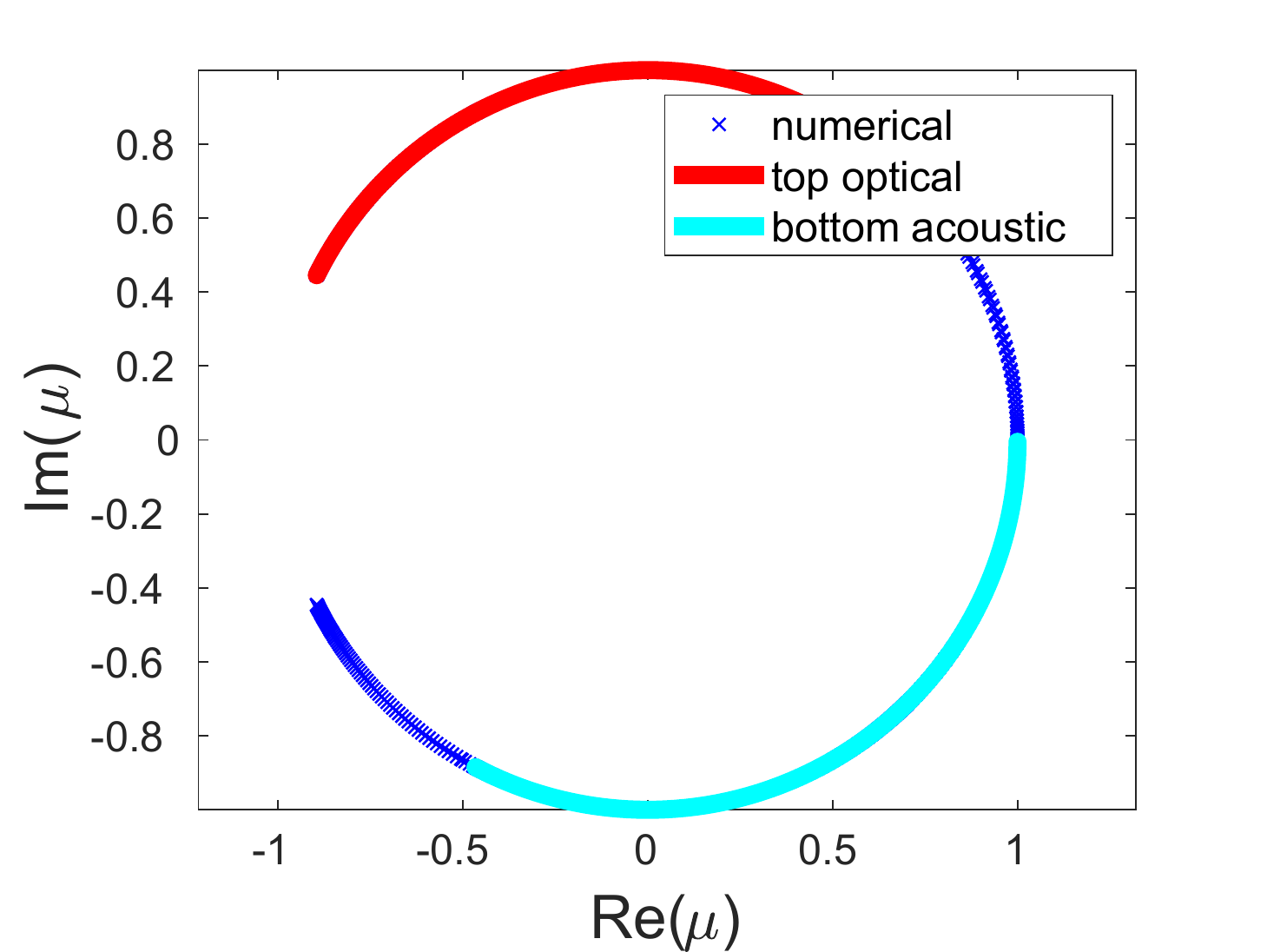}}
\subfloat[]
{\includegraphics[width=0.33\textwidth]{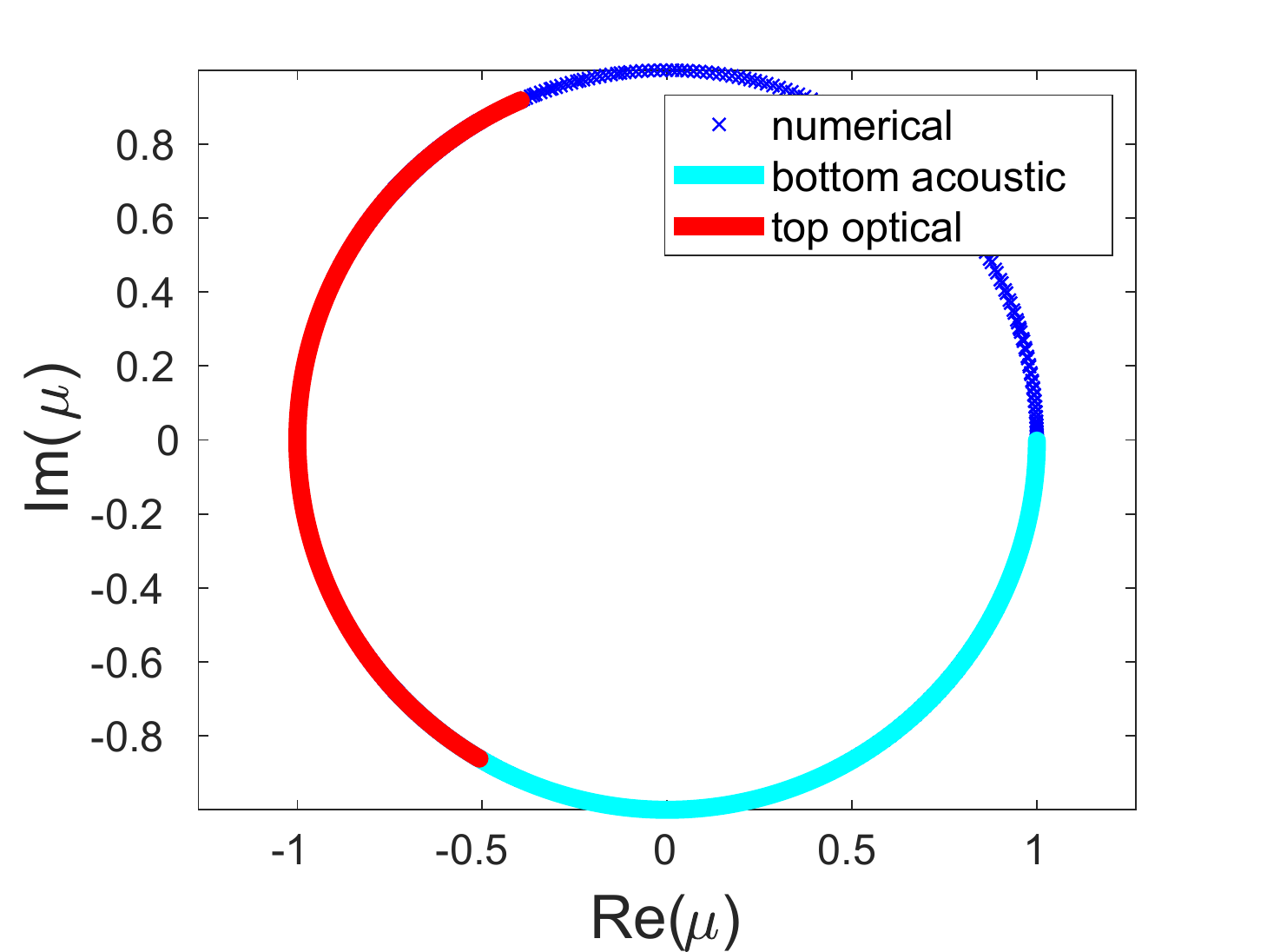}}
\caption{\footnotesize  Numerically computed Floquet multipliers (dark blue crosses) and arcs of Floquet multipliers \eqref{eq:BackgroundFloquet} corresponding to top optical ($\omega_{+}(k)$, red arc) and bottom acoustic ($-\omega_{-}(k)$, light blue arc) dispersion bands at (a) $\omega=1.57$; (b) $\omega = 1.4$; (c) $\omega=1.201$.
}
\label{fig:AcousticOpticalContribution}
\end{figure}

The enlarged view of the Floquet multiplier curve for the single-period solution branch and the insets shown in Fig.~\ref{fig:Multipliers_phi026pi}(c) reveal that the onset of the period-doubling instability is preceded by small-magnitude \emph{oscillatory} instabilities associated with pairs of multipliers colliding on the unit circle and then moving slightly off it in the form of a quartet as discussed above.
Note also that the Floquet multipliers $\mu$ form an arc on or near the unit circle. Using the linearization \eqref{eq:linearized} of Eq.~\eqref{eq:EoM} about the uniform equilibrium state for an infinite chain, one can show \cite{chaunsali2021} that the background state of the breather with period $T$ contributes the Floquet multipliers
\beq
\mu = e^{\pm i\omega_{\pm}(k) T},
\label{eq:BackgroundFloquet}
\eeq
where we recall from Sec.~\ref{sec:dispersion} that $\omega_{+}(k)$ and $\omega_{-}(k)$ are the optical and acoustic branches of the dispersion relation. As we vary $k$ from $0$ to $\pi$, we obtain arcs of multipliers along the unit circle. Such arcs corresponding to the upper optical ($\omega_{+}(k)$, red arc) and the bottom acoustic ($-\omega_{-}(k)$, light blue arc) bands are depicted in panels (a), (b) and (c) of Fig.~\ref{fig:AcousticOpticalContribution} for different values of $\omega$ (and hence different $T=2\pi/\omega$ in \eqref{eq:BackgroundFloquet}) along with the numerically computed Floquet multipliers (dark blue crosses) for the obtained DB solutions. There are also symmetric arcs (not shown in the figure) corresponding to the bottom optical ($-\omega_{+}(k)$) and the upper acoustic ($\omega_{-}(k)$) bands.

Under the mapping given by \eqref{eq:BackgroundFloquet}, the left ends of the arcs corresponding to the top optical and bottom acoustic bands, respectively, seen in Fig.~\ref{fig:AcousticOpticalContribution}, are associated with $\omega_+(\pi)$ and $-\omega_-(\pi)$. As $\omega$ is decreased, the two ends approach each other along the unit circle and eventually coincide when
\[
e^{i2\pi \omega_+(\pi)/\omega} = e^{-i2\pi \omega_-(\pi)/\omega},
\]
which yields
\[
\frac{\omega_+(\pi)+\omega_-(\pi)}{\omega} = n,
\]
where $n$ is a positive integer. We find that the first such collision takes place when $n=2$, which together with \eqref{eq:k_pi} yields
\[
\omega = \frac{2+\alpha\sqrt{2(K_{\theta}+2K_s\cos^2\phi_0)}}{2} \approx 1.2293.
\]
This predicted value of $\omega = 1.2293$ is close to the first significant peak shown in Fig.~\ref{fig:Multipliers_phi026pi}(c), although there are also two smaller peaks to the right of it at $\omega = 1.231$ and $\omega = 1.239$. This discrepancy between predicted and actual collision frequency values may be attributed to numerical accuracy of computing the Floquet multipliers, as well as possible
effects of weak nonlinearity.

\begin{figure}[!htb]
\centering
{\includegraphics[width=0.7\textwidth]{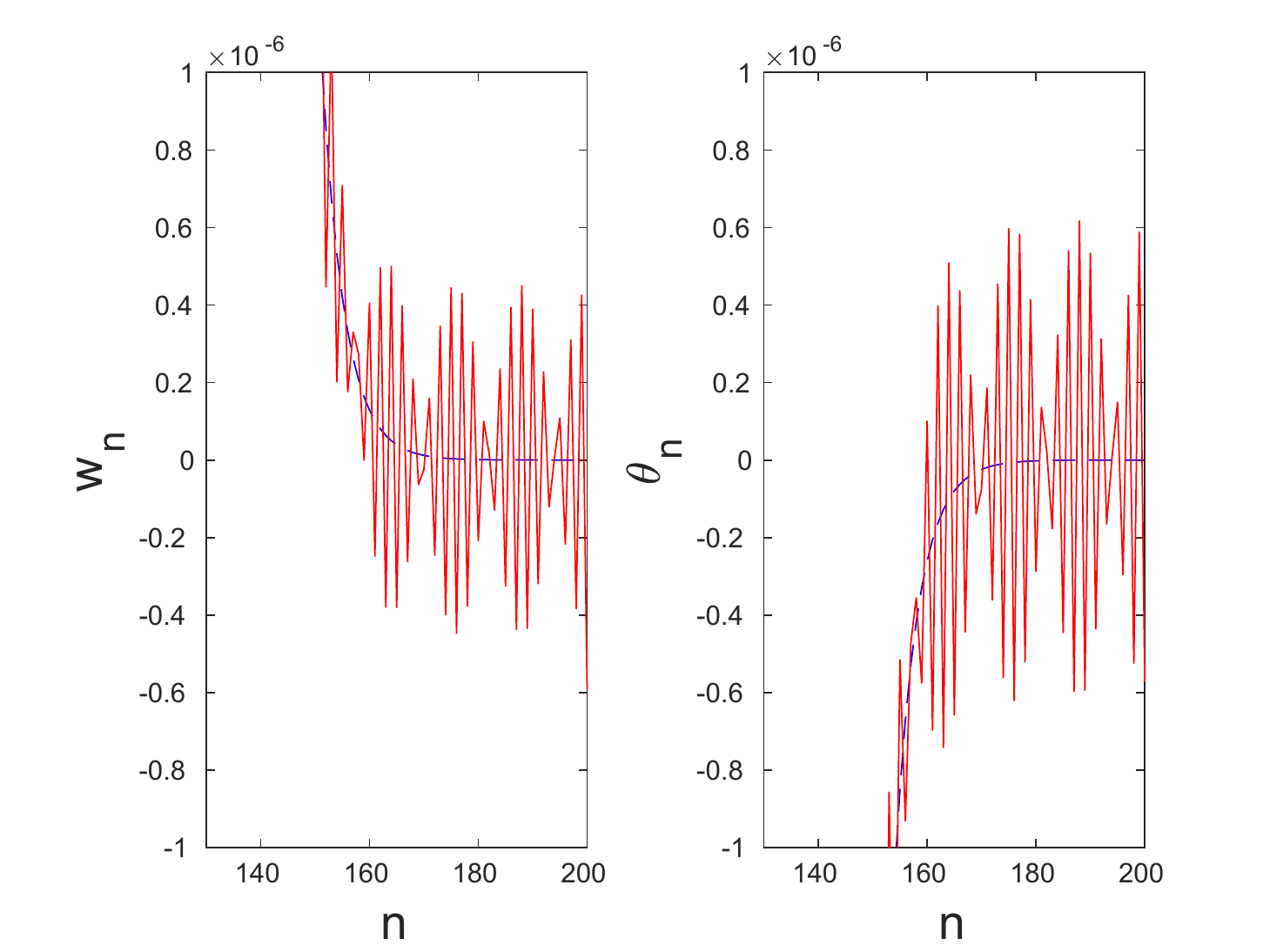}}
\caption{\footnotesize The angle and strain variables near the right end of the chain for the phantom breather with frequency $\omega = 0.9972$ (solid red) and the regular (localized) discrete breather with frequency $\omega = 1.02$ (dashed blue).}
\label{fig:tail}
\end{figure}

The solution curve shown in Fig.~\ref{fig:EnergyFloquet_phi026pi}(a) was continued until the frequency $\omega = 0.9972$, and thus includes solutions with frequencies $\omega \leq 1$. As noted in Sec.~\ref{sec:dispersion}, these frequencies are associated with second harmonic resonances of the DB solution with the linear waves that have frequencies in the optical band. As a result, the corresponding solutions are no longer localized and instead possess non-decaying oscillatory wings. Such solutions are known as \emph{phantom breathers}
\cite{Morgante02} or \emph{nanoptera} \cite{Boyd90,SegurKruskal87}.  The latter term stems from their non-vanishing tails given the resonance
with the linear modes. An example of a phantom breather with frequency $\omega = 0.9972$ (red curve) is shown in Fig.~\ref{fig:tail} along with the regular (localized) DB solution at $\omega = 1.02$ (dashed blue).

\begin{figure}[!htb]
\centering
\subfloat[]
{\includegraphics[width=0.5\textwidth]{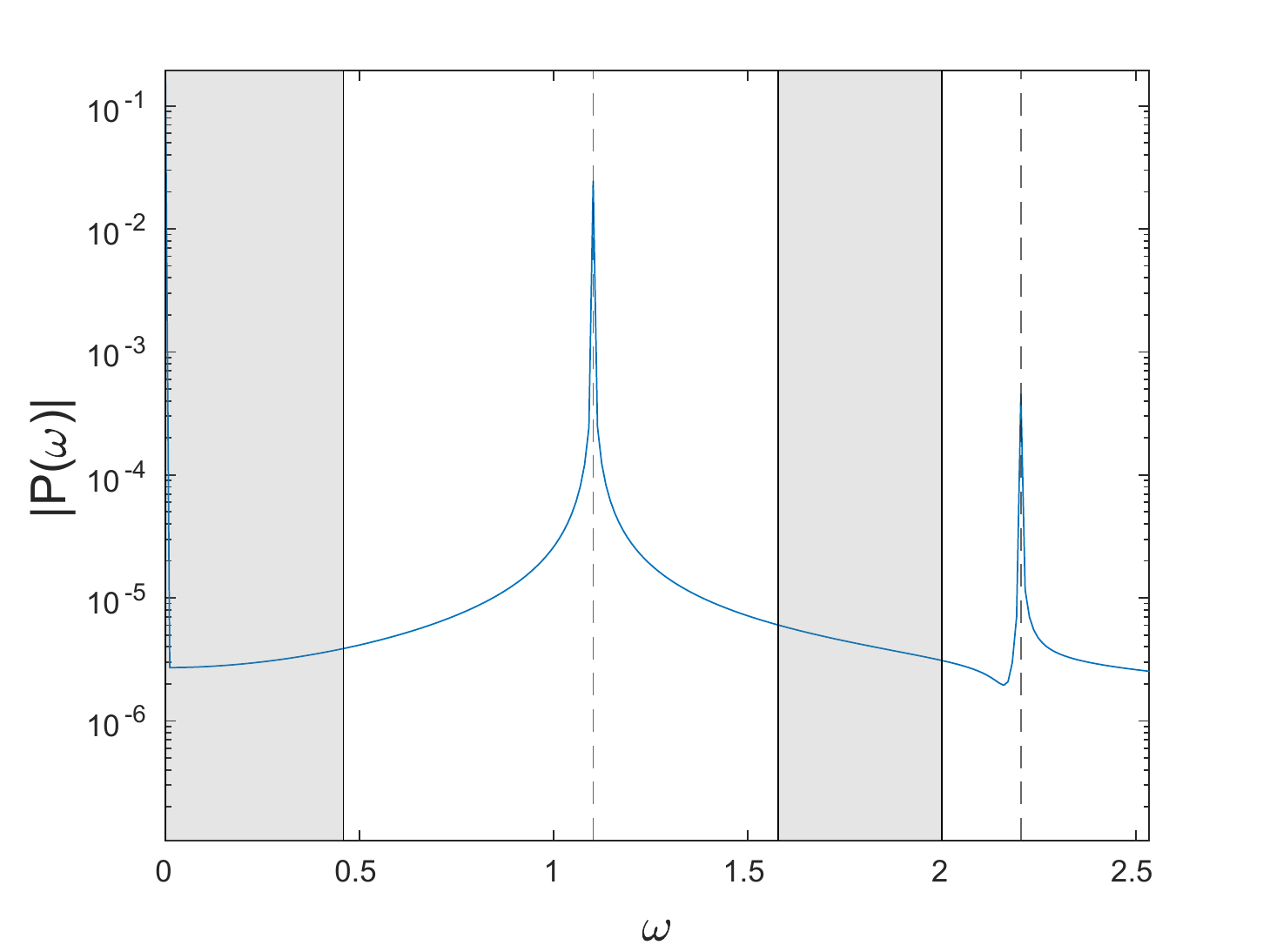}}
\subfloat[]
{\includegraphics[width=0.5\textwidth]{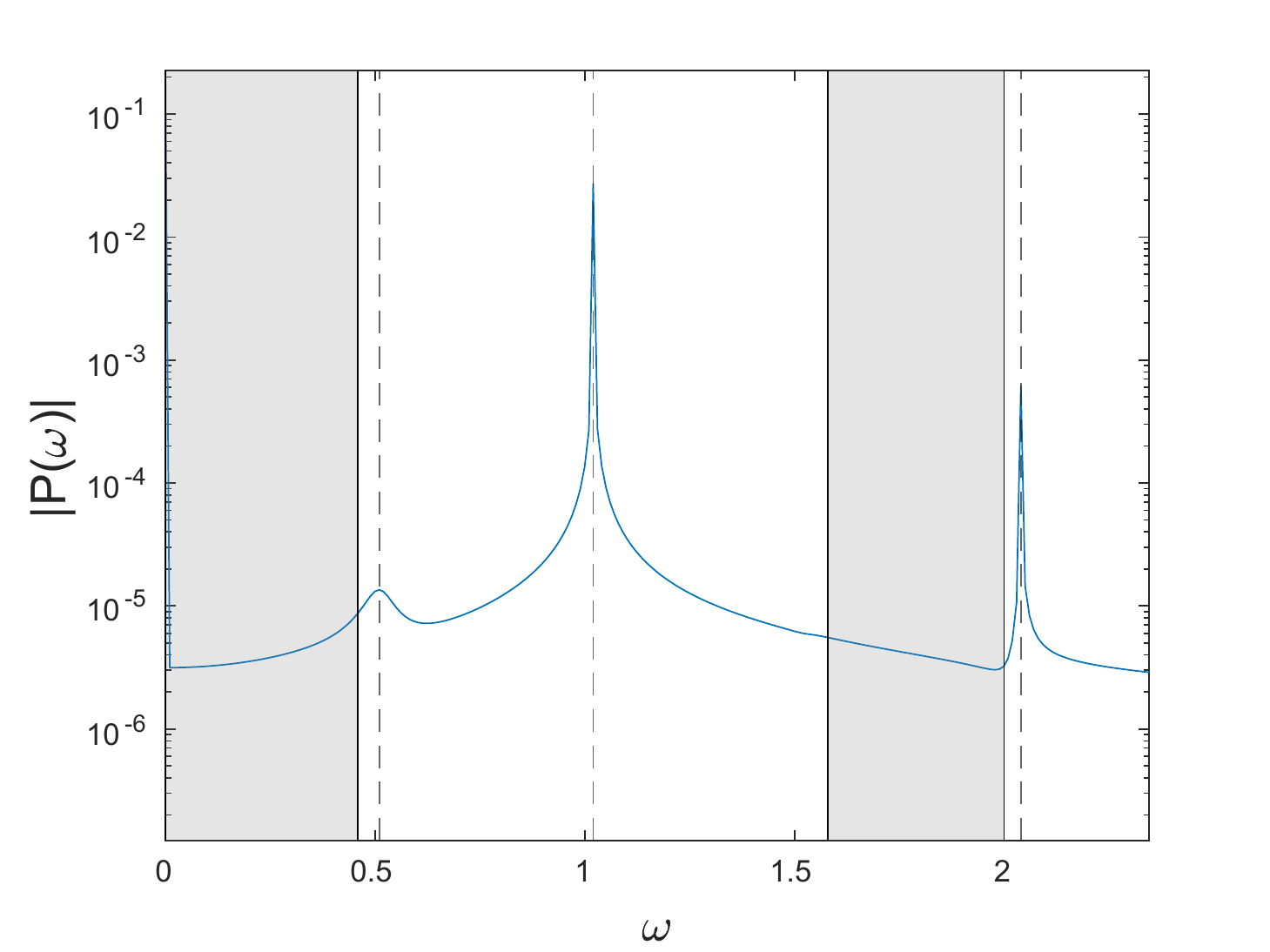}}
\caption{\footnotesize  The amplitude spectrum $P(\omega)$ obtained using the FFT for different values of the prescribed breather frequency $\tilde{\omega}$: (a) $\tilde{\omega} = 1.1$; (b) $\tilde{\omega} = 1.02$. The left and right shaded stripes in each of the bottom panels indicate the acoustic and optical dispersion bands, respectively. The dashed vertical lines indicate $\tilde{\omega}$ and $2\tilde{\omega}$ in both panels and $\tilde{\omega}/2$ in panel (b). It is clear that the frequencies
  associated with the breather do not resonate with the linear spectral bands
in the cases shown.}
\label{fig:FFT_phi026pi}
\end{figure}

We now consider the Fourier spectrum associated with the dynamic evolution of the obtained breathers with prescribed frequency $\tilde{\omega}$. Fig.~\ref{fig:FFT_phi026pi} shows the Fast Fourier Transform (FFT) results involving the dynamics simulated over a course of $100$ oscillation periods for two different values of $\tilde{\omega}$, along with the acoustic and optical bands shaded in gray. In the case $\tilde{\omega} = 1.1$ (panel (a)), there are only two peaks at nonzero frequencies for the displayed range, at $\tilde{\omega}$ and $2\tilde{\omega}$, and the latter is clearly above the top of the optical band (the right shaded strip) at $\omega=2$. When $\tilde{\omega} = 1.02$ (panel (b)), one can see a third nonzero-frequency peak in addition to $\tilde{\omega}$ and $2\tilde{\omega}$. This peak is at $\tilde{\omega}/2$ and is associated with the period-doubling instability, which is present at this frequency. Note that $2\tilde{\omega}$ is above the optical band (the right shaded strip), and $\tilde{\omega}/2$ is above the acoustic band (the left shaded strip), so there are no resonances with either optical or acoustic linear waves. In contrast, in the case $\tilde{\omega} = 0.9972$ (not shown), the peak at $2\tilde{\omega}$ is just inside the optical band, and the second-harmonic resonance results in the phantom breather structure shown in Fig.~\ref{fig:tail}.

\section{Snake-like solution branches}
\label{sec:snaking}
As we have seen, the existence of DB solutions with frequencies inside the band gap requires rather large angles $\phi_0$ (above $16^\circ$) for the set of model parameters used in the previous subsection. Since large offset angles may render the present description of the system with only two degrees of freedom somewhat less accurate \cite{Guo2018}, we consider in what follows the parameters $\alpha=5$, $K_s=0.02$, $K_\theta=0.01$, which allow breather existence at smaller values of $\phi_0$.

\subsection{Branches associated with the $k=\pi$ mode}
\label{sec:pi_mode}
We start by considering solutions that exist when the bottom of the optical band is at $k=\pi$, which, as shown in Sec.~\ref{sec:dispersion}, can occur when the angle $\phi_0$ is above $\phi_0''$. Recalling that $\phi_0''=0.1588$ for the chosen parameter values, we set $\phi_0 = 10\pi/180 \approx 0.1745$. The corresponding dispersion relation plot is shown in Fig.~\ref{fig:OpticalAcousticGap}(c).

\begin{figure}[!htb]
\centering
\subfloat[]
{\includegraphics[width=0.5\textwidth]{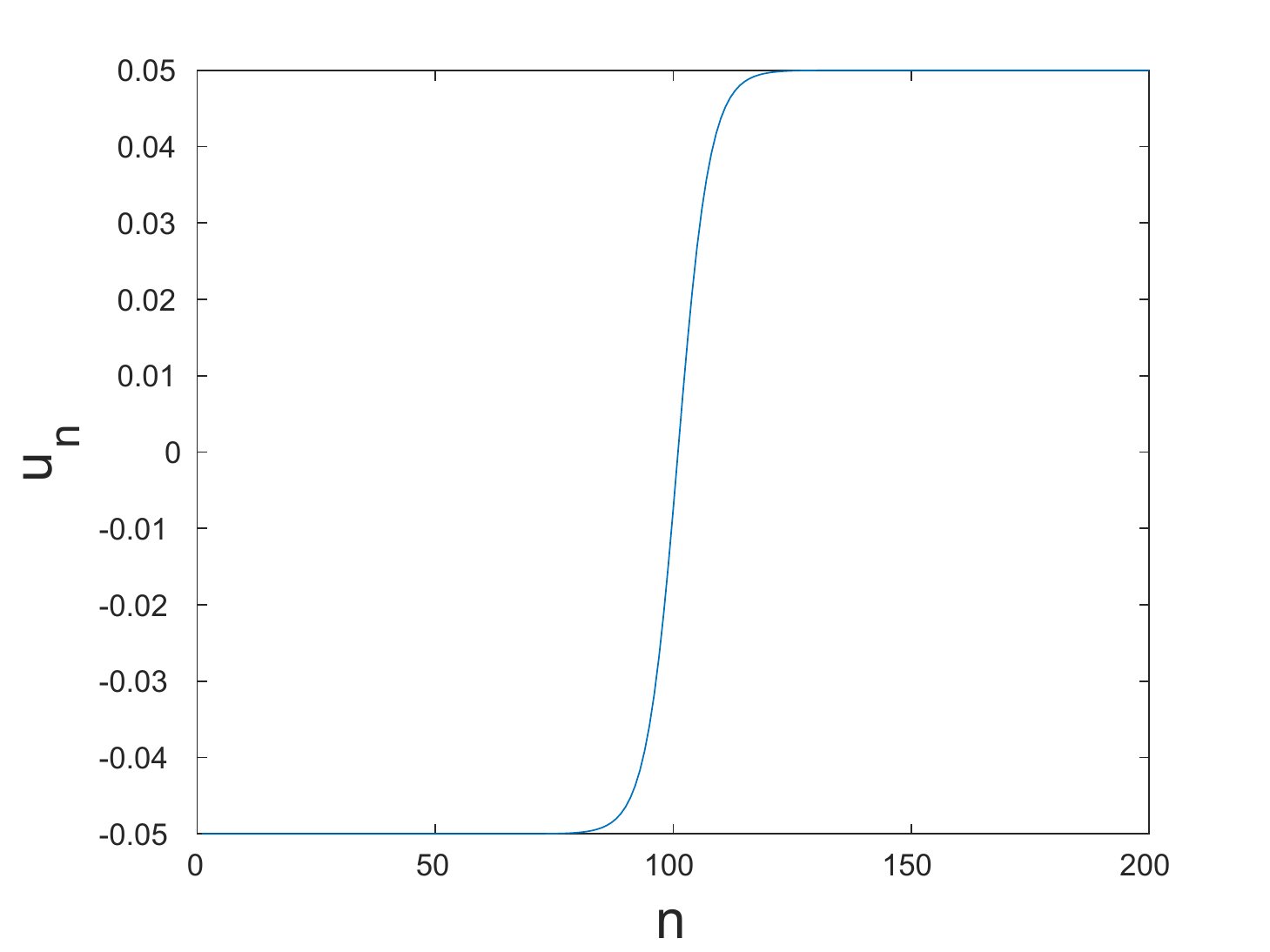}}
\subfloat[]
{\includegraphics[width=0.5\textwidth]{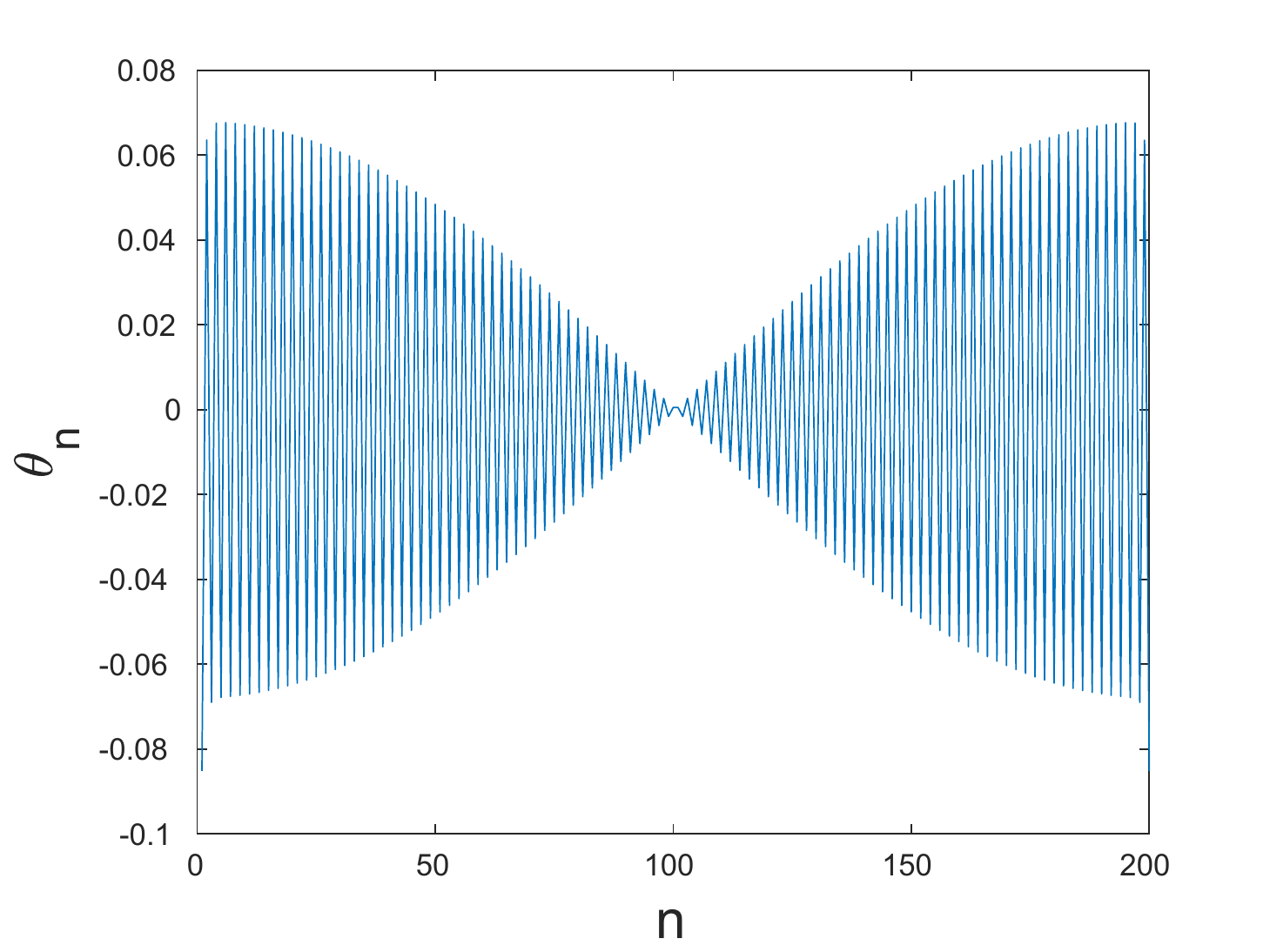}}
\caption{\footnotesize Initial guess for (a) displacement $u_n=\varepsilon_u\tanh(\delta(n - N/2))$; (b) angle $\theta_n$ obtained from the $\pi$-mode eigenvector (see the text for details). Here $\varepsilon_u = 0.05$ and $\delta = 0.15$.}
\label{fig:piguess}
\end{figure}

To compute solutions associated with the $k=\pi$ mode, we modify our initial guess as follows. To obtain the initial guess for the angle variable $\theta_n$, we solve the linear problem \eqref{eq:linearized} for the finite chain of size $N=200$ with zero strain and zero angle
prescribed at the boundaries, observing that the eigenvalues $\nu$ are equal to the negative of the square of the frequencies that make up the optical and acoustic bands obtained for the linearized problem, and selecting the angle-related part of the eigenvector associated with the eigenvalue $\nu=-\omega_{+}^2(\pi)=-4$.
Selecting the corresponding displacement part of the eigenvector did not yield nontrivial solutions, and thus we used the same form of the initial guess for $u_n$ as in \eqref{eq:initial_guess_k0}. Fig.~\ref{fig:piguess} shows the initial guess we used in the computation.

\begin{figure}[!htb]
\centering
{\includegraphics[width=0.75\textwidth]{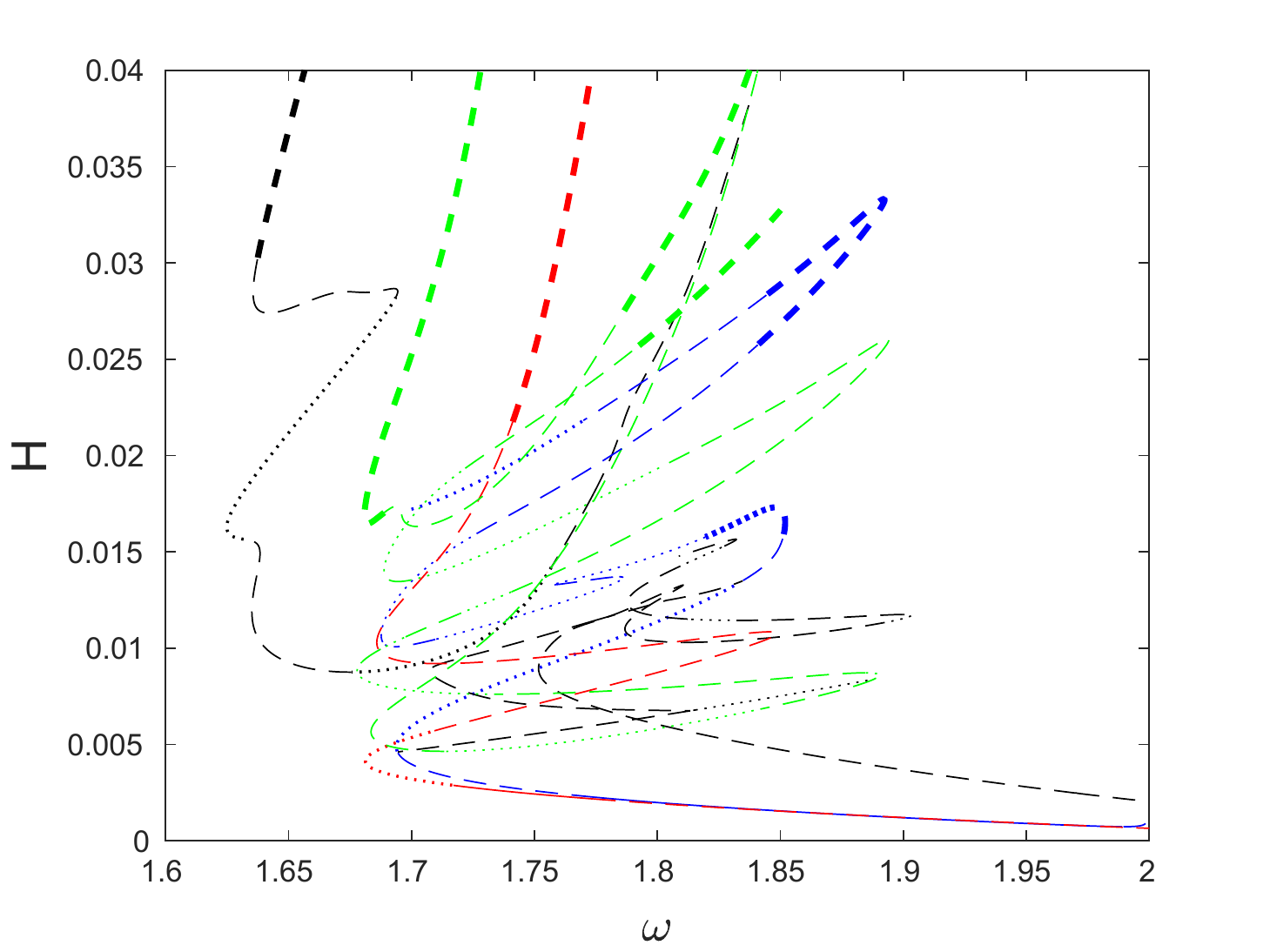}}
\caption{\footnotesize Energy $H$ of the computed DB solutions as a function frequency $\omega$. Blue, red and green curves are branches of solutions that have even symmetry, while the asymmetric solution curves are shown in black. Thin dashed portions of the curves indicate the presence of the real multiplier pairs $(1/\mu,\mu)$ with $\mu > 1$. Along the thick dashed segments there are also real multipliers $(1/\mu,\mu)$ with $\mu<-1$. Parts of the curve where there are only oscillatory instabilities with the maximum modulus of the Floquet multipliers exceeding $1.009$ are indicated by thin dotted segments. Solutions that also have real multiplier pairs $(1/\mu,\mu)$ with $\mu<-1$ are along the thick dotted parts. Solid curves indicate the portions where there are no exponential instabilities, and the maximum modulus of the Floquet multipliers is below $1.009$. Here and in the remainder of this subsection we have $\alpha = 5$, $K_s = 0.02$, $K_{\theta} =0.01$, $N = 200$, and $\phi_0 = 10\pi/180$.}
\label{fig:EnergyvFrequency_All}
\end{figure}

The results of our computations are summarized in Fig.~\ref{fig:EnergyvFrequency_All}, which shows the energy of the obtained solution branches as a function of frequency. Blue, red and green curves show branches of DB solutions that have even symmetry, while the black curves indicate asymmetric solution branches. For each solution branch, thin dashed portions of the curve indicate the existence of real Floquet multipliers satisfying $\mu > 1$, along with the corresponding real multipliers $1/\mu$ inside the unit circle along
the real line. Thick dashed segments indicate the additional presence of real Floquet multipliers $\mu$ and $1/\mu$ satisfying $\mu < -1$ and thus corresponding to a period-doubling instability akin the one discussed in Sec.~\ref{sec:period_doubling}. Parts of the curve where there are only oscillatory instabilities with the maximum modulus of the Floquet multipliers exceeding $1.009$ are shown by thin dotted segments, while along the thick dotted portions there are also real multipliers $\mu$ and $1/\mu$ with $\mu<-1$. Solid curves indicate the portions where there are no exponential instabilities, and the maximum modulus of the Floquet multipliers is below the threshold value $1.009$. Small-magnitude oscillatory instabilities along the solid portions are similar to the ones observed in Sec.~\ref{sec:period_doubling} and can be neglected, so that the associated solutions can be considered effectively
(i.e., practically, for long-time simulations) stable. The threshold of $1.009$ is
(by necessity) somewhat arbitrary and is connected with observations over
the time horizons selected for our numerical simulations of the breather dynamics.

\begin{figure}
\centering
\subfloat[]
{\includegraphics[width=0.4\textwidth]{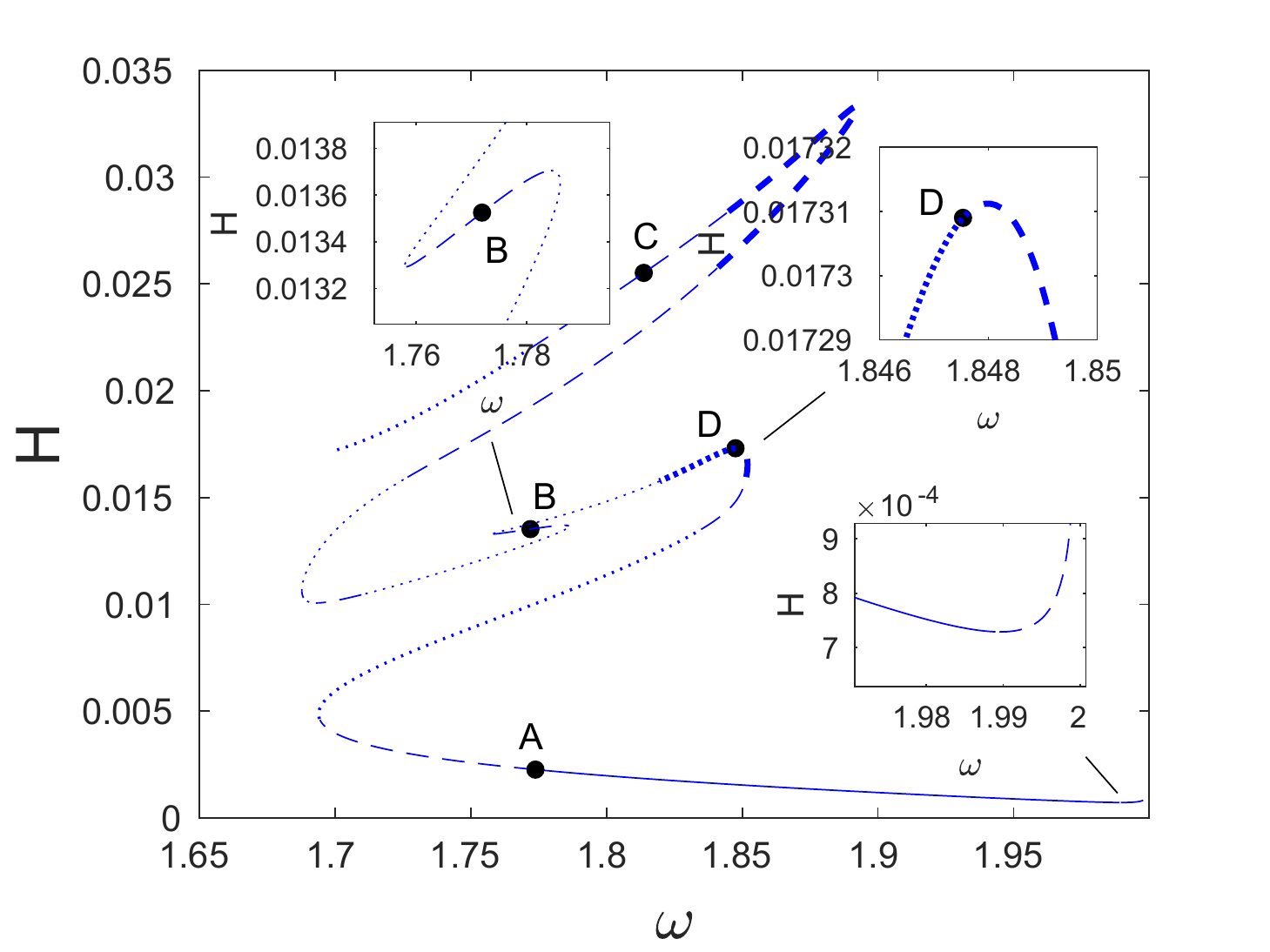}}
\subfloat[]
{\includegraphics[width=0.4\textwidth]{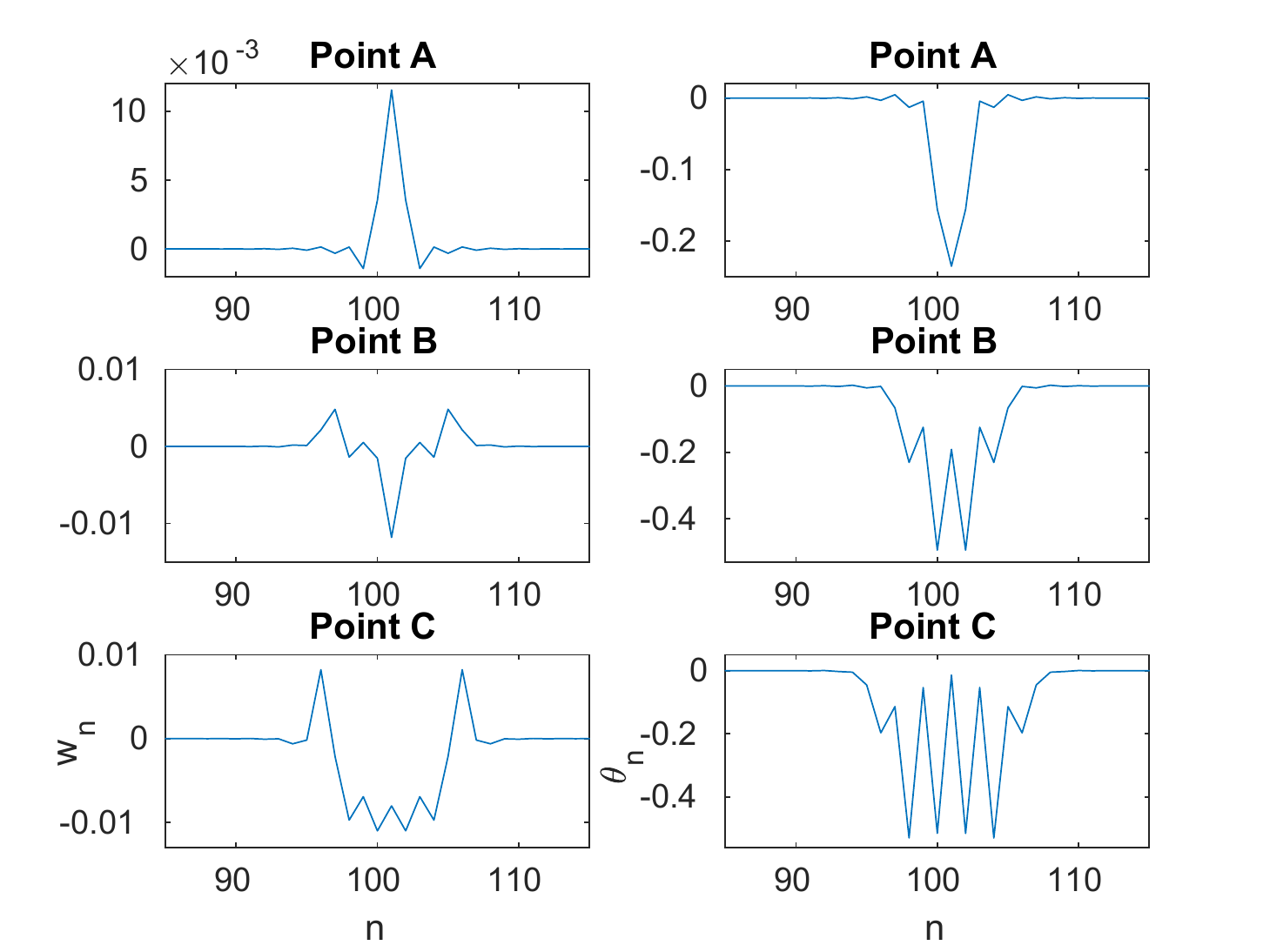}}\\
\subfloat[]
{\includegraphics[width=0.4\textwidth]{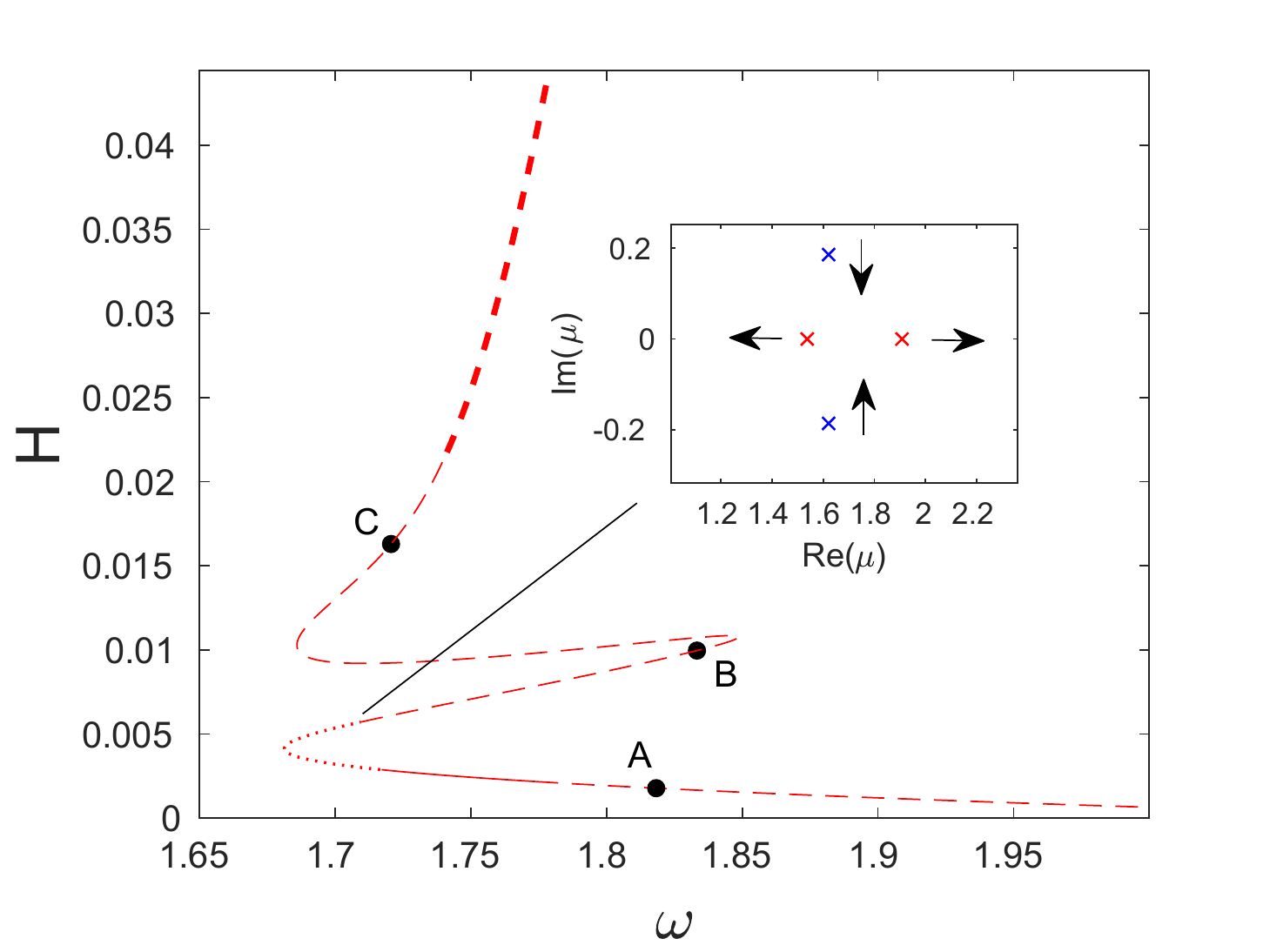}}
\subfloat[]
{\includegraphics[width=0.4\textwidth]{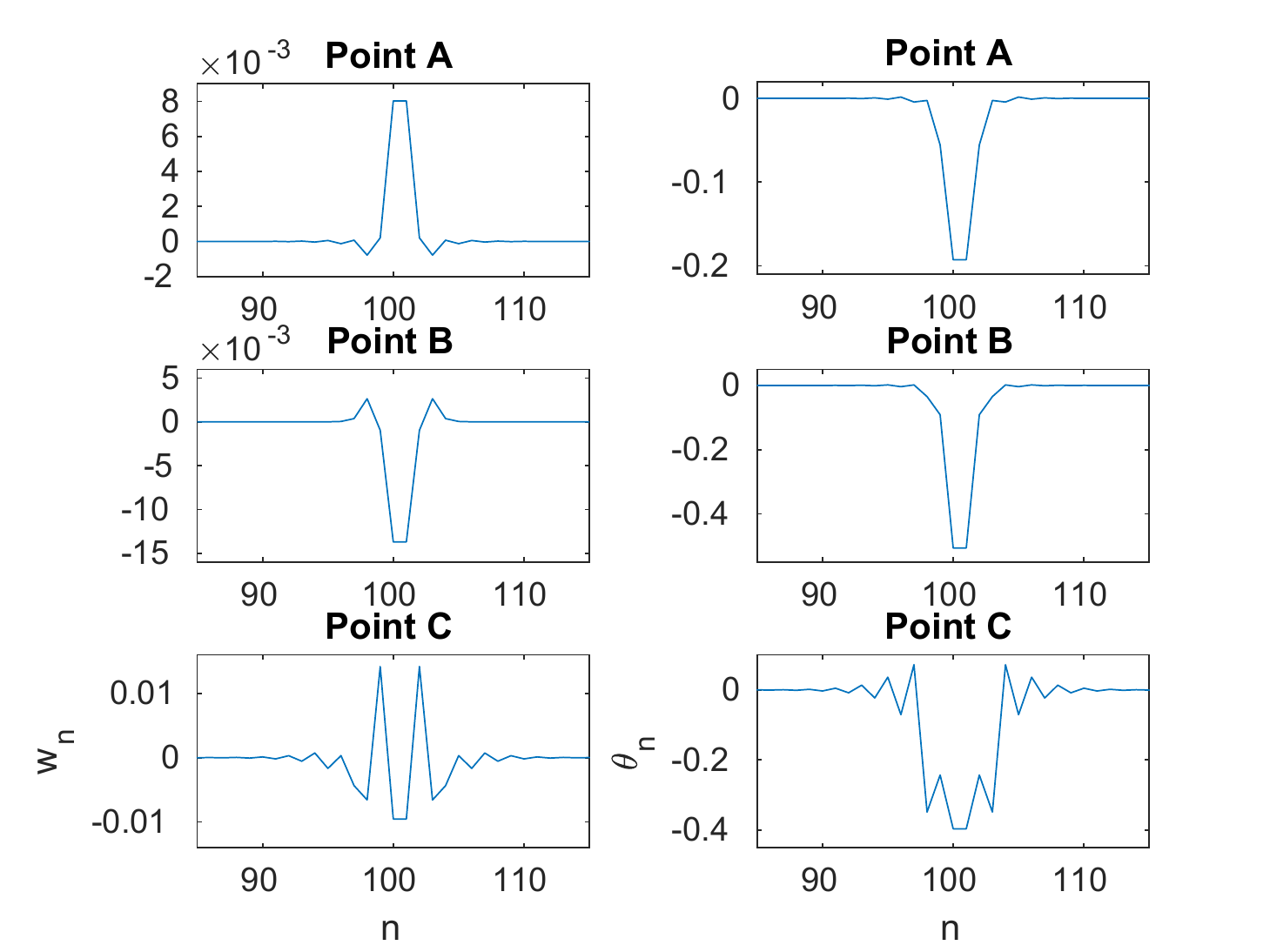}}\\
\subfloat[]
{\includegraphics[width=0.4\textwidth]{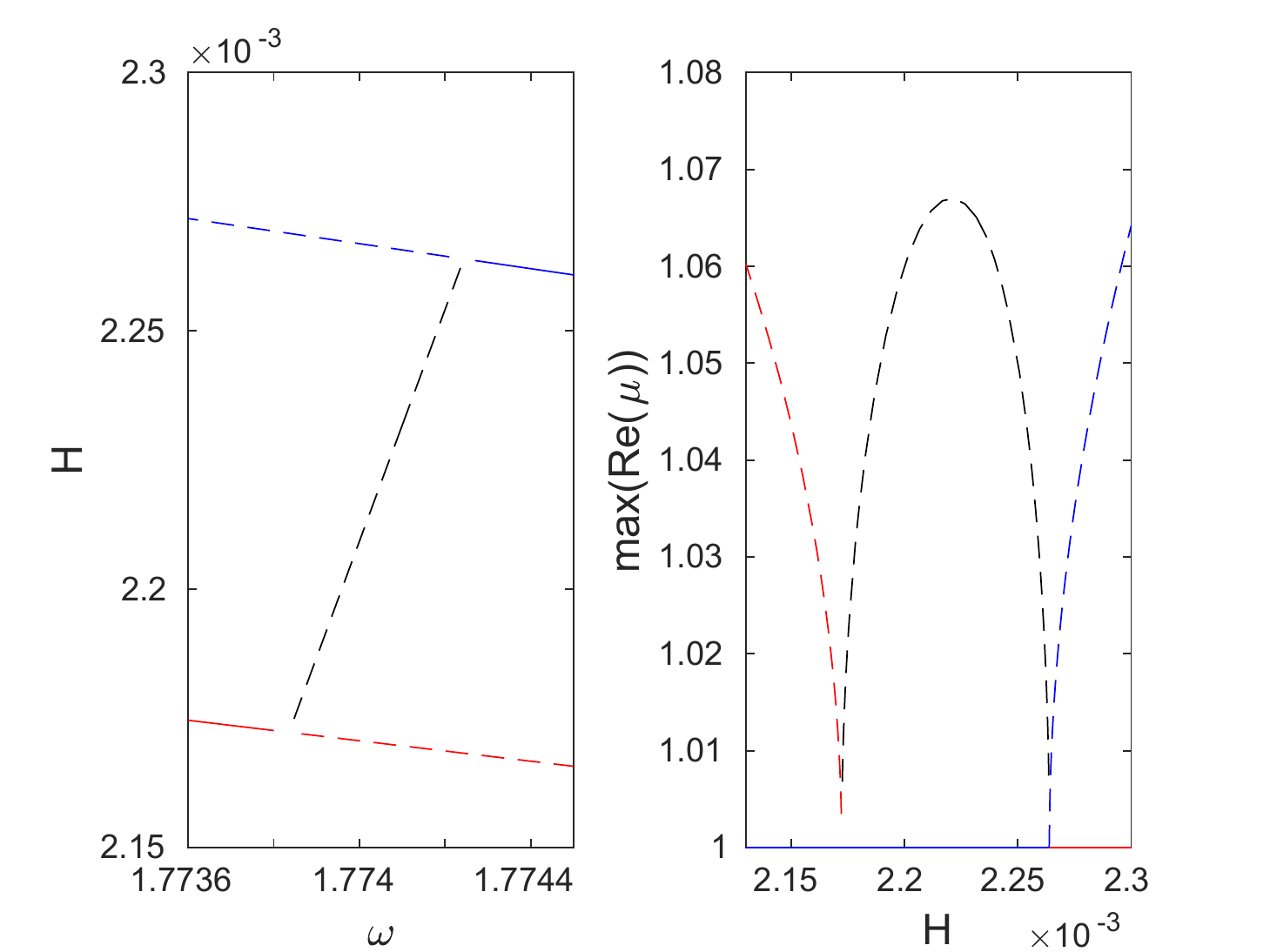}}
\caption{\footnotesize  (a) Energy $H$ as a function of frequency $\omega$ along the blue symmetric solution branch. The insets provide a enlarged view of the turning points. (b) Strain and angle variables for the solutions at the points $A$, $B$, and $C$ in (a). (c) $H(\omega)$ along the red symmetric solution branch. The inset showing Floquet multipliers illustrates the emergence of an exponential instability. A pair of complex Floquet multipliers (blue crosses) associated with a solution before the transition collides to form two positive real multipliers (red crosses) associated with the solution after the collision. The corresponding symmetric multipliers inside the unit circle are not shown. (d) Strain and angle variables for the solutions at the points $A$, $B$, and $C$ in (c). (e) Left panel: the unstable asymmetric branch connecting the red and blue symmetric branches (left panel). Right panel: maximum real Floquet multiplier as a function of energy for the three branches. All solution profiles are shown at the time instances of maximal amplitude.}
\label{fig:FirstSecondSym}
\end{figure}

We first consider the blue and red symmetric solution curves shown in panels (a) and (c), respectively, of Fig.~\ref{fig:FirstSecondSym}. Panels (b) and (d) of the same figure show strain and angle variables for the solutions at selected points along the corresponding curves in panels (a) and (c) at the time instances of maximal amplitude.
Near $\omega = 2$, the solutions for the blue curve have only a single trough in the angle $\theta_n$. As the curve is traversed, this single trough evolves first into a double trough, as can be seen at points $A$ and $B$ in Fig.~\ref{fig:FirstSecondSym}(b), and later into a quadruple trough at point $C$. Meanwhile, the strain $w_n$ evolves from a single initial peak at point $A$ into a single trough at point $B$ in Fig.~\ref{fig:FirstSecondSym}(b), and finally into a quadruple trough at point $C$. The solutions along the red curve near $\omega=2$ have a single minimum in $\theta_n$, which is maintained at points $A$ and $B$ in Fig.~\ref{fig:FirstSecondSym}(d). However, as can be seen at point $C$ in Fig.~\ref{fig:FirstSecondSym}(d), these solutions also evolve from having a single minimum to multiple extrema.  As before, in the strain component we see an inversion of an initial peak to a single trough as seen at points $A$ and $B$ in Fig.~\ref{fig:FirstSecondSym}(d). A key distinction between the blue and red solution curves is that the solutions along the blue branch are site-centered, and the solutions along the red branch are bond-centered.

We remark that although both the energy and the amplitude of solutions along the blue and red branches decreases as the frequency approaches the edge of the optical band, they do not appear to tend to zero in the limit. This suggests that instead of bifurcating from the band edge, these DB branches retain a finite amplitude as their frequency approaches the band edge, akin the large-amplitude bright breathers computed in \cite{Sanchez04} for the Fermi-Pasta-Ulam lattices.

Examining now the stability of the solutions along the two branches, we note first that as shown in the left panel of Fig.~\ref{fig:FirstSecondSym}(e), the two exchange an
effective stability via a connecting unstable asymmetric solution branch.
This is reminiscent of a similar phenomena observed in different settings
(yet still connecting the bifurcations from site-centered and bond-centered solution branches)~\cite{Vicencio06}; see also the discussion of \cite{Aubry06}, where asymmetric solution curves carry instabilities between neighboring symmetric solutions.
The blue curve has a real Floquet multiplier pair $(1/\mu,\mu)$ with $\mu>1$ until the bifurcation point at $\omega = 1.7742$ and $H=2.264 \times 10^{-3}$, where it becomes effectively stable (modulo small-amplitude oscillatory instabilities), while the emerging asymmetric branch is exponentially unstable; in other words, this
is a subcritical pitchfork bifurcation. The asymmetric branch then connects to the red curve, where a similar stability exchange (i.e., another subcritical pitchfork bifurcation) takes place at $\omega=1.7738$ and $H=2.172 \times 10^{-3}$. The stability exchange is further illustrated in the right panel of Fig.~\ref{fig:FirstSecondSym}(e), where we plot the maximum real Floquet multiplier $\mu$ as a function of the energy $H$.

Next, we note that the exponential instability that emerges from the oscillatory instability in the solutions along the red curve, indicated by the inset in Fig.~\ref{fig:FirstSecondSym}(c), is due to the collision of two complex pairs of Floquet multipliers $\mu$ (only the multipliers outside the unit circle are shown in the inset). A similar collision is responsible for the transition to exponential instability near the first local maxima in the blue curve, which is indicated in the inset containing the point $D$ in Fig.~\ref{fig:FirstSecondSym}(a).

\begin{figure}[!htb]
\centering
\subfloat[]{\includegraphics[width=0.5\textwidth]{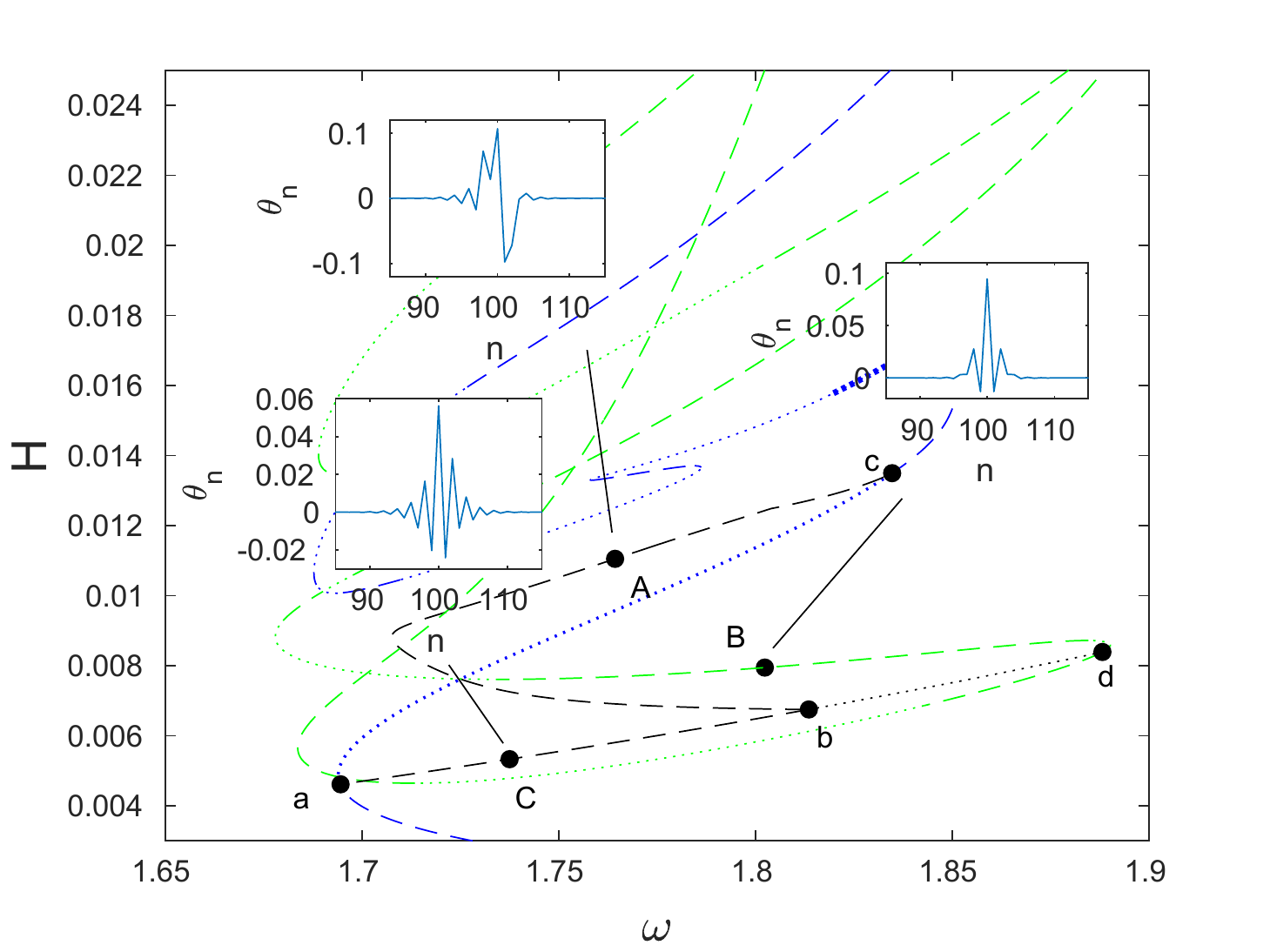}}
\subfloat[]
{\includegraphics[width=0.5\textwidth]{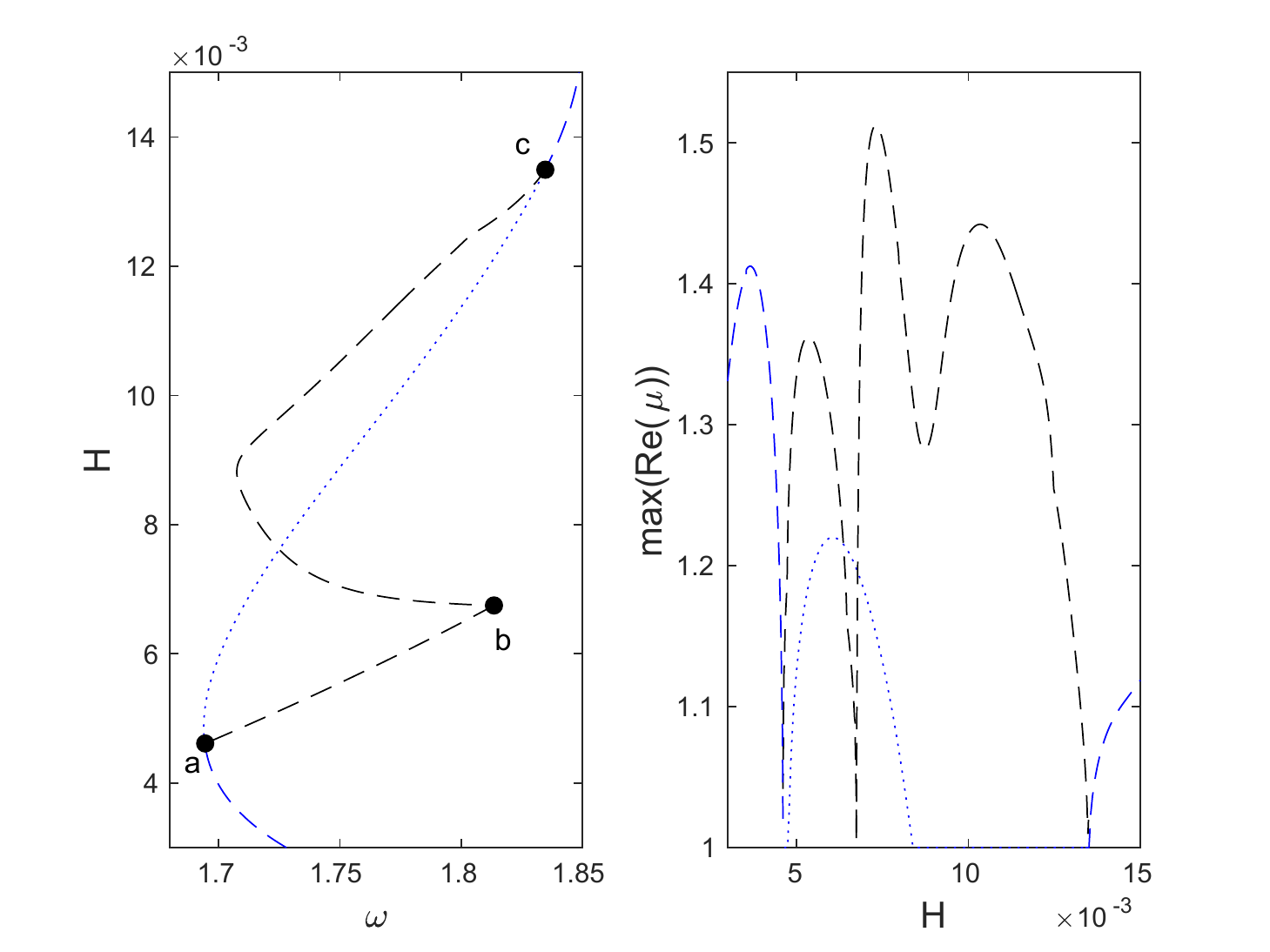}}\\
\subfloat[]
{\includegraphics[width=0.5\textwidth]{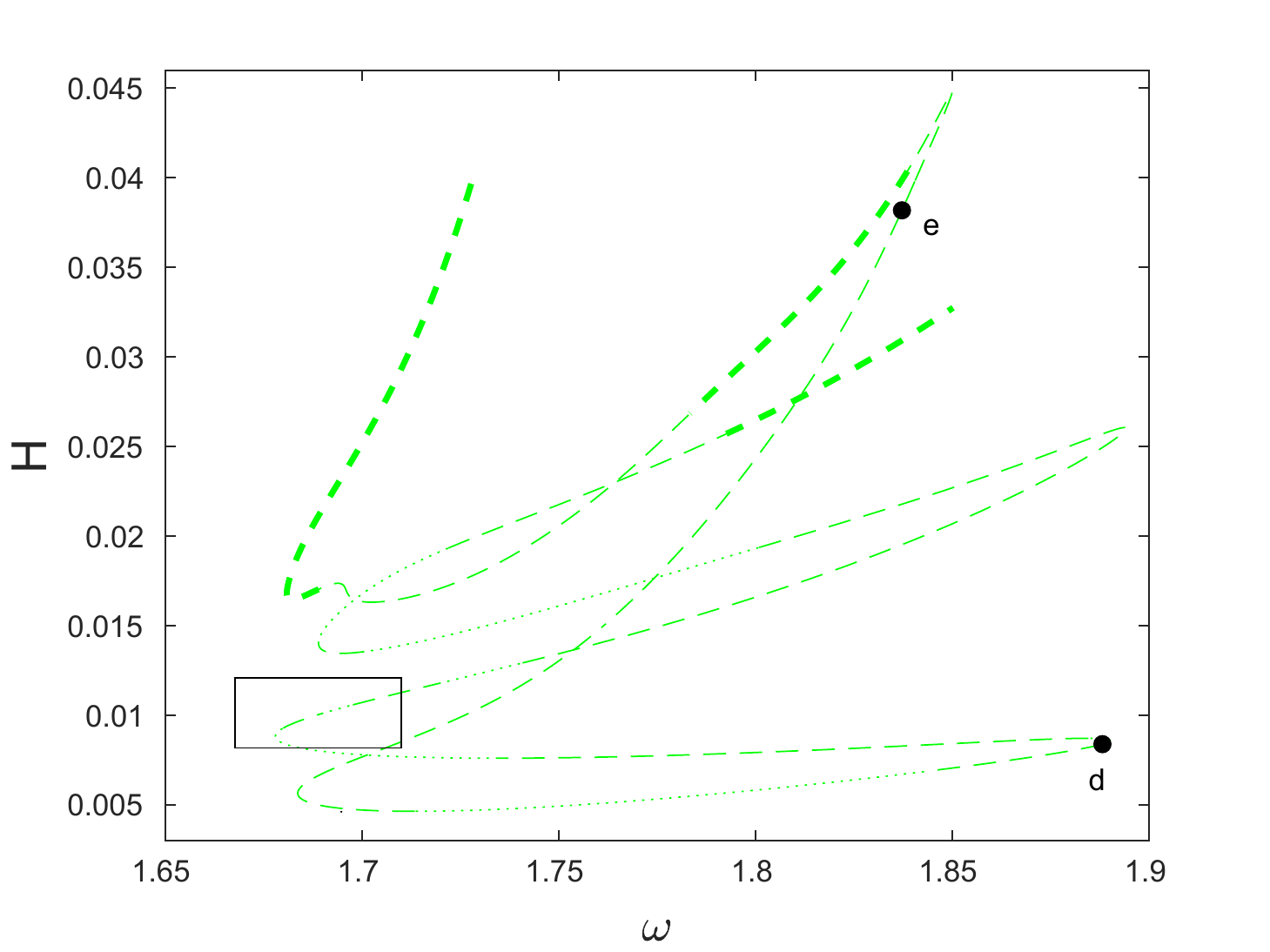}}
\subfloat[]
{\includegraphics[width=0.5\textwidth]{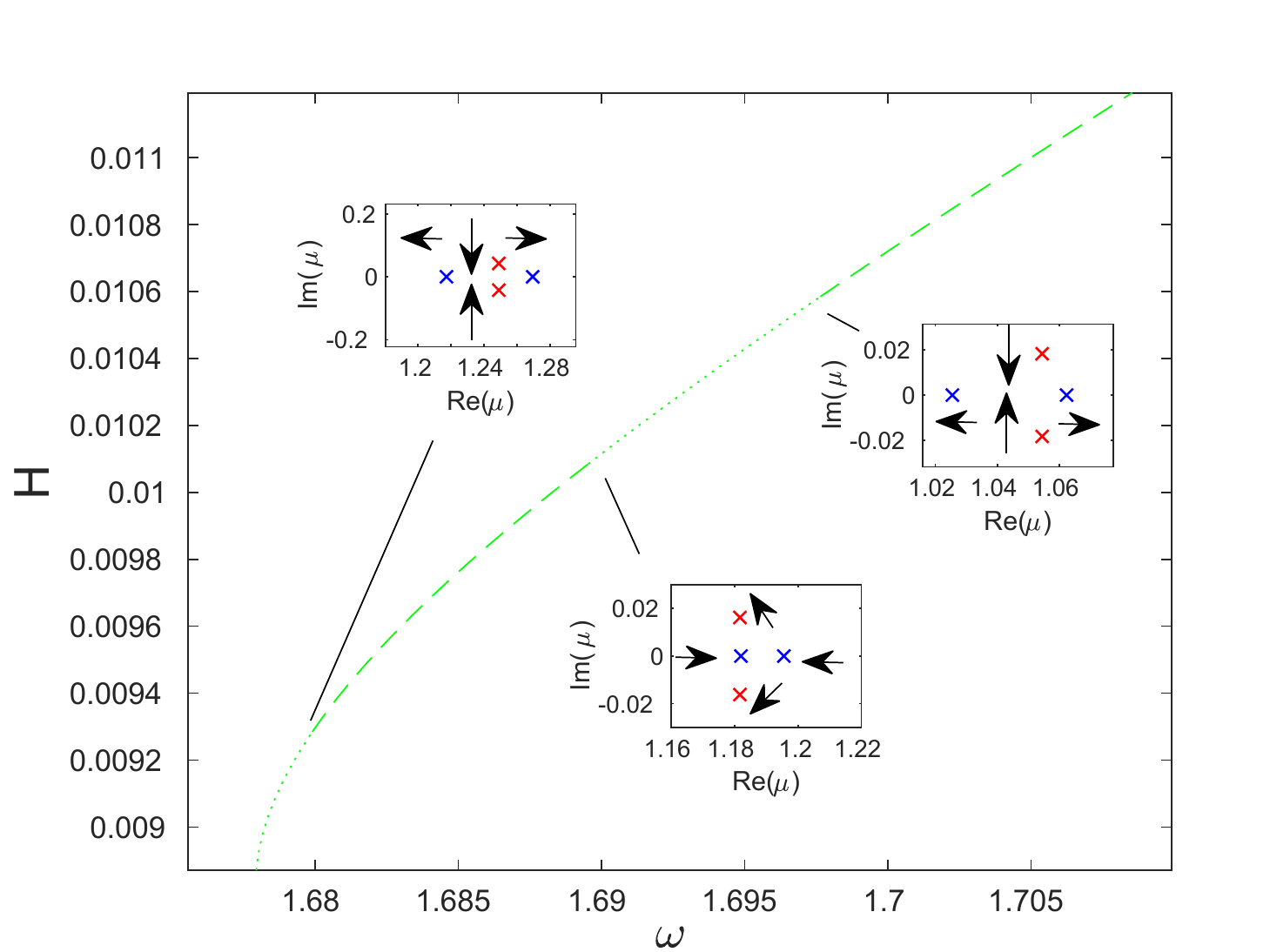}}
\caption{\footnotesize (a) Energy $H$ as a function of frequency $\omega$ along the blue and green symmetric solution branches and bifurcating branches of asymmetric solutions, with $a$, $b$, $c$, and $d$ marking the bifurcation points. The insets show the solutions of the asymmetric and symmetric branches at the points $A$, $B$, and $C$. (b) The stability exchange
  between the symmetric (blue) and the asymmetric (black) branch. Both the
  associated portion of the bifurcation diagram and the dominant multiplier
  of each branch associated with the instability growth rate are shown.
  (c) Energy $H$ versus frequency $\omega$ for the green symmetric solution branch with $d$ and $e$ marking the bifurcation points (see Fig.~\ref{fig:WanderingAsym} for the asymmetric branch bifurcating from $e$). (d) The enlarged view of the region inside the rectangle in (c). The insets show the transition from exponential to oscillatory instabilities and vice versa that take place over the green symmetric curve. The red and blue crosses indicate Floquet multipliers $\mu$ outside the unit circle that correspond to solutions before and after the transition point, respectively.
}
\label{fig:StabSnake}
\end{figure}

Panels (a) and (b) of Fig.~\ref{fig:StabSnake} show a bifurcation at the point $a$ along the blue curve, at which point the blue curve loses its
exponential instability (while still retaining oscillatory instability modes).
The instability is transferred to an asymmetric solution branch
(again through a subcritical pitchfork bifurcation). Another exponentially unstable asymmetric branch bifurcates at the point $b$ from this branch and at the point $c$ from the blue curve.
The resulting part of the bifurcation diagram, depicted in the right panel of Fig.~\ref{fig:StabSnake}(b), is reminiscent of the ``snaking" behavior that has been observed in other systems \cite{Beck2009,Chong2009}. Further exploration of such snaking
features and associated asymmetric branches
in the present metamaterial setting is a potentially interesting topic
for future studies.

We also observe that stability changes at the points where $H'(\omega)$ changes sign are associated with the emergence of a pair of real Floquet multipliers from $\mu=1$. The multiplier $\mu>1$ then corresponds to an exponential instability. One such example is shown in the inset of Fig.~\ref{fig:FirstSecondSym}(a) zooming in on a sharp turning point. The initial stability change happens at a local minimum, and the second
saddle-center bifurcation at a local maximum. This change in multiplicity of the unit Floquet multiplier at the extrema of the energy-frequency curve is similar to the one we observed earlier in Sec.~\ref{sec:period_doubling} and again consistent with the stability criterion in \cite{Kevrekidis2016}. The same mechanism is responsible for the onset of exponential instability at a local minimum of $H(\omega)$ near $\omega=2$ (see the bottom right inset of Fig.~\ref{fig:FirstSecondSym}(a)). Another example of such change in multiplicity takes place at the local maximum near the point $D$ in Fig.~\ref{fig:FirstSecondSym}(a) (see the inset). At this point, a second pair of real Floquet multipliers emerges from the unit circle, and this new pair subsequently collides at the point $D$ with an already existing pair of real multipliers forming a complex quartet of Floquet multipliers. A similar emergence of a pair of real Floquet multipliers from $\mu=1$ is observed at the local extrema of energy along the red curve.

As discussed above, a secondary asymmetric branch bifurcates from a primary asymmetric branch at the point $b$ in panels (a) and (b) of Fig.~\ref{fig:StabSnake}. The primary branch continues on past this bifurcation point to intersect with a symmetric solution curve at the point $d$, shown in green color in panel (a). Following this green curve, shown in its entirety in Fig.~\ref{fig:StabSnake}(d), upward from point $d$, we observe the sequence of events illustrated in Fig.~\ref{fig:StabSnake}(d). Two pairs of real multipliers $(1/\mu,\mu)$ with $\mu>1$ emerge due to two pairs of complex multipliers colliding on the real axis (only the multipliers outside the unit circle are shown in the insets). The real multipliers then collide to form complex ones anew, and subsequently reemerge again due to another collision of the oscillatory multipliers. Eventually, the real multipliers rejoin the unit circle. This provides a sense of the complexity of the associated bifurcation
diagram.

\begin{figure}[!htb]
\centering
\subfloat[]{\includegraphics[width=0.5\textwidth]{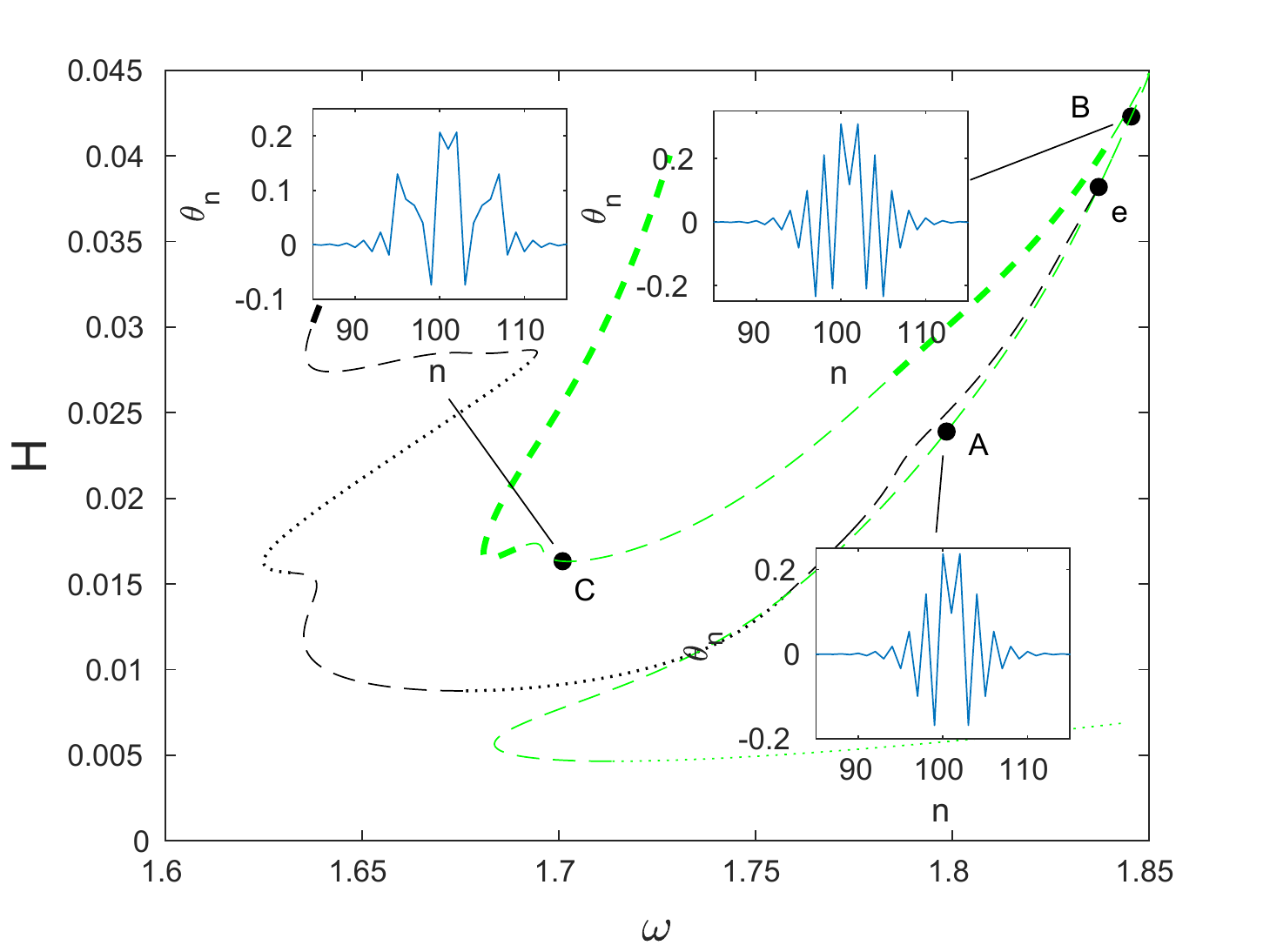}}
\subfloat[]
{\includegraphics[width=0.5\textwidth]{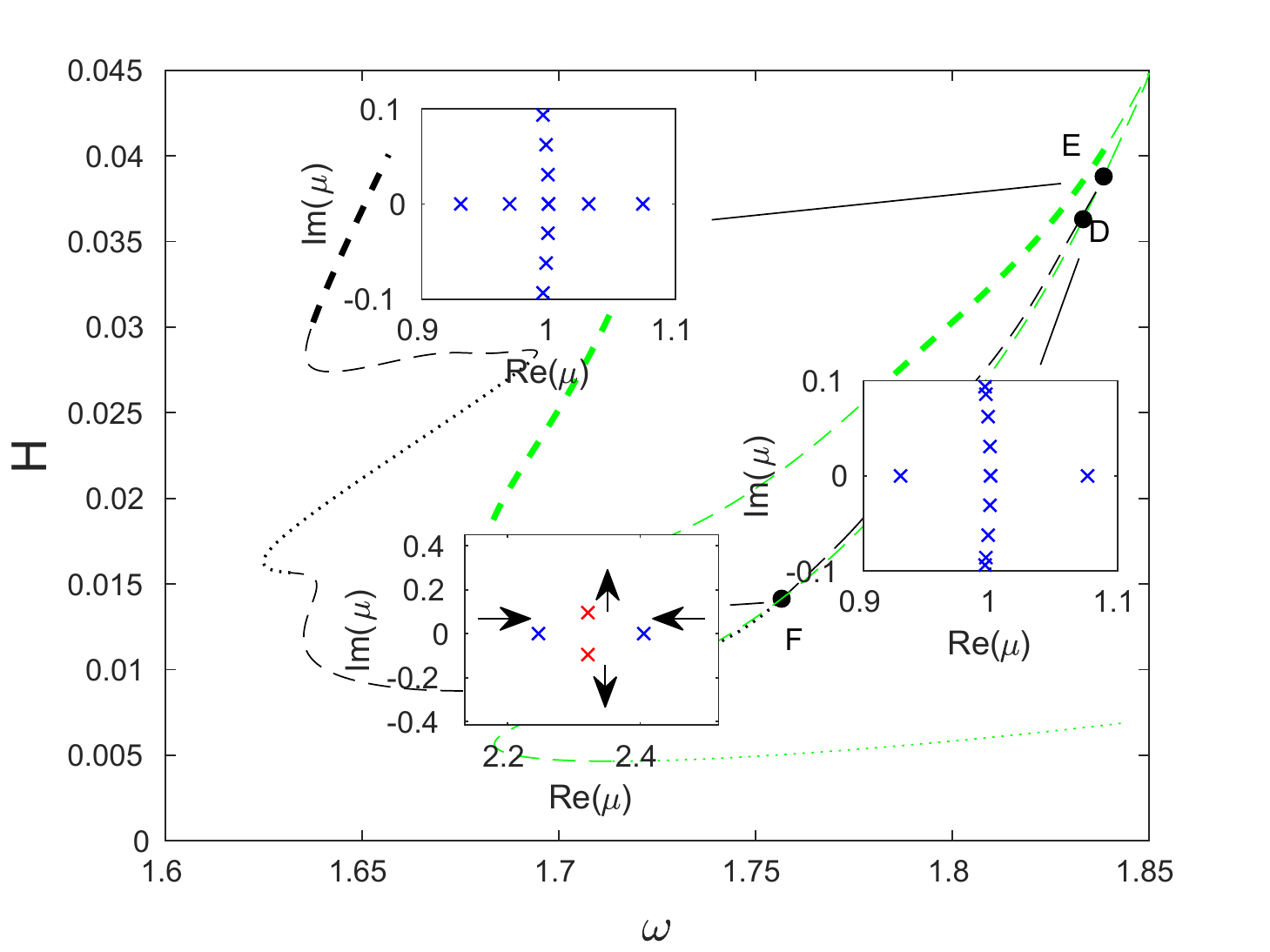}}
\caption{\footnotesize (a) Energy $H$ as a function of frequency $\omega$ along the green symmetric solution branch and a branch of asymmetric solutions (different from the ones discussed earlier) bifurcating at the point $e$. The insets include profiles of the angle variable at the points $A$, $B$, and $C$. (b) The insets pointing toward points $D$ and $E$ show the emergence of a third pair of real Floquet multipliers.
  In both insets, an additional pair of real multipliers is present but not shown due to its larger magnitude. The inset pointing toward the point $F$ illustrates the collision of two pairs of real multipliers to form two  complex pairs. The red and blue crosses indicate Floquet multipliers outside the unit circle that correspond to solutions before and after the transition point, respectively. }
\label{fig:WanderingAsym}
\end{figure}

Traveling downward now from the point $d$ along the green curve, we eventually arrive at another bifurcation of an asymmetric solution branch at the point $e$. This bifurcation is shown in Fig.~\ref{fig:WanderingAsym} and appears not to be associated with any stability change. A closer examination shows that this is due to the prior existence of two pairs of real Floquet multipliers (one is not included due to its larger magnitude), shown in the inset zooming in around the point $D$. After the bifurcation, a third pair of real Floquet multipliers joins the other two, as shown in the inset of Fig.~\ref{fig:WanderingAsym}(b) zooming in around the point $E$, indicating the emergence of a new exponential instability.
It is important to note that in both insets of panel (b) around points $D$ and $E$, an additional exponential instability is present but not shown due to its larger magnitude.
As before, we also observe changes in stability due to collisions of complex pairs, as shown in the inset zooming in around the point $F$, as well as due to turning points in energy, e.g., near the local minimum of the black asymmetric solution curve of Fig.~\ref{fig:WanderingAsym}.

\begin{figure}[!htb]
\centering
\subfloat[]{\includegraphics[width=0.5\textwidth]{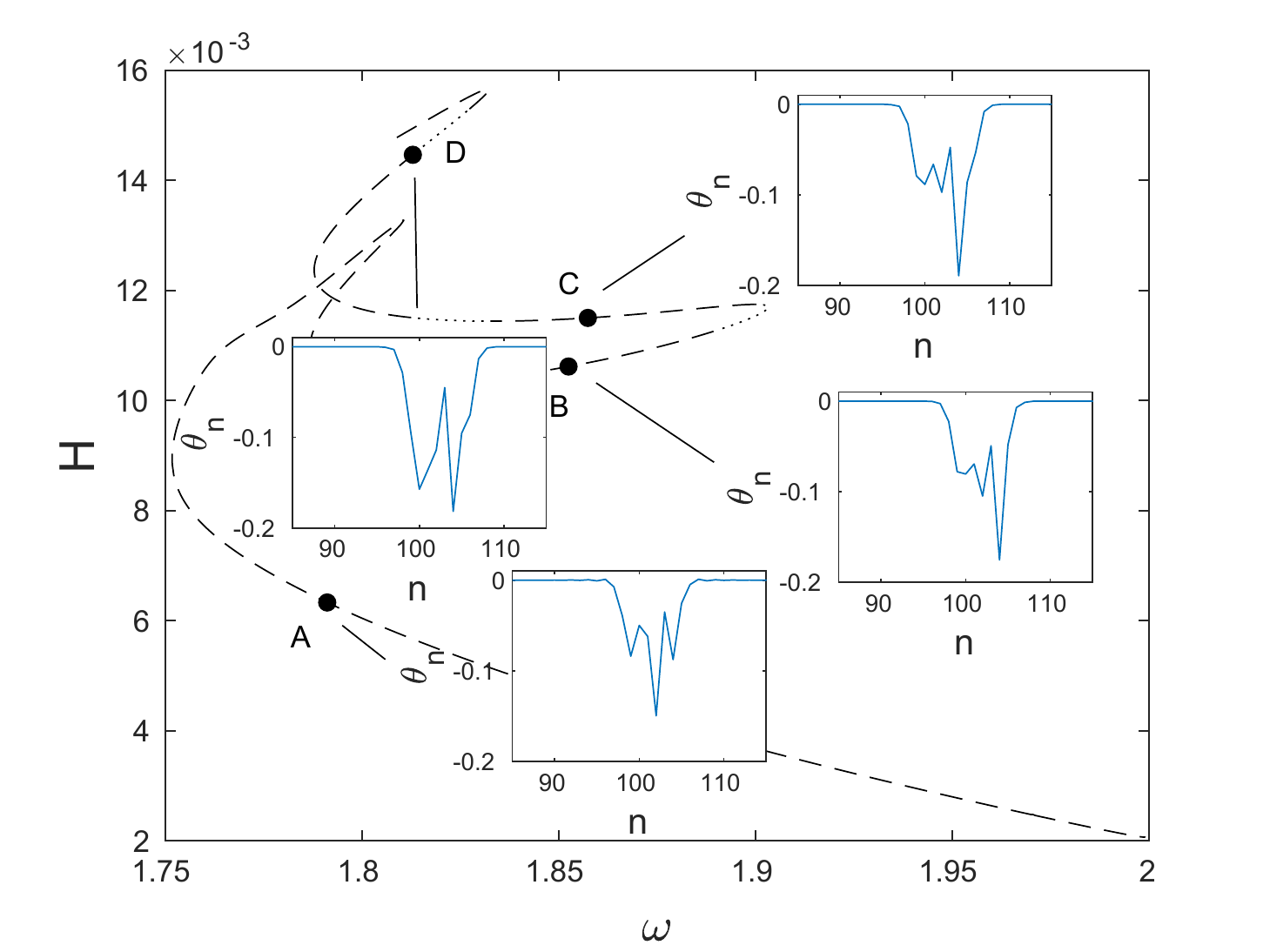}}
\subfloat[]
{\includegraphics[width=0.5\textwidth]{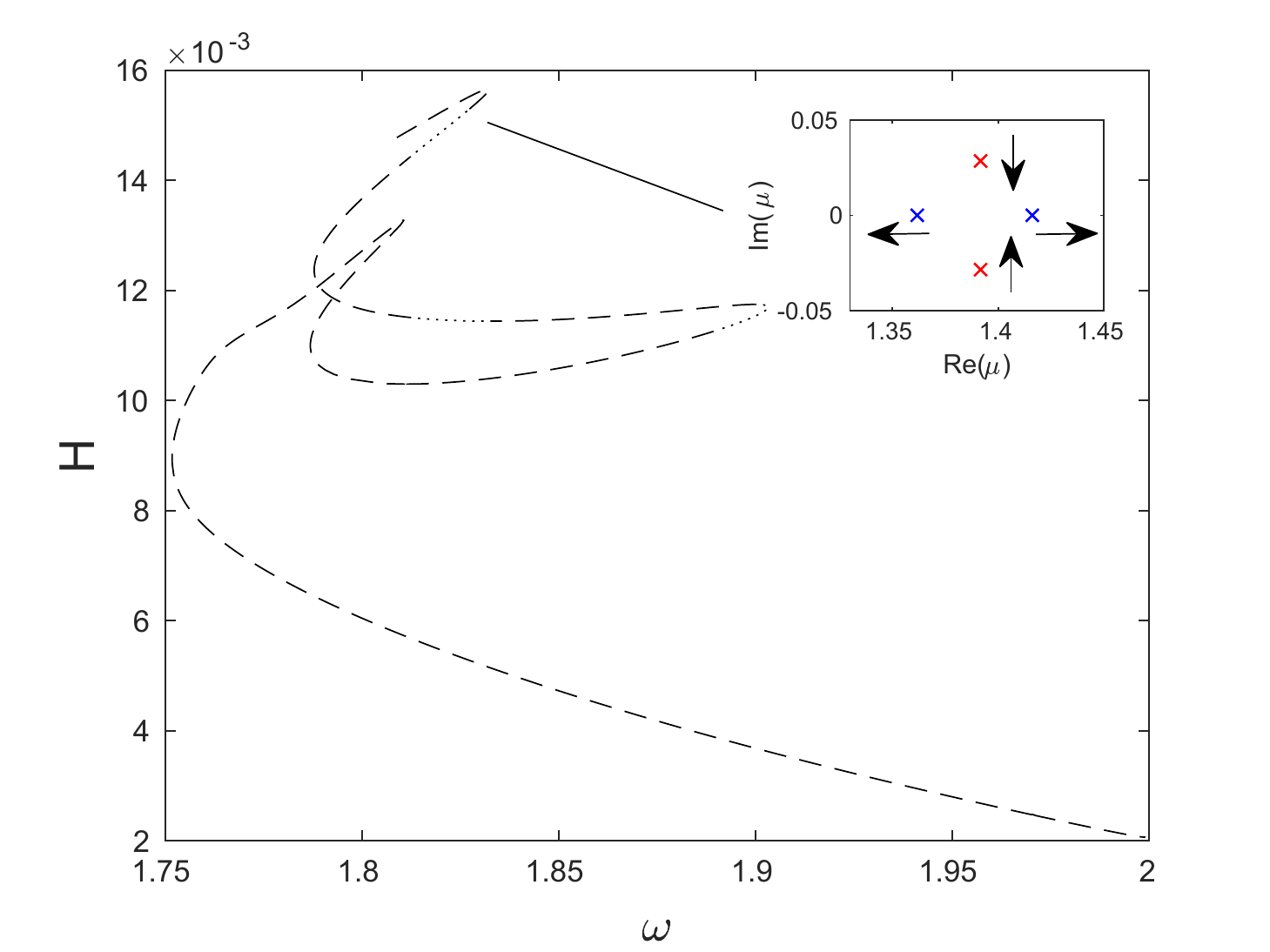}}
\caption{\footnotesize (a) Energy $H$ as a function of frequency $\omega$ along an asymmetric solution branch that exists near the $k=\pi$ edge of the optical branch. The insets include profiles of the angle variable at the points $A$, $B$, $C$, and $D$. (b) The same branch, with the inset showing the collision of two pairs of complex Floquet multipliers to form two real pairs. Blue and red crosses show the pairs outside the unit circle that correspond to solutions before and after the transition, respectively. }
\label{fig:PiAsym}
\end{figure}

Finally, we consider the asymmetric branch in Fig.~\ref{fig:EnergyvFrequency_All} that has not yet been discussed. This branch is unique among the other asymmetric branches in that it comes near the $\pi$-mode edge of the optical branch. However, that similar to the blue and red branches, it does not appear to bifurcate from the edge.
As in the previous cases, we observe the emergence or collision of real Floquet multipliers at the turning points in energy. In Fig.~\ref{fig:PiAsym}(a), we show the evolution of the solutions as the branch is traversed, and in Fig.~\ref{fig:PiAsym}(b), one can see the emergence of pairs of real multipliers from complex ones;
once again these are signaled by transitions from dotted lines to dashed ones.

\begin{figure}[!htb]
\centering
\subfloat[]
{\includegraphics[width=0.5\textwidth]{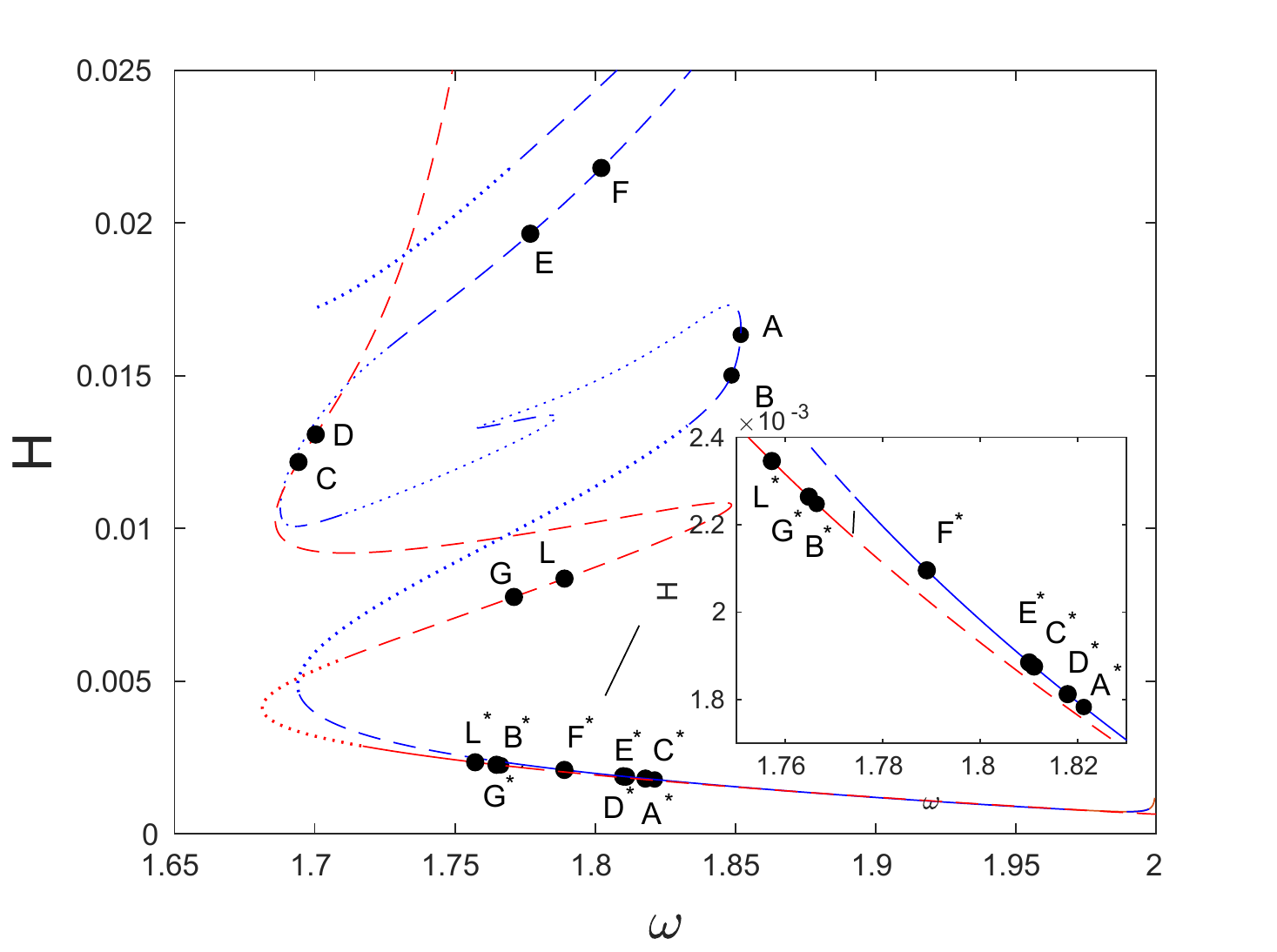}}
\subfloat[]
{\includegraphics[width=0.5\textwidth]{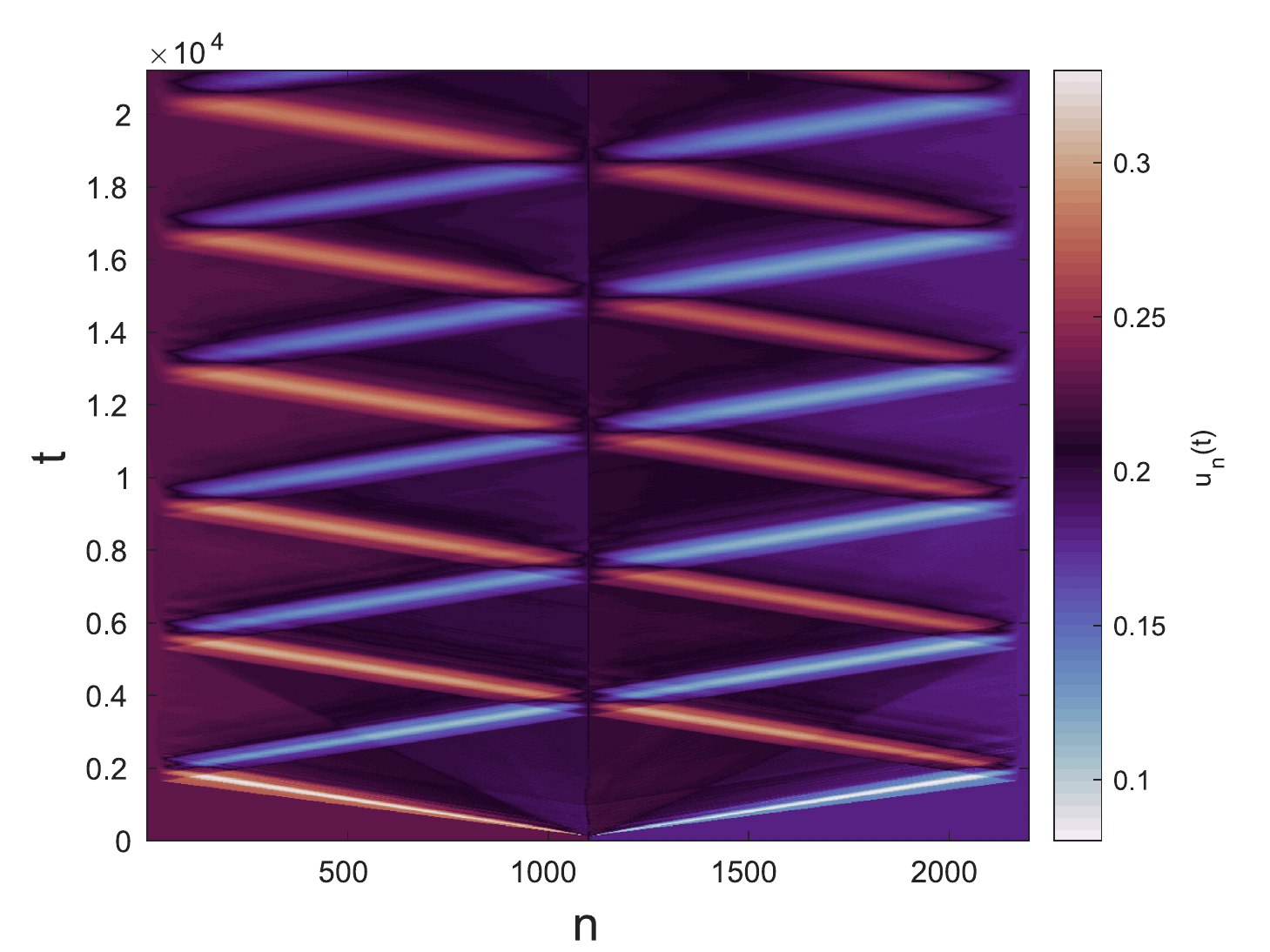}}\\
\subfloat[]
{\includegraphics[width=0.5\textwidth]{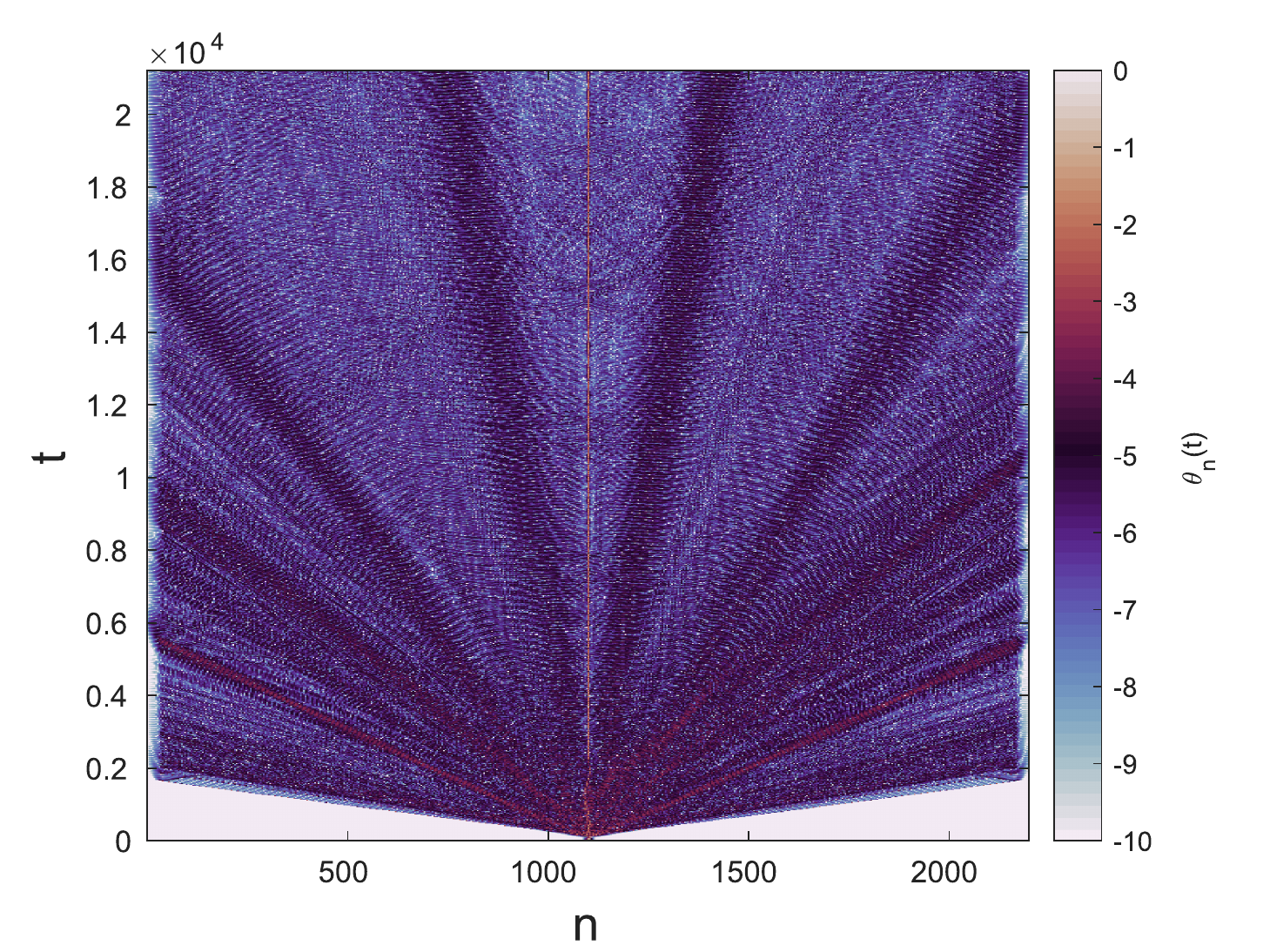}}
\subfloat[]{\includegraphics[width=0.5\textwidth]{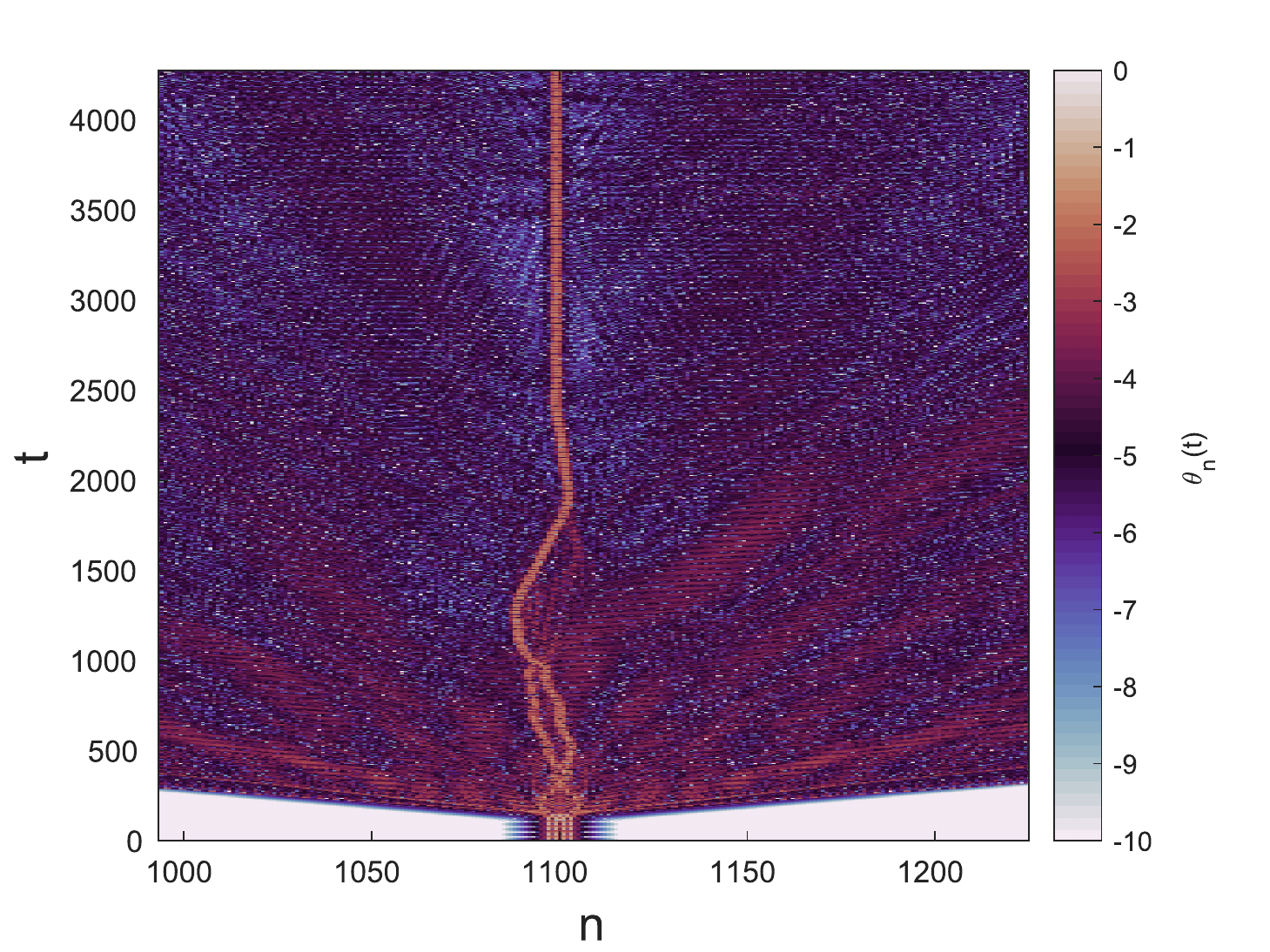}}
\caption{\footnotesize (a) Energy $H$ as a function of frequency $\omega$ along the red and blue symmetric solution curves. The points $A$-$L$ indicate the perturbed unstable solutions, while the points $A^*$ - $L^*$ mark the corresponding final states. The inset zooms in on the region including the end points. (b) Space-time plot of the displacement $u_n(t)$ for the solution corresponding to point $E$. Here $\epsilon = 10^{-5}$ is the strength of the perturbation, and $\mu = 1.3596$ is the largest real Floquet multiplier. (c) Space-time plot of the angle $\theta_n(t)$. (d) Enlarged view of (c). Both (c) and (d) are shown in a logarithmic plot to facilitate the visualization
of the small scales involving dispersive wave radiation as a result of the instability.}
\label{fig:EndPertSpaceTime}
\end{figure}

To examine the consequences of an instability associated with real Floquet multipliers $\mu>1$ along the blue and red symmetric solution branches, we perturb unstable solutions at various points featuring such an exponential instability
along the corresponding eigenmodes and simulate the resulting dynamics. In Fig.~\ref{fig:EndPertSpaceTime}(a), these points on the blue and red dashed portions of the curves are labeled $A$ - $L$. The corresponding final states are indicated by the points $A^*$ - $L^*$. As can be seen in the inset of Fig.~\ref{fig:EndPertSpaceTime}(a), in all cases, the perturbed solution eventually settles onto one of the two effectively stable regions of the blue and red solution curves, with an apparent preference toward the blue curve, which is effectively stable for a much larger interval of frequencies than the red curve.

As an example, we consider the point $E$ in Fig.~\ref{fig:EndPertSpaceTime}(a) and show the dynamic evolution of the perturbed solution in Fig.~\ref{fig:EndPertSpaceTime}(b-d). Here $\epsilon = 10^{-5}$, and the largest real Floquet multiplier is $\mu = 1.3596$. The space-time plots of the displacement and angle are shown in panels (b) and (c), respectively, while panel (d) zooms in on the dynamic evolution of the angle variable at smaller times. Both (c) and (d)
are shown on a logarithmic scale. This facilitates the last plot to show the
nontrivial amount of radiation that is emitted by the perturbed wave as it develops, as well as its temporary mobility. Eventually, this perturbed wave settles into a stable breather, associated with the point $E^*$, as can be verified by comparing its properties (once it settles) with those of the latter solution.

\subsection{Zero-mode optical and $\pi$-mode acoustic branches}
\label{sec:zero_mode}
We now consider breather solutions bifurcating from the bottom of the optical band at $k=0$, as well as solutions that exist near the top of the acoustic branch at $k=\pi$. To ensure that the optical branch has a minimum at $k=0$, we choose $\phi_0 = 8\pi/180 \approx 0.1396$, which is below $\phi_0^{''} = 0.1588$. The corresponding dispersion relation plot is shown in Fig.~\ref{fig:OpticalAcousticGap}(b).

\begin{figure}[!htb]
\centering
{\includegraphics[width=0.75\textwidth]{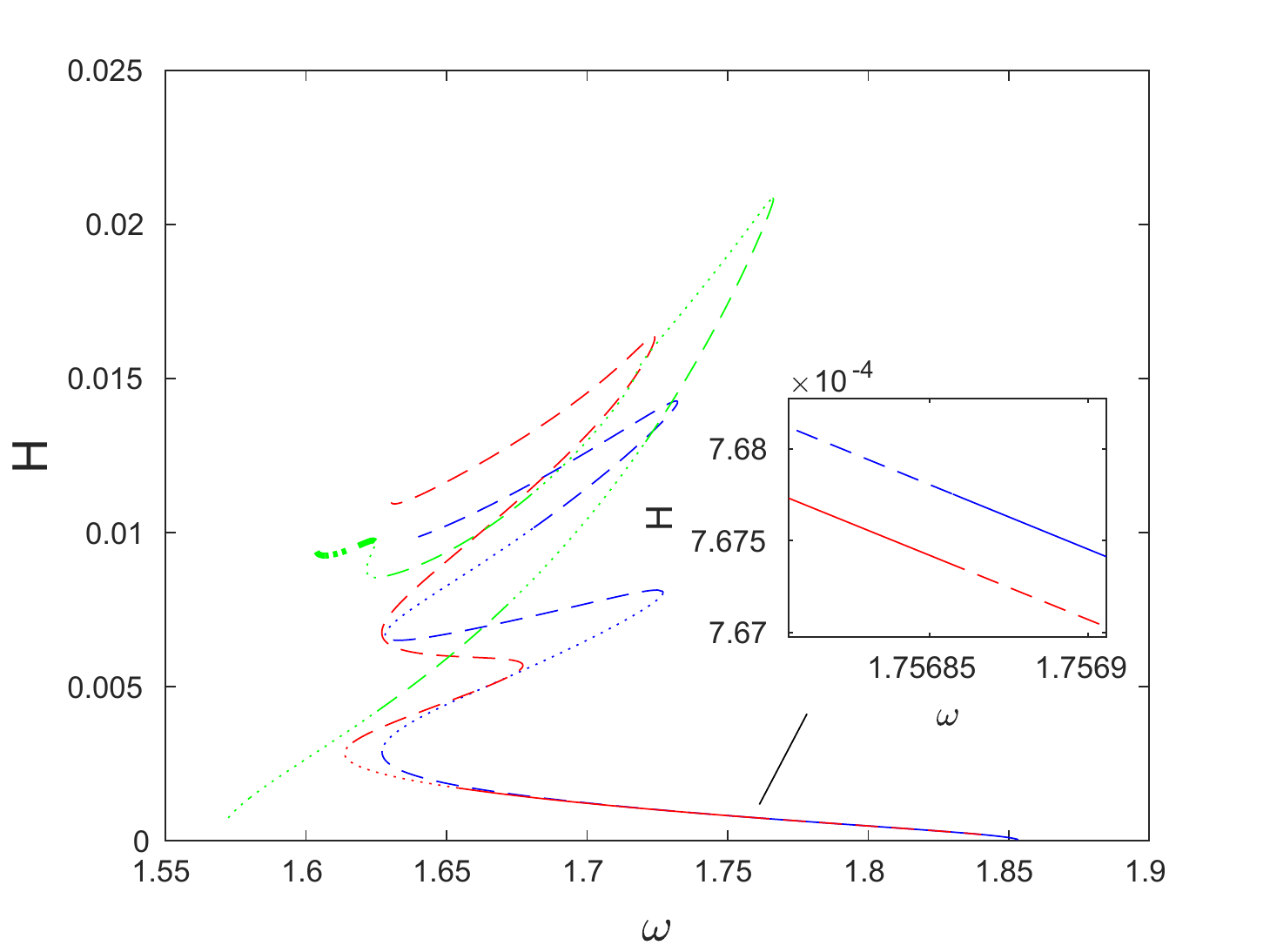}}
\caption{\footnotesize Energy $H$ of the computed DB solutions as a function frequency $\omega$. Blue and red curves bifurcate from the optical band at $k=0$, while the green curve is associated with the acoustic $\pi$-mode. All of the branches shown contain solutions with even symmetry. Thin dashed portions of the curves indicate the presence of the real multiplier pairs $(1/\mu,\mu)$ with $\mu > 1$. Along the thick dashed segments there are also real multipliers $(1/\mu,\mu)$ with $\mu<-1$. Parts of the curve where there are only oscillatory instabilities with the maximum modulus of the Floquet multipliers exceeding $1.009$ are indicated by thin dotted segments. Solutions that also have real multiplier pairs $(1/\mu,\mu)$ with $\mu<-1$ are along the thick dotted parts. Solid curves indicate the portions where there are no exponential instabilities, and the maximum modulus of the Floquet multipliers is below $1.009$. Here and in the remainder of this subsection we have $\alpha = 5$, $K_s = 0.02$, $K_{\theta} =0.01$, $N = 200$, and $\phi_0 = 8\pi/180$.}
\label{fig:EnergyvFrequency_All_zero}
\end{figure}

Using the continuation procedure with the initial guess of the form \eqref{eq:initial_guess_k0}, we obtained the blue and red branches
of symmetric DB solutions shown in Fig.~\ref{fig:EnergyvFrequency_All_zero} that are site-centered and bond-centered, respectively, and bifurcate from the edge of the optical band at $k=0$. The green solution branch of site-centered breathers shown in the same figure extends from near the top of the acoustic band at $k=\pi$ and was obtained using the initial guess that was constructed as described in Sec.~\ref{sec:pi_mode}.
As in the previous case discussed in Sec.~\ref{sec:pi_mode}, we expect there to be other solution branches emanating from the band edges, as well as secondary branches that bifurcate from the primary ones. However, the discussion below is limited to the three branches included in Fig.~\ref{fig:EnergyvFrequency_All_zero}.

\begin{figure}[!htb]
\centering
\subfloat[]
{\includegraphics[width=0.4\textwidth]{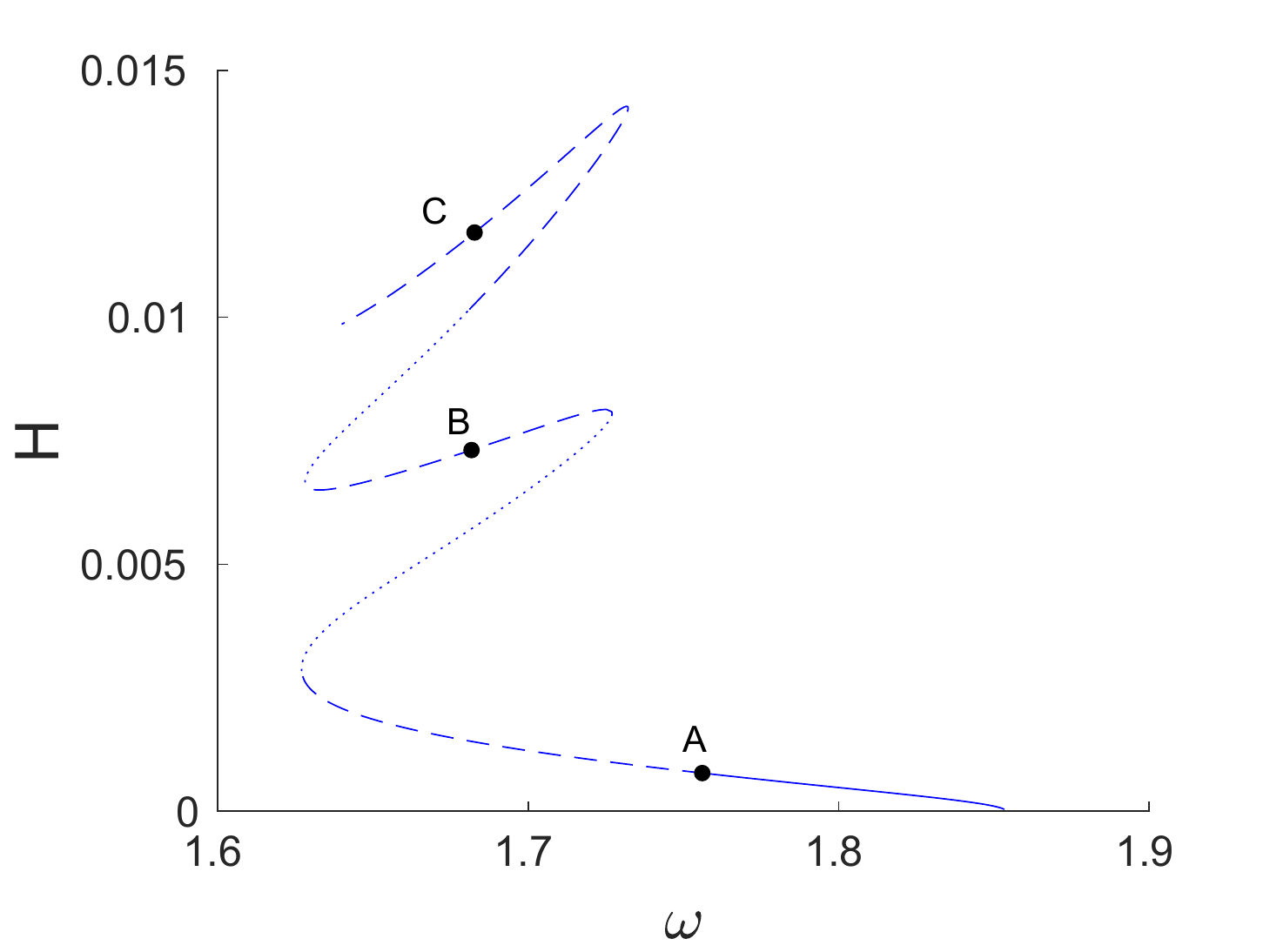}}
\subfloat[]
{\includegraphics[width=0.4\textwidth]{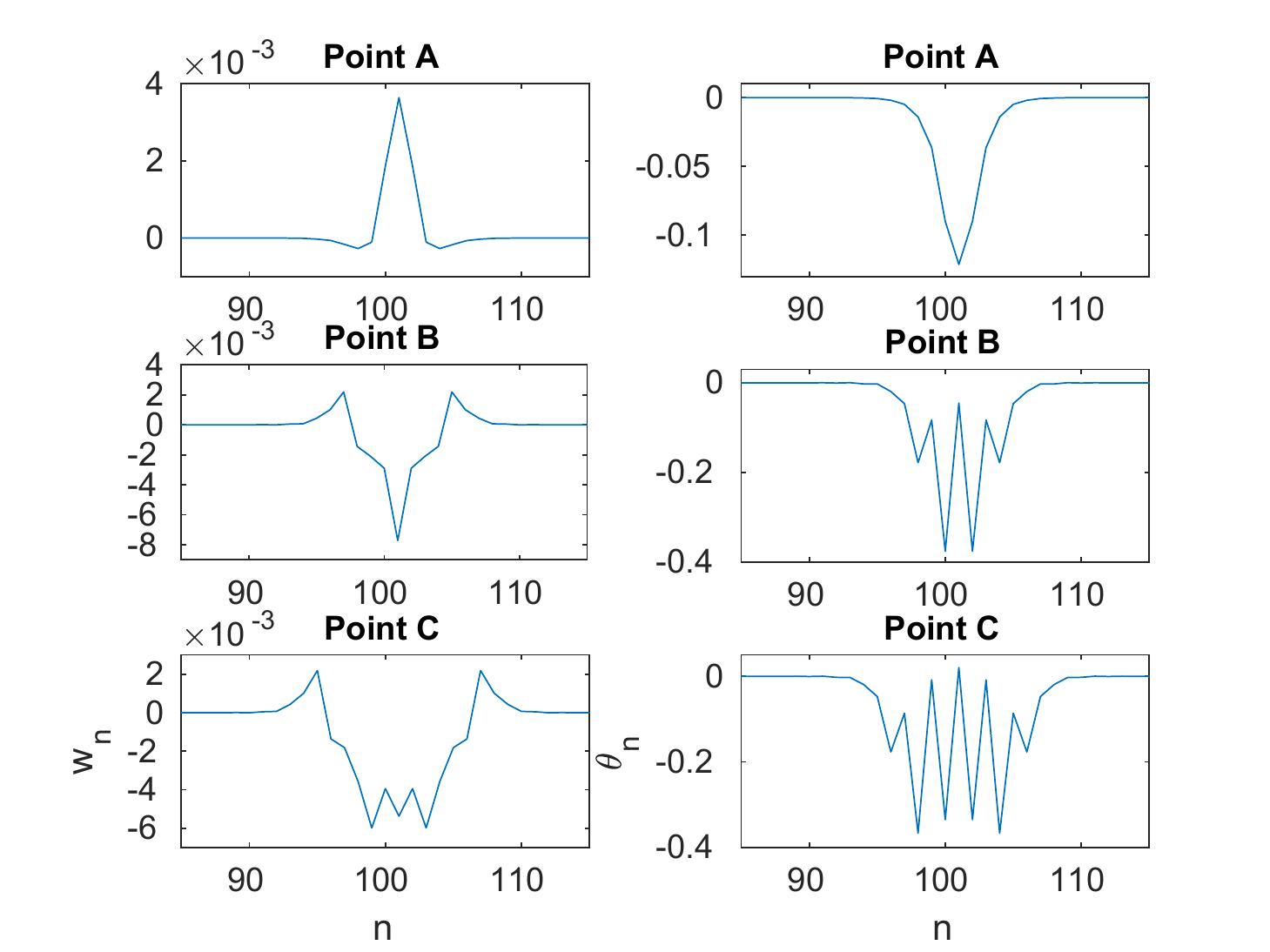}}\\
\subfloat[]
{\includegraphics[width=0.4\textwidth]{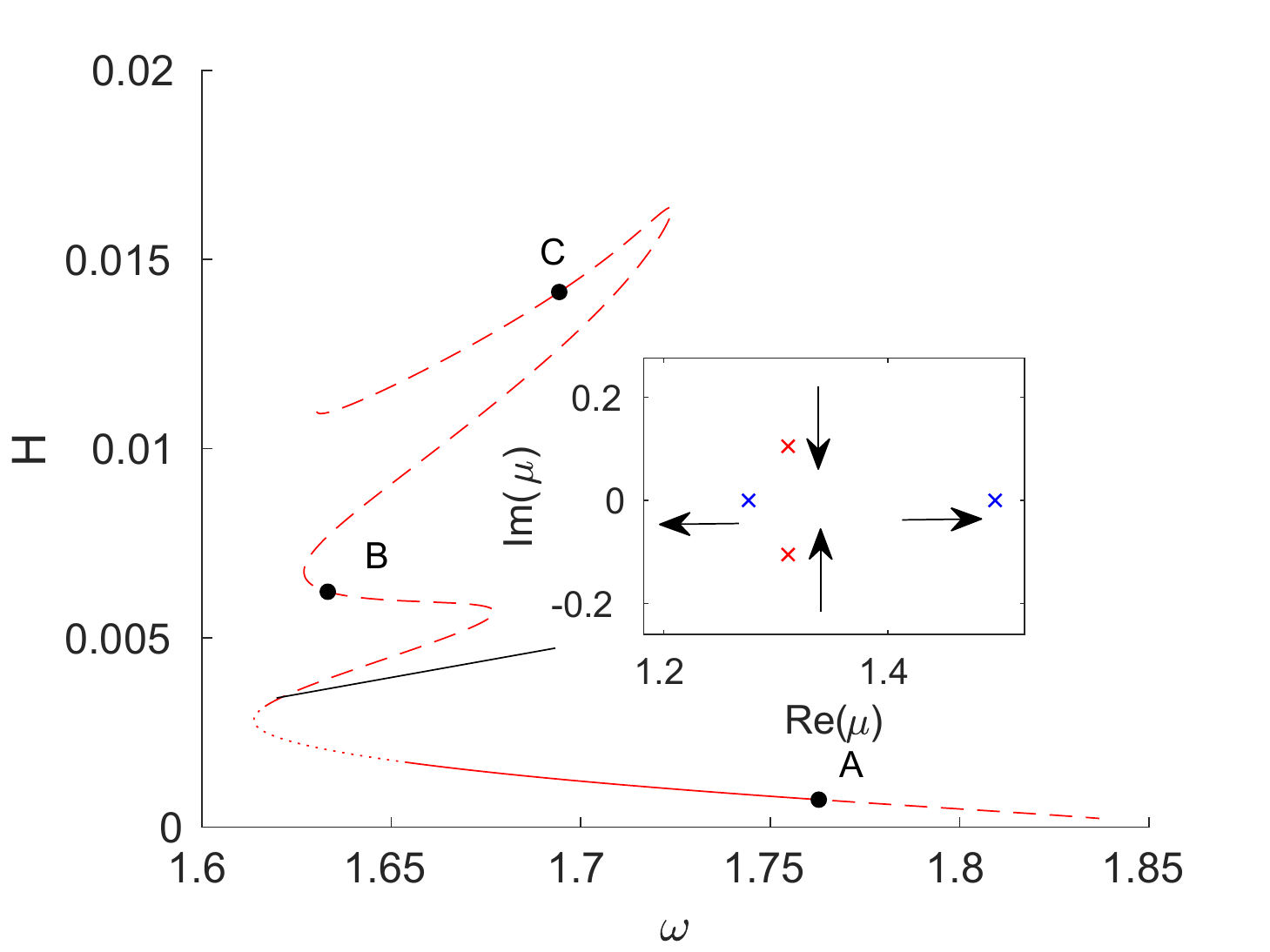}}
\subfloat[]
{\includegraphics[width=0.4\textwidth]{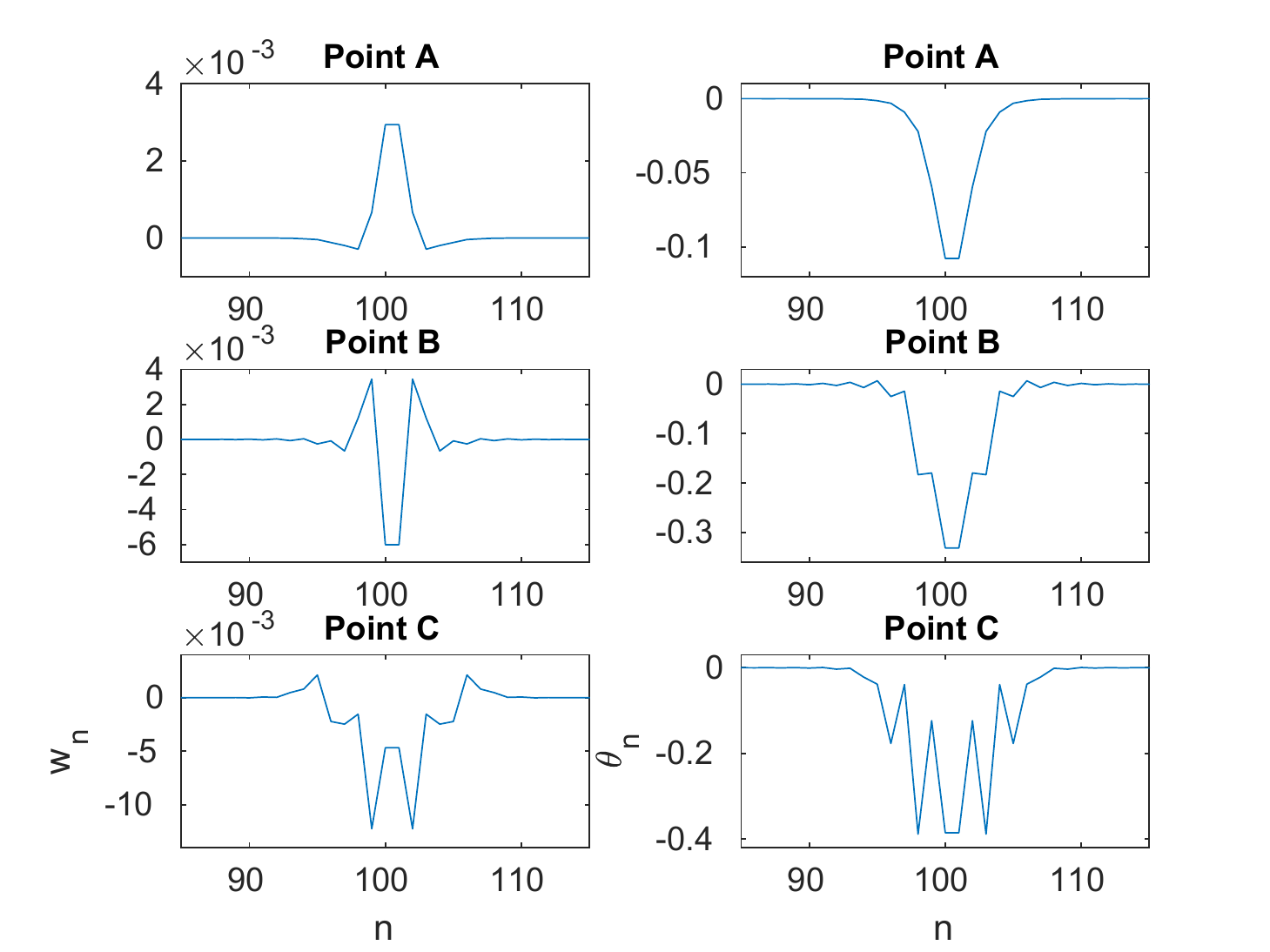}}\\
\subfloat[]
{\includegraphics[width=0.4\textwidth]{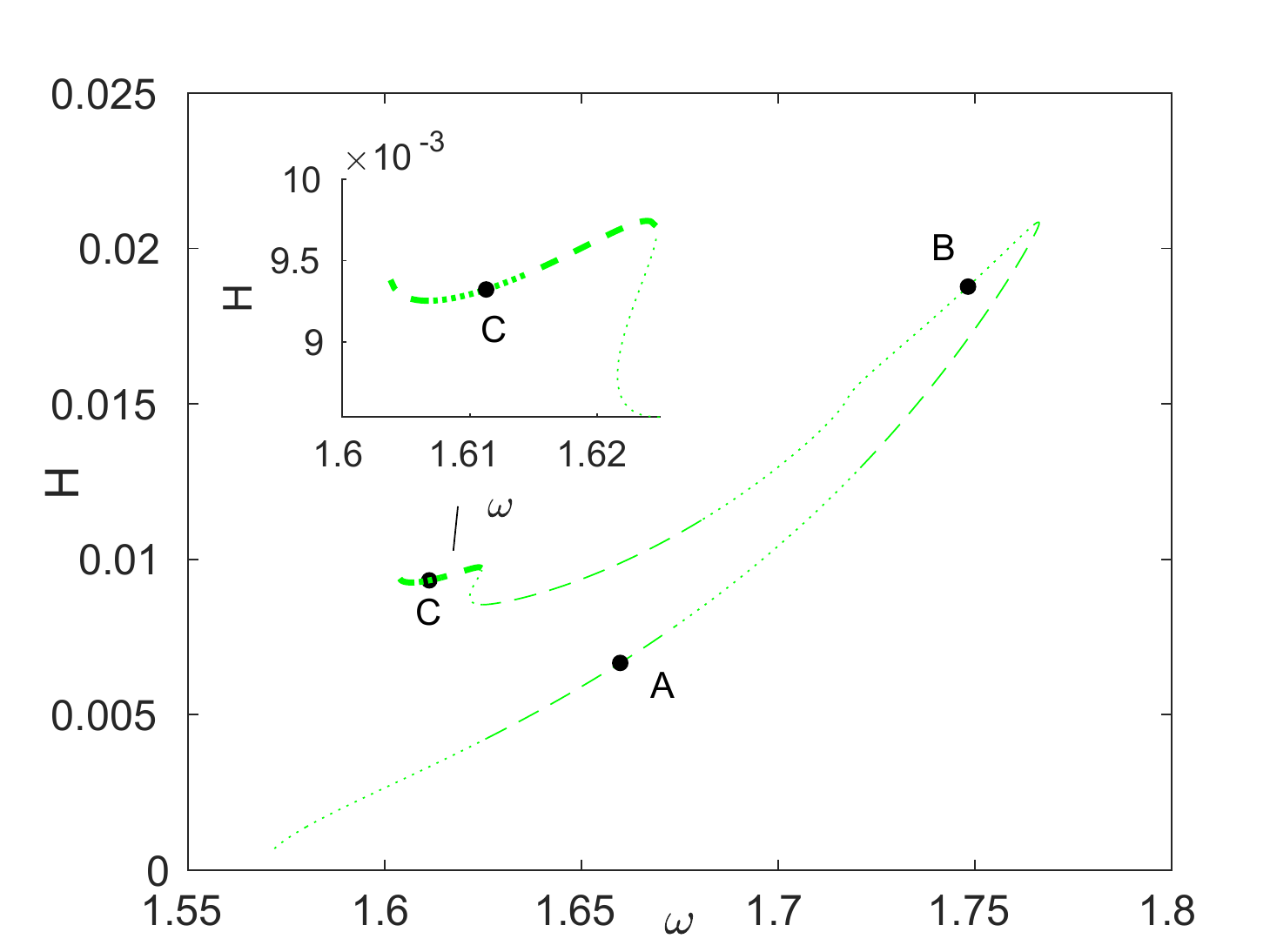}}
\subfloat[]
{\includegraphics[width=0.4\textwidth]{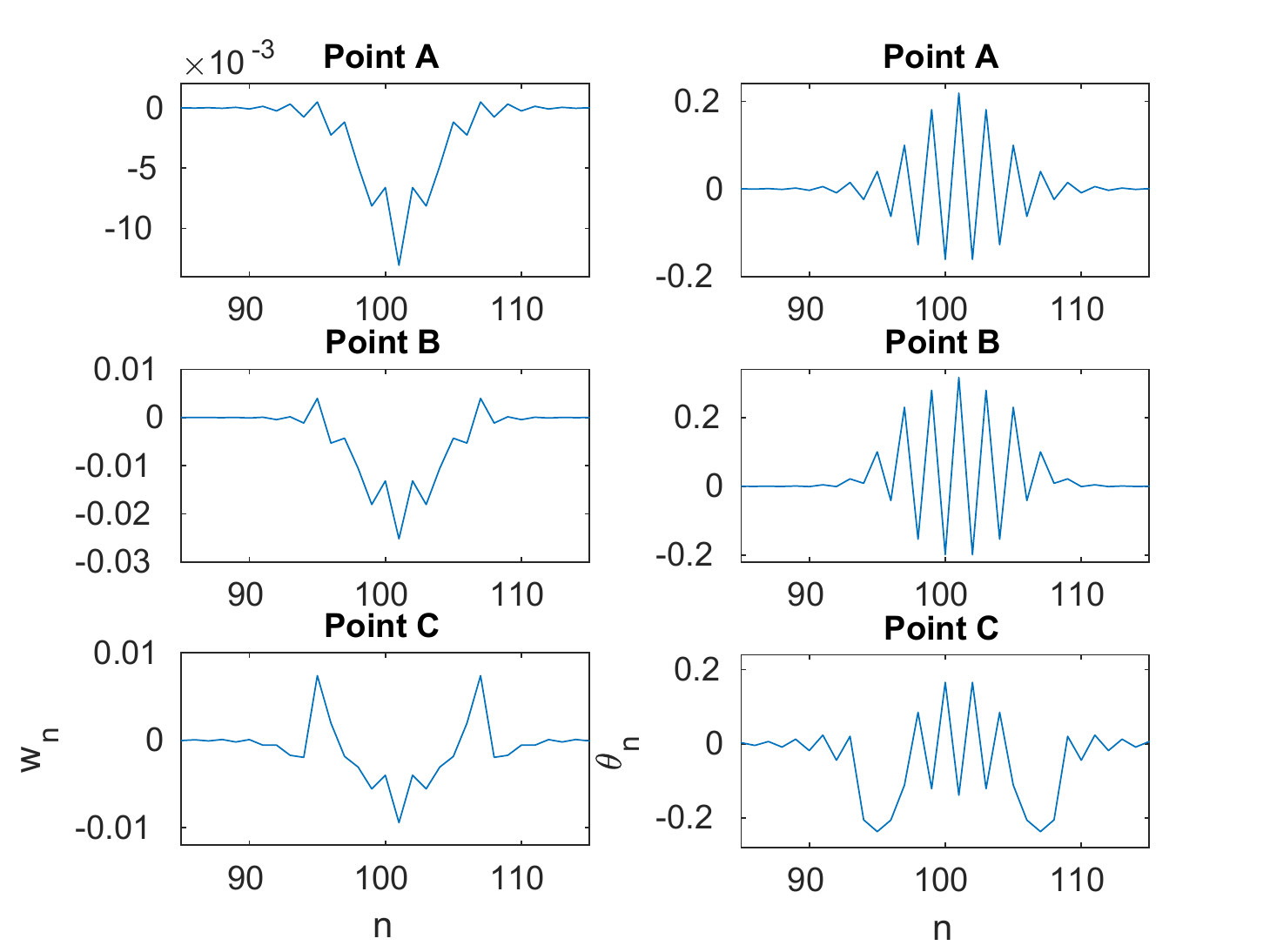}}
\caption{\footnotesize  (a) Energy $H$ as a function of frequency $\omega$ along the blue symmetric solution branch. (b) Strain and angle variables for the solutions at the points $A$, $B$, and $C$ in panel (a). (c) $H(\omega)$ along the red symmetric solution branch. The inset showing Floquet multipliers illustrates the emergence of an exponential instability. A pair of complex Floquet multipliers (red crosses) associated with a solution before the transition collides to form two positive real multipliers (blue crosses) associated with the solution after the collision. The corresponding symmetric multipliers inside the unit circle are not shown. (d) Strain and angle variables for the solutions at the points $A$, $B$, and $C$ in panel (c). (e) $H(\omega)$ along the green symmetric solution branch. The inset shows the enlarged view near the end of the computed branch. (f) Strain and angle variables for the solutions at the points $A$, $B$, and $C$ in panel (e). All solution profiles are shown at the time instances of maximal amplitude.}
\label{fig:SolutionEvolution_zero}
\end{figure}

Fig.~\ref{fig:SolutionEvolution_zero}, shows each of the branches (left panels) along with the evolution of the strain and angle variables along each curve (right panels). Along the blue branch shown in panel (a), the strain variable shown in panel (b) has a single peak at point $A$, which evolves to a single trough at point $B$, and then to a triple trough at point $C$. Meanwhile, the angle variable changes from a single trough at point $A$ to a double trough at point $B$, and finally to a quadruple trough at point $C$.

In the case of the red symmetric branch (panel (c)), the strain variables shown in panel (d) initially has a single peak at point $A$, which then evolves into a single trough at point $B$ and later to a double trough at point $C$. Meanwhile, the angular variable has a single trough at point $A$ and develops steps at point $B$, which subsequently evolve into a triple trough at point $C$. In this case too, as is the case for the blue branch, the expansion of the
solution to more sites bearing high amplitudes is associated with higher energies
along the snake-like solution branch.

For the green solution branch that extends to near the top of the acoustic band (panel (e)), we find that as we move from point $A$ to point $C$, the strain variable shown in panel (f) develops two peaks.
Notice that in this case, the point $A$ illustrates the provenance of this mode
from a $k=\pi$ band edge, since adjacent sites are out of phase with each other
at the starting point of the relevant branch in panel (f). In the angular variable, we observe a widening of the core from point $A$ to point $C$ along with the emergence of two troughs at point $C$.

As in the previous case discussed in Sec.~\ref{sec:pi_mode}, we expect the existence of an asymmetric solution branch connecting the red and blue branches and facilitating an exchange of the exponential instability shown in the inset in Fig.~\ref{fig:EnergyvFrequency_All_zero}. Due to the extremely narrow frequency and energy intervals over which this exchange takes place, we were unable to accurately compute the asymmetric solutions. Similar stability exchange through symmetry-breaking bifurcations is expected at other points where the emergence of an exponential instability is not caused by a collision of complex multipliers, as depicted in the inset of Fig.~\ref{fig:SolutionEvolution_zero}(c), or associated with splitting of a pair of real multipliers at $\mu=1$ when $H'(\omega)$ changes sign.

\section{Concluding remarks}
\label{sec:conclusions}

In this work we have revisited a dynamical system that constitutes a prototypical, experimentally tractable
example of a nonlinear mechanical metamaterial.
While earlier work \cite{Deng2018,Deng19,Deng21} on this system focused on the possibility of
its featuring propagating nonlinear excitations in the form of traveling
waves, the emphasis in this paper has been on the dynamics of discrete breathers
with parameters allowed by the experimental setting (in accordance, e.g.,
with the Supplemental material in~\cite{Deng2018}). To explore the DB
waveforms, we started with a systematic analysis of the linear spectrum
of the system. We ensured the presence of a gap between the acoustic
and optical branches of the linear dispersion relation. In addition, we
ensured the avoidance of resonances involving the second harmonic,
in order for the DBs to exist~\cite{macaub}. When the relevant
conditions applied, we were able to identify a rich set of families
of discrete breathers, both symmetric and asymmetric. This includes DB solutions bifurcating from or existing near the lower edge of the optical
band, as well as solution branches that extend to the upper edge of the acoustic
band. Utilizing the energy-vs-frequency representation of the
associated bifurcation diagrams, we were able to showcase
numerous solution branches, and importantly identified the
wealth of bifurcations emerging between them. These included
saddle-center bifurcations (leading to exponential instabilities),
symmetry-breaking bifurcations (involving asymmetric branches)
and finally Hamiltonian-Hopf bifurcations associated with the emergence
of complex multipliers. We also briefly discussed the nonlinear
evolution dynamics associated with different branch instabilities and
showed how these could lead to a restructuring of the waveforms towards
stable DB patterns, while shedding some dispersive wave radiation
as a result of the dynamical instability.

Naturally, we believe that this work paves the way for further
explorations of nonlinear wave structures in this class of metamaterial
lattices. The relevant possibilities emerge at different levels of experiment,
computation and theory. Experimentally, it remains to be seen
whether parametric regimes considered in this work allow for the identification of the discrete breather
waveforms examined in this work. Theoretically, we showed that
some of the obtained solutions bifurcate from the band edges of
the dispersion relation. This is a feature that calls for the analysis
of such a bifurcation via multiple-scale expansions and the possible
derivation of a nonlinear Schr{\"o}dinger type model to describe it, an effort
that is already underway \cite{Demiquel2022}. Lastly, it would be particularly
interesting to extend the relevant considerations of breathing
waveforms to (numerically) exact computations of discrete traveling
solutions along the lines of recent connections between the two types
of structures~\cite{Cuevas17}. Such studies are currently
in progress and will be reported in future publications.\\

\noindent {\bf Acknowledgements.} This work was supported by the U.S. National Science Foundation (DMS-1808956, AV and DMS-1809074, PGK).
JCM acknowledges support from EU (FEDER program2014-2020) through both Consejer\'{\i}a de Econom\'{\i}a, Conocimiento, Empresas y Universidad de la Junta de Andaluc\'{\i}a (under the projects P18-RT-3480 and US-1380977), and MCIN/AEI/10.13039/501100011033 (under the projects PID2019-110430GB-C21 and PID2020-112620GB-I00). The authors thank G. Theocharis and A. Demiquel for helpful discussions.


\end{document}